\documentclass[12pt,english]{report}
\usepackage{amssymb, amsmath, amsthm}
\usepackage{float}
\usepackage{xypic}
\usepackage[numbers]{natbib}
\usepackage{graphicx} %

\usepackage[left=1.7in, top=1.3in, right=1.3in, bottom=1.2in]{geometry}

\usepackage{setspace}
\usepackage{multirow}
\usepackage{threeparttable}

\usepackage{siunitx}
\usepackage{booktabs}
\usepackage{array}
\newcolumntype{C}[1]{>{\centering\arraybackslash}p{#1}}
    
\theoremstyle{definition}

\theoremstyle{remark}

\numberwithin{equation}{chapter}
\numberwithin{figure}{chapter}

\usepackage[T1]{fontenc}
\usepackage[latin9]{inputenc}
\usepackage{geometry}
\usepackage{color}
\usepackage{amssymb,amsmath}
\usepackage{graphicx}
\usepackage{setspace}
\PassOptionsToPackage{normalem}{ulem}
\usepackage{ulem}
\usepackage{secdot}

\usepackage{tikz}
\usetikzlibrary{positioning}
\usetikzlibrary{arrows}
\usetikzlibrary{shapes.arrows}
\usetikzlibrary{matrix}
\usepackage{cellspace}

\makeatletter

\@ifundefined{showcaptionsetup}{}{%
 \PassOptionsToPackage{caption=false}{subfig}}
\usepackage{subfig}
\makeatother

\usepackage{babel}
    \makeatletter
    \def\thebibliography#1{\chapter*{References\@mkboth
      {REFERENCES}{REFERENCES}}\list
      {[\arabic{enumi}]}{\settowidth\labelwidth{[#1]}\leftmargin\labelwidth
	\advance\leftmargin\labelsep
	\usecounter{enumi}}
	\def\newblock{\hskip .11em plus .33em minus .07em}
	\sloppy\clubpenalty4000\widowpenalty4000
	\sfcode`\.=1000\relax}
    \makeatother
 
\newcommand{\myfigure}[3]{ \begin{figure}[tbp]
		\begin{center}
			#1
		\end{center}
		\caption{#2} \label{#3} \vskip -0.14in \end{figure} }

\newcommand{\mytable}[3]{ \begin{table}[tbp]
		\begin{center}
			\caption{#2}
			#1
			\label{#3} \vskip -0.01in \end{center} \end{table} }

\begin{document}

\pagenumbering{roman}
{

{\singlespacing

\newpage
\thispagestyle{empty}
\begin{center}
{ 
UNIVERSITY OF OKLAHOMA
\par
\vspace{0.16in}
GRADUATE COLLEGE
\par
\vspace{1.2in}
\begin{doublespace}
RESILIENCE-BASED PERFORMANCE MODELING AND DECISION \par
 OPTIMIZATION FOR TRANSPORTATION NETWORK \par
\end{doublespace}
\vspace{0.17in}
\par
\vspace{1.2in}
A DISSERTATION
\par
\vspace{0.17in}
SUBMITTED TO THE GRADUATE FACULTY
\par
\vspace{0.17in}
in partial fulfillment of the requirements for the
\par
\vspace{0.17in}
Degree of
\par
\vspace{0.17in}
DOCTOR OF PHILOSOPHY
\par
\vfill
By
\par
\begin{singlespace}
	WEILI ZHANG\\
	Norman, Oklahoma\\
	2017
\end{singlespace}

}
\end{center}

\clearpage
\setcounter{page}{1}
\newpage
\thispagestyle{empty}
\begin{center}
{
RESILIENCE-BASED PERFORMANCE MODELING AND DECISION OPTIMIZATION FOR TRANSPORTATION NETWORK \par
\par

\vspace{0.9in}
\begin{singlespace}
	A DISSERTATION APPROVED FOR THE \\
	SCHOOL OF INDUSTRIAL AND SYSTEMS ENGINEERING
\end{singlespace}

\vspace{1.7in}
BY
\vspace{0.3in}
\begin{flushright}
\begin{tabular}{cr}
\hline
\  \  \  \  \  \  \  & Dr. Charles D. Nicholson, Chair\\
 \\ &  \\
 \\ &  \\
\hline
 & Dr. Naiyu Wang \\
\\   & \\
\\ &  \\
\hline
  & Dr. Bruce Ellingwood \\
\\ & \\
\\ &  \\
\hline
 & Dr. Theodore Trafalis \\
\\   & \\
\\ &  \\
\hline
& Dr. Andrés González \\
\end{tabular}
\end{flushright}
}
\end{center}

\newpage
\thispagestyle{empty}
\   \
\par
\vfill
\begin{center}
\copyright\ Copyright by WEILI ZHANG 2017

All Rights Reserved.
\end{center}
}

\newpage
\thispagestyle{empty}

\vspace*{\fill}
\begin{center}
\textit{This dissertation is dedicated to my wife, Shengjie Li, for her endless love, support and understanding.} 
\end{center}
\vspace*{\fill}
\newpage
\chapter*{Acknowledgments}
This work could not have been completed without the members of my doctoral committee. First and foremost, I would like to express my deepest gratitude to my advisor, Dr.Charles D. Nicholson, and  co-advisor, Dr.Naiyu Wang, for their excellent guidance, patience, encouragement and providing me with an excellent laboratory for doing research. It has been a privilege for me to work with them. They taught me all the necessary skills to be a good scientific researcher. I appreciate all their contributions of time and ideas that made my research journey productive and fascinating. Not only do their recommendations contribute to my research over the course of my studies, but their recommendations have also been a source of inspiration in many other aspects of my life. In addition, I gratefully acknowledge the funding sources provided by Dr.Nicholson and Dr.Wang that made my graduate studies possible. My research would not have been possible without their kind support.  

Many other people contributed to the development of this research. Particularly, I would
like to acknowledge Dr.Bruce Ellingwood (Colorado State University) and Dr.Theodore Trafalis for their support. I also like to thank Dr.Andrés González for his expertise.    

My time at the University of Oklahoma was made enjoyable in large part due to the many friends and groups that became a part of my life. I am grateful to my fellow students in the NIST-funded Center for Risk-Based Community Resilience Planning, to my friends at the University of Oklahoma, and especially to Peihui Lin, Xianwu Xue, Mohammad Tehrani, Yingjun Wang, Alexander Rodríguez Castillo. My time at the University of Oklahoma was also enriched by getting along with absolutely nice faculty members, and amazing graduate and undergraduate students. 

Finally, I would like to express by deepest and most heart felt gratitude to my wife,
Shengjie Li. No road is void of stumbling
block and pot holes, but my wife has always been there to pull me up whenever
I fell. This experience has required a lot of personal sacrifice from Shengjie, including
financial, physical and emotional. It was more than I had right to ask but something she
was willing to give. To her I not only give my thanks but also my love.

\begin{spacing}{0.1}
\tableofcontents{}
\listoftables{}
\listoffigures{}
\end{spacing}

\chapter*{Abstract}

The economy and social well-being of a community heavily rely on the availability and functionality of its critical infrastructure systems, including power, water, gas, and transportation.  Roadway networks are a fundamental component of transportation systems and, in the event of an extreme hazard, play a critical role during and after the event. Consequently, quantifying the performance of  transportation infrastructures and optimizing decisions to mitigate, prepare for, respond to, and recover from the potential hazards.
This research presented a novel resilience-based framework to support resilience planning regarding pre-disaster mitigation and post-disaster recovery. 

First, the author proposes a new performance metric for transportation network, weighted number of  independent pathways (WIPW), integrating the network topology, redundancy level, traffic patterns, structural reliability of network components, and functionality of the network during community's post-disaster recovery in a systematical way. To the best of our knowledge, WIPW is the only performance metric that permits risk mitigation alternatives for improving transportation network resilience to be compared on a common basis. Based on the WIPW, a decision methodology of prioritizing transportation network retrofit projects is developed. 

Second, our studies extend from pre-disaster mitigation to post-hazard recovery, in which this research presents two metrics to evaluate the restoration over the horizon after disasters . That is, total recovery time and the skew of the recovery trajectory. Both metrics are involved in the multi-objective stochastic optimization problem of restoration scheduling. The metrics provided a new dimension to evaluate the relative efficiency of alternative network recovery strategies.  The author then develops a restoration scheduling methodology for network post-disaster recovery that minimizes the overall network recovery time and optimizes the recovery trajectory, which ultimately will reduce economic losses due to network service disruption.
The WIPW, pre-disaster mitigation, and post-disaster recovery are illustrated in the same hypothetical bridge network with 30 nodes and 37 bridges subjected to a scenario seismic event.

Finally, a comprehensive stage-wise decision framework is introduced. The entire resilience planning is separated into three stages, pre-disaster mitigation, post-disaster emergency response, and long-term recovery. The WIPW is decomposed to three specific decision metrics to measure the performance of a network regarding robustness, redundancy, and recoverability, respectively. Decision support models for mitigation and recovery developed in the previous studies are revised to accommodate the stage-wise metrics. The proposed stage-wise framework is applied to a real-world roadway network of Shelby County, TN, USA subjected to seismic hazards.

{\singlespacing
}
\newpage

} 


\pagenumbering{arabic}

\doublespacing
\chapter{Introduction}
\label{ch_introduction}

\section{Background}
\label{sec_background}

Transportation networks play a vital role in ensuring the economic and social well-being of a community and the condition of such networks following the occurrence of an extreme hazard (e.g. earthquake, extreme wind storms, flood, terrorism, etc.) has a significant impact on the recovery of the community. The resilience of robust, large-scale, interdependent civil infrastructure networks, including transportation systems, utilities, telecommunication facilities, and social networks, individually and collectively play a major role in determining the resilience of a community as a whole.   The performance of transportation networks, in particular, is critical because post-disaster restoration of virtually all other facilities and lifelines in a community depends on people and equipment being able to move to the sites where damage has occurred.  Highway bridges typically are the vulnerable links in road transportation systems and require especially effective risk mitigation strategies aimed at improving the overall resilience of transportation systems against future natural disasters. 

 Highway bridges are vulnerable components in road transportation system, and robustness and recovery of the transportation network as a whole highly depends on their performance. Large-scale hazards can damage many bridges in a transportation system simultaneously, and the loss that results from this damage can be classified into two categories: initial direct loss caused by structural damage and indirect loss caused by downtime of the network before its full recovery. The initial loss is determined by the vulnerability of the network to the hazard event \citep{bruneau2003framework, chang2004measuring, zhang2016prediction, zhang2017bridge, nicholson2016optimal, zhang2018probabilistic, zhang2010lattice, zhang2016multi, zhang2010reformed, zhang2017resilience, zhang2016resilience}. The indirect loss due to downtime, which often is long-term and equally significant, largely depends on the overall recovery time and trajectory of the network. This chapter investigates bridge-road transportation network restoration schedules that minimize the recovery time and optimize the recovery trajectory of the network as a whole; such optimal schedules ultimately lead to reduced indirect economic losses resulting from the downtime of damaged road systems.

The economy and social well-being of a community heavily rely on the availability and functionality of its critical infrastructure systems, including power, water, gas, and transportation.  Roadway networks are a fundamental component of transportation systems and, in the event of an extreme hazard, play a critical role during and after the event.  For example, prior to the landfall of a hurricane, roadways are crucial for population evacuation; during a flood event, available roadways may provide key means of rescue; and during the longer-term recovery, the accessibility of schools, businesses, centers of government and commerce, etc. are significant elements of the economic recovery and social well-being of a community. However, the components of a roadway network, i.e. roads and bridges, are directly vulnerable to extreme hazard events; for instance, the Wenchuan earthquake in 2008 damaged 1,657 bridges in China \citep{zhuang2009earthquake}; Hurricane Irene in 2011 affected over 500 miles of highways, 2,000 miles of roadways, and 300 hundred bridges in Vermont \citep{vermont2012irene}. Such physical damage can lead to extensive and expensive functionality losses to the impacted community. For example, Hurricane Sandy in 2012 caused \$7.5 billion loss in direct damage to the New York transportation infrastructure (WABC-TV/DT, http://abc7ny.com/archive/8911130/), which does not included the indirect loss of lives, commerce, or other losses associated with the inability to effectively access emergency facilities or services, to route repair crews, or provide access to places of business. The flooding caused by hurricane Harvey prevented 9,000 victims from evacuation in Houston.  Enhancing transportation network resilience to these natural hazards has become a national imperative \citep{newman2011structure}.

The resilience of robust, large-scale, interdependent civil infrastructure networks, including transportation systems, utilities, telecommunication facilities, and social networks, individually and collectively play a major role in determining the resilience of a community as a whole.   The performance of transportation networks, in particular, is critical because post-disaster restoration of virtually all other facilities and lifelines in a community depends on people and equipment being able to move to the sites where damage has occurred.  Highway bridges typically are the vulnerable links in road transportation systems and require especially effective risk mitigation strategies aimed at improving the overall resilience of transportation systems against future natural disasters. 

In the dissertation, a new resilience-based performance metric of transportation is developed, which can be either used directly as an individual performance indicator for any transportation network or flexible to be decomposed to measure one of the aspects (i.e., robustness, redundancy, resourcefulness, and rapidity) commonly used in resilience. According to different scenarios, different metrics will be constructed and the corresponding optimization models will be proposed to resolve problems of mitigation optimization, project ranking mechanism for prioritizing transportation network retrofit projects, emergency response, post-disaster recovery scheduling, respectively. Finally, all the proposed mathematical models are illustrated on a virtual bridge network and a real-world transportation network of Shelby County, TN, US.

\section{Principal Goals}
\label{sec_goals}

The principal goals of the research are to develop sound resilience-based metric for transportation infrastructures, optimization models for risk mitigation, emergency response, and recovery. To achieve the goals, the specific research
tasks are given as follows: 

\begin{itemize}
	\item Perform literature review in each chapter about the corresponding topic.
	\item  Formulate a resilience-based metric to evaluate the transportation infrastructures.
	\item Develop an efficient optimization model to search the optimal retrofit planning in terms of improving the resilience-based metric.
	\item Formulate a two-dimension metrics to evaluate the post-disaster recovery trajectories.
	\item Model the recovery scheduling problem using stochastic optimization to find the most rapid and efficient trajectories.
	\item Build the stage-wise metric system to flexibly evaluate the performance in terms of the occurrence of  hazards.
	\item Develop the corresponding stage-wise decision framework to support pre-disaster planning, emergency response, and post-disaster recovery. 
	\item Integrate individual bridge fragility curve, network science, and Monte Carlo simulation in decision making.
	\item Demonstrate the proposed methodologies with both hypothetical and real-world regional transportation networks.
\end{itemize}

The study has important academic contribution and implications in measuring resilience of transportation systems and pre-disaster mitigation, disaster response and
long-term recovery for transportation systems under extreme events. With the proposed methodology, the researchers and practitioners are able to prepare strategic mitigation plans for transportation infrastructure systems, and to
model post-earthquake performance of transportation systems. The findings are beneficial for government agencies and emergency managers to evaluate the performance of transportation
systems and estimate losses induced from damaged bridges or road closures, to improve the systems' disaster resilience under economic constraints, and to evaluate the contingency plans
for transportation management.

\section{Organization of dissertation}
\label{sec_organization}

The remaining dissertation is organized as follows. 
In Chapter \ref{ch_WIPW}, the author proposes a resilience-based framework for mitigating risk to surface road transportation networks.  This study utilizes recent developments in modern network theory to introduce a novel metric based on system reliability and network connectivity to measure resilience-based performance of a road transportation network. The formulation of this resilience-based performance metric (referred in the chapter as WIPW), systematically integrates the network topology, redundancy level, traffic patterns, structural reliability of network components (i.e., roads and bridges) and functionality of the network during community's post-disaster recovery, and permits risk mitigation alternatives for improving transportation network resilience to be compared on a common basis.  Using the WIPW as a network performance metric, the author proposes a project ranking mechanism for identifying and prioritizing transportation network retrofit projects that are critical for effective pre-disaster risk mitigation and resilience planning.  The author further presents a decision methodology to select optimal solutions among possible alternatives of new construction, which offer opportunities to improve the resilience of the network by altering its existing topology.  Finally, this research concludes with an illustration that uses the WIPW as the performance metric to support resilience-based risk mitigation decisions using a hypothetical bridge network susceptible to seismic hazards.

Chapter \ref{ch_recovery} presents a novel resilience-based framework to optimize the scheduling of the post-disaster recovery actions for road-bridge transportation networks. The methodology systematically incorporates network topology, redundancy, traffic, damage level and available resources into the stochastic processes of network post-hazard recovery strategy optimisation. Two metrics are proposed for measuring rapidity and efficiency of the network recovery: total recovery time (TRT) and the skew of the recovery trajectory (SRT). The TRT is the time required for the network to be restored to its pre-hazard functionality level, while the SRT is a metric defined for the  first time in this chapter to capture the characteristics of the recovery trajectory that relates to the efficiency of those restoration strategies considered. Based on this two-dimensional metric, a restoration scheduling method is proposed for optimal post-disaster recovery planning for bridge-road transportation networks. To illustrate the proposed methodology, a genetic algorithm is used to solve the restoration schedule optimisation problem for a hypothetical bridge network with 30 nodes and 37 bridges subjected to a scenario seismic event. A sensitivity study using this network illustrates the impact of the resourcefulness of a community and its time-dependent commitment of resources on the network recovery time and trajectory.

Chapter \ref{ch_stage_wise} introduces a comprehensive stage-wise decision framework to support resilience planning for roadway networks regarding pre-disaster mitigation (Stage I), post-disaster emergency response (Stage II) and long-term recovery (Stage III).  Three decision metrics are first defined, each based on a derivation of the number of independent pathways (IPW) within a roadway system, to measure the performance of a network in term of its robustness, redundancy, and recoverability, respectively.   Using the three IPW-based decision metrics, a stage-wise decision process is then formulated as a stochastic multi-objective optimization problem, which includes a project ranking mechanism to identify pre-disaster network retrofit projects in Phase I, a prioritization approach for temporary repairs to facilitate immediate post-disaster emergency responses in Phase II, and a methodology for scheduling network-wide repairs during the long-term recovery of the roadway system in Phase III.  Finally, this stage-wise decision framework is applied to the roadway network of Shelby County, TN, USA subjected to seismic hazards,  to illustrate its implementation in supporting community network resilience planning. 

Finally, Chapter \ref{ch-conlusion} summarizes the entire dissertation and future work.







\chapter{Resilience-based risk mitigation for road networks}
\label{ch_WIPW}

\section{Introduction}
\label{sec_mitigation_introduction}

The resilience of a system is its ability to withstand or adapt to external shocks and to recover from such shocks efficiently and effectively \citep{timmerman1981vulnerability, pimm1984complexity}. In the case of civil infrastructure, resilience is often associated with four attributes \citep{bruneau2003framework, chang2004measuring}: robustness - the ability to withstand an extreme event and deliver a certain level of service even after the occurrence of that event; rapidity - the ability to recover the desired functionality as quickly as possible; redundancy - the extent to which elements and components of a system can be substituted for one another; and resourcefulness - the capacity to identify problems, establish priorities, and mobilize personnel and financial resources after an extreme event.  These attributes are illustrated in Figure \ref{concept_of_resilience}; all are characterized by considerable uncertainties.  Many research studies have discussed the resilience of systems other than civil infrastructure, including ecosystems \citep{holling1973resilience,walker2012resilience,leichenko2011climate}, computer networks \citep{simonis2006constraint}, communication networks\citep{sterbenz2010resilience,zhu2015game}, and socio-economic systems \citep{rose2007economic,martin2012regional}. 

\myfigure{
	\centering
	\includegraphics[width=\textwidth]{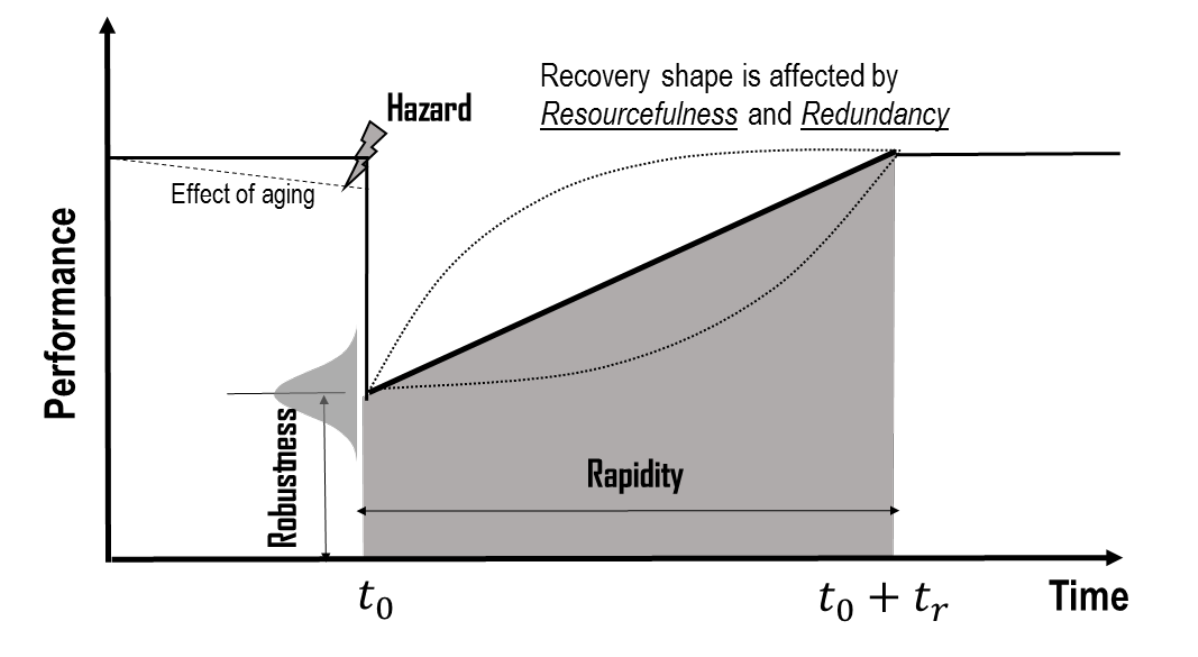}
}{Illustration of the Concept of Resilience}{concept_of_resilience}

\section{Literature review}
\label{sec_mitigation_literature}

As illustrated in Figure \ref{effect_mitigation}, any resilience-based analysis and decision require a quantitative measure of the system performance and condition (i.e. the vertical coordination in Figure \ref{effect_mitigation}).  While structural reliability (failure probability) is a well-accepted performance measure for individual roads and bridges to natural hazards, the performance of a transportation network must be measured by different metrics. Many researchers have quantified network performance based on network service functions, e.g., flow capacity \citep{nagurney2007network,lee2011post}, connectivity \citep{chen2002capacity,clark2005modelling,guikema2009reliability,bocchini2011connectivity,kurtz2015seismic}, and travel time \citep{asakura1995traffic,chen2007alpha, zhang2015travel}.  However, these metrics are mainly used to measure network performance under normal service conditions and are not effective in reflecting the network susceptibility to disruptive, low-probability high-consequence natural and man-made hazards or its resilience (earthquakes, floods, terrorist attacks, etc.).  More recently, \citet{peeta2010pre} used post-disaster connectivity and traversal cost between multiple origin-destination pairs in a network as the basis for pre-disaster investment decisions; \citet{morlok2004measuring} proposed capacity flexibility to reflect a transportation systems ability to adapt to changes in traffic patterns caused by natural disasters; \citet{chang2001measuring} introduced the notion of network coverage and transport accessibility as the performance measures for post-disaster network recovery; and \citet{ip2011resilience} suggested that pathway redundancy between all origin-destination pairs be used as a resilience measure for transportation networks.  These performance metrics all have their merits in quantifying the network performance under hazardous conditions, but none of them directly reflect the network resilience-based performance in terms of its ability to provide functionality to community following a disaster and to support the community recovery from hazard-induced interruptions.  Furthermore, none of these studies have attempted to quantify the uncertainties associated with these performance metrics.   

\myfigure{
	\centering
	\includegraphics[width=\textwidth]{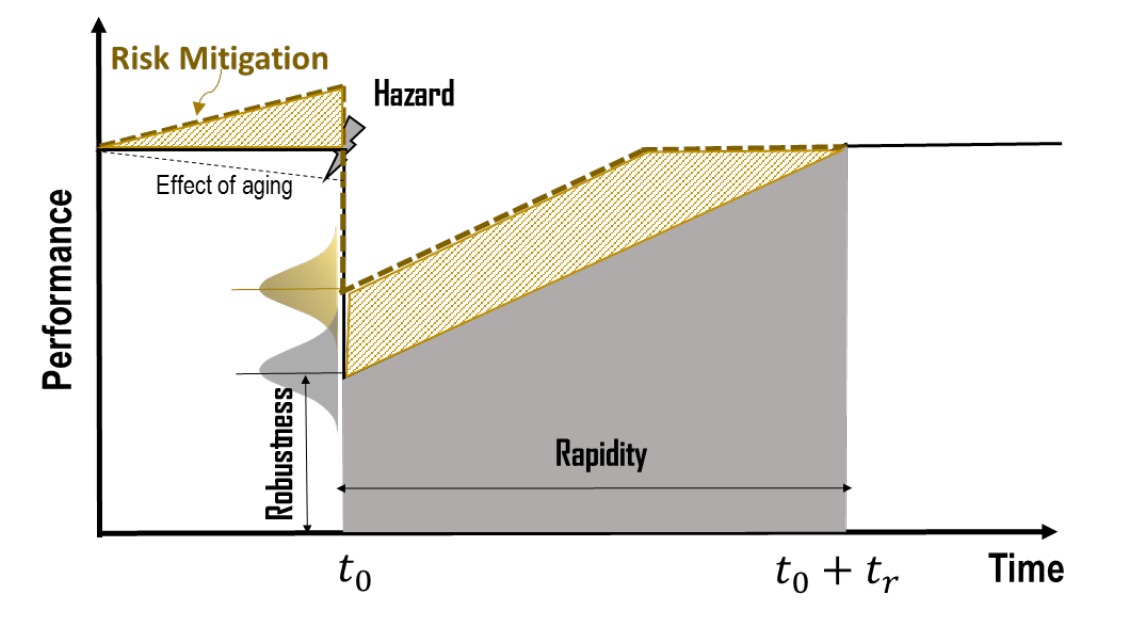}
}{
}{effect_mitigation}

\section{Highlights}
\label{sec_mitigation_highlights}

This chapter proposes a novel resilience-based performance metric for road transportation networks, which allows resilience-based risk mitigation alternatives to be measured and compared on a common basis.  The performance metric is based on graph theory, in a formulation which systematically integrates the network topology, system redundancy, traffic patterns, reliability (failure probability) of network components (i.e. bridges and roads) and the network functionality in a community's immediate post-disaster recovery period.  Based on this resilience-based performance metric, the author next introduces a project ranking mechanism for identifying and prioritizing bridge retrofit projects that are critical for effective pre-disaster risk mitigation of road transportation networks.  This investigation provides a decision methodology to select optimal solutions among possible alternatives of new construction which offer opportunities to improve network resilience by altering its existing topology.  This chapter concludes with an illustration of a resilience-based risk mitigation framework, considering a hypothetical networked system of 37 bridges that are susceptible to seismic hazard.

\section{Resilience-based network performance metric, WIPW}
\label{sec_WIPW}

The fundamental purpose of a transportation system is to carry traffic from origins to destinations. The resilience of such a system is reflected in its ability to continue to fulfill this purpose in the event of natural or man-made disasters. Extreme hazard events can damage many bridges and roads simultaneously in a local transportation network, and financial and human resources required to restore the network function often are not immediately available following the disaster. Thus, the existence of redundant alternative paths between network origin-destination (O-D) pairs is crucial for the continued function of the transportation system during the period of emergency response immediately following the disaster as well as the long-term recovery of the community, and is an essential characteristic of a resilient transportation network.

Accordingly, by extending the concept suggested by Ip and Wang (2011), this author defines a resilience-based performance metric of a transportation system as the weighted average number of reliable independent pathways (WIPW) between any network O-D pairs. A pathway between an O-D pair usually consists of several links that represent roads, with or without a bridge, which are connected in series. Two pathways between the same O-D pair are considered as independent pathways (IPW) if they do not share any common road links. Although the number of pathways between an O-D pair can be very large, the number of IPWs is often very limited.  The process of identifying all IPWs in a transportation network will be discussed later.   Note that IPWs for different O-D pairs or even between a same O-D pair may not have identical impacts on the network performance. The resilience-based performance metric, as formulated subsequently and referred as WIPW, includes a novel and systematic weighting mechanism to quantitatively reflect the contribution of each IPW to the overall network performance under hazard consideration.    

Introducing the terminology of graph theory \citep{gibbons1985algorithmic}, let $G=(V,A)$ denote the road network, where $V=\{1,2,...,n\}$ is the set of nodes that represents major road intersections and economic hubs and key destinations in a community, and $A=\{1,2,...,m\}$ is the set of arcs that represent roads either without a bridge or with a maximum of one bridge. The network performance metric, $WIPW$, as defined above, can be written as,

\begin{align}
WIPW(G) = \sum_{i=1}^{n}w_{i}r_{i}, \label{eq_wipw}
\end{align} 
where $w_{i}$ is the weighting factor applied to individual node $i\in V$, $\sum_{i=1}^{n}w_{i}=1$, and $r_i$ denotes the average number of reliable IPWs between node $i$ and any other $n-1$ nodes in the network, as expressed in Eq. \eqref{eq_ipw},

\begin{align}
r_i = \frac{1}{n-1}\sum_{j=1,j\neq i}^{n}\sum_{k=1}^{K_{(i,j)}}w_{k}(i,j)R_{k}(i,j) \label{eq_ipw}
\end{align}
in which $K_{(i,j)}$ represents total number of IPWs between nodes $i$ and $j$; $R_{k}(i,j)$ is the reliability of IPW $P_{k}(i,j)$; the $k$th IPW between node $i$ and $j$; $w_{k}(i,j)$ denotes the weighting factor applied to IPW $P_{k}(i,j)$, and for all $K_{(i,j)}$ IPWs between nodes $i$ and $j$, $\sum_{k=1}^{K_{(i,j)}}w_{k}(i,j)=K_{(i,j)}$. Weighting facotrs $w_i$ and $w_{k}(i,j)$ will be discussed in details below. Each IPW, $P_{k}(i,j)$, usually consists of several road links connected in series. Let $l$ denote the individual road links and $q_l$ denote the reliability of $l$; thus, for a series that consists of independent links, the system reliability of the series is the product of the reliabilities of all road links included in $P_{k}(i,j)$,
\begin{align}
R_{k} = \prod_{\forall l \in P_{k}(i,j)}q_{l} \label{eq_reilibilityofipw}
\end{align}

For a system in which the component performance are positively correlated, which is likely the case for transportation networks (further discussion is provided in Section \ref{sec_bridge_reliability}), Eq. \eqref{eq_reilibilityofipw} provides a lower bound on system reliability. The exact system reliability $R_{k}(i,j)$  can be estimated through Monte Carlo simulation using a Gaussian Copula to model correlation. Combining Eqs. \eqref{eq_wipw} and \eqref{eq_ipw}, the resilience-based performance metric of the road network, $WIPW$, becomes,
\begin{align}
WIPW(G) = \sum_{i=1}^{n}w_{i}\frac{1}{n-1}\sum_{j=1,j\neq i}^{n}\sum_{k=1}^{K_{(i,j)}}w_{k}(i,j)R_{k}(i,j). \label{eq_wipw_full}
\end{align}

Two weighting factors, $w_{i}$ and $w_{k}(i,j)$, appear in Eq. \eqref{eq_wipw_full}. Factor $w_{i}$ applies to nodes; it is inversely proportional to the shortest distance from node $i$ to the nearest emergency response facility in the community, reflecting the relative importance of the node $i$ being connected in the context of community post-disaster emergency response. Let $E$, a subset of $V$, denote the set of nodes in the network where emergency response facilities are located, $N$ the set of nodes that do not belong to $E$, and $L_{P_{k}(i,j\in E)}$ is equal to the length of $P_{k}(i,j)$ where $j\in E$. $w_{i}$ is evaluated as,

\begin{align}
w_{i}=\frac{\omega_{i}}{\sum_{j=1}^{n}\omega_{i}} \label{eq_weight}
\end{align}
where
\begin{align}
\omega_{i} = \begin{cases}
\frac{1}{min\{L_{P_{k}(i,j\in E)}|k=1,2...K_{(i,j\in E)}\}}, i\in N\\
1, i\in E
\end{cases}
\forall i\in V. \label{weigh_assign}
\end{align}
As noted previously, the sum of $w_i$ for all the nodes in the network equals 1.

The other weighting factor $w_{i}$ in Eq. \eqref{eq_wipw_full} applies to IPWs; it is related to both the average daily traffic (ADT) and the length of the IPW, and reflects the relative impact that this pathway has on people's normal life activities and the local economy. Pathways between any given O-D pair that has shorter length and carries larger traffic flow contribute more to the network functionality and should be weighted more heavily in quantifying the network resilience.  ADT data is often readily available with federal, state or local bridges owners, or can be estimated using traffic assignment models.  Let $T_{i}$ denote the ADT of road link $l\in P_{k}(i,j)$. Define $T_{P_{k}(i,j)}$, the ADT of IPW $P_{k}(i,j)$, as the minimum ADT of all road links on that pathway:
\begin{align}
T_{P_{k}(i,j)} = \text{min}[T_{l}|l\in P_{k}(i,j)] \label{ADT_path}
\end{align}
The normalized ADT of the path is then defined as,
\begin{align}
T'_{P_{k}(i,j)} = \frac{T_{P_{k}(i,j)}}{\sum_{k=1}^{K_{(i,j)}}T_{P_{k}(i,j)}}K_{(i,j)} \label{eq_normalized_ADT}
\end{align}

Note that for any node pair $(i,j)$, $\sum_{k=1}^{K_{ij}}T'_{P_{k}(i,j)} = K_{(i,j)}$. Similarly, let $L_l$ denote the length of the road link $l$; then the length of the IPW $P_{k}(i,j)$ is  simply the summation of the lengths of all road links within that path,
\begin{align}
L'_{P_{k}(i,j)} = \sum_{l\in P_{k}(i,j)}L_{l}\label{normalized_length}
\end{align}
Finally, let $L_{max}(i,j)$ denote the maximum of all $L_{P_{k}(i,j)}$ for a given O-D pair $(i,j)$; the the normalized length of the path is,
\begin{align}
L'_{P_{k}(i,j)} = \frac{L_{max}(i,j)}{L_{P_{k}(i,j)}\sum_{k=1}^{K(i,j)}(L_{max}(i,j)/L_{P_{k}(i,j)})}K_{(i,j)} \label{eq_normalzied_length}
\end{align}

Note that for any node pair $(i,j)$, $\sum_{k=1}^{K_{(i,j)}}L'_{P_{k}(i,j)}=K_{(i,j)}$. Using Eqs. \eqref{eq_normalized_ADT} and \eqref{eq_normalzied_length}, the aggregated pathway weighting factor $w_{k}(i,j)$ is defined as, 
\begin{align}
w_{k}(i,j) = uL'_{P_{k}(i,j)} + (1-u)T'_{P_{k}(i,j)}\label{weighted_weight}
\end{align}
where $u\in [0,1]$ is a weighting factor to impose the relative importance between the pathway length and its ADT. A community (or government decision makers) can assign different valus to $u$ based on their preferences in order to obtain the ``best'' measure to their specific situation. For the illustration presented in the subsequent section, uniform weights are applied, i.e., $u=0.5$. Note that the summation of all $w_{k}(i,j)$ for a given O-D pair $(i,j)$ equals $K_{(i,j)}$, the total number of IPWs between nodes $i$ and $j$.

The node weighing factor  and IPW weighting factor, so defined, not only ensure that all nodes and links in the network are properly weighted in the resilience-based performance metric based on their individual attributes (topology, traffic patterns, and functionality during a community's post-disaster recovery as well as structural reliability of individual bridges), but also preserve the physical meaning of the metric WIPW - the weighted average number of reliable IPWs between all O-D pairs in the road network.  The role that a transportation system plays before, during and after a disaster varies uniquely for every community. Communities of different size, population and social-economic attitudes and vulnerabilities are likely to show different values and preferences when evaluating the performance of their transportation systems. The weighting mechanism, formulated as above, provide a transparent framework to incorporate and properly weigh other network attributes in addition to those discussed herein that might be valuable to a specific community. 

Figure \ref{evaluation_wipw} displays the algorithm for computing WIPW. The identification of IPWs between many O-D pairs is non-unique, depending on the algorithm or process used to search for IPWs. To mitigate this problem, the author applies Dijkstra's algorithm \citep{skiena1990dijkstra}, as highlighted in Figure \ref{evaluation_wipw} with a dashed-line box, to search for a succession of independent shortest paths. 

\myfigure{
	\centering
	\includegraphics[width=\textwidth]{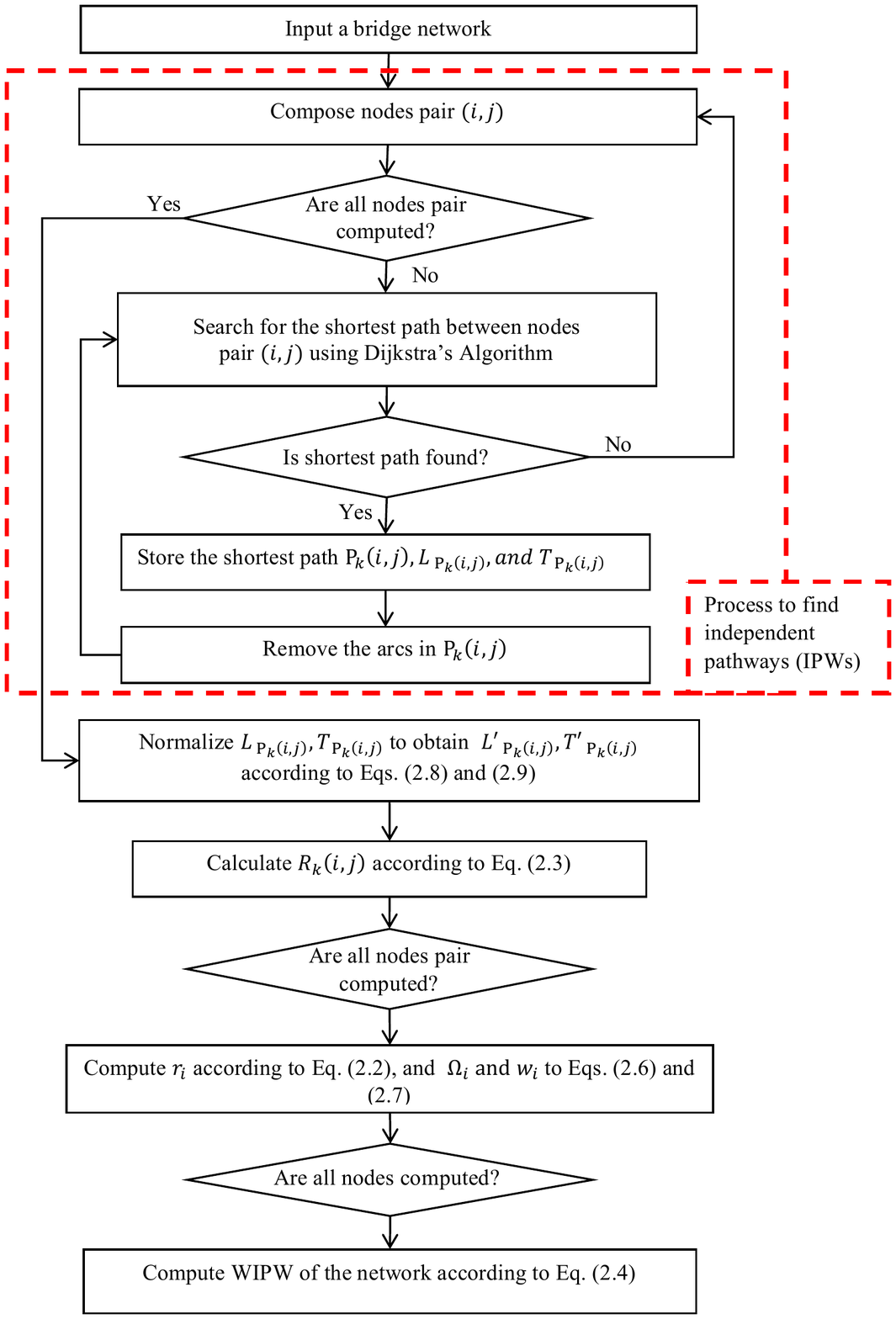}
}{Evaluation of network resilience-based performance metric, WIPW}{evaluation_wipw}

\section{Reliability of bridges in a community transportation system}
\label{sec_bridge_reliability}

An important step to quantify the WIPW is to properly assign link (i.e. road segments with or without a bridge) reliabilities.  The reliability of an individual bridge or road segment can be evaluated using fragility analyses associated with damage states of interest (e.g. using platform HAZUS-MH MR4 \citep{dhs2009mr4}).  The calculated WIPW is therefore hazard-specific because the link reliabilities are hazard specific.  It should be recognized that hazardous events with large footprints introduce spatial and temporal correlations to the demands on community infrastructure systems \citep{adachi2008serviceability,jayaram2009correlation}. Common construction practices and code enforcement within a community also introduce positive correlation in structural response above and beyond that introduced by the hazard \citep{vitoontus2013role,bonstrom2015building}.  Such correlation structures depend on the stochastic variability in the hazard demand, the locations of bridges and roads, and their susceptibility to damage if the hazardous event occurs, and need to be taken into consideration in the probabilistic evaluation of the network performance \citep{lee2007uncertainty,bocchini2011connectivity}.

Bridge fragility assessment is an important ingredient for seismic risk assessment of transportation infrastructure \citep{padgett2009retrofitted}.  The current research is focused on transportation network performance assessment.  Because such networks may contain many different types of bridges,  bridge fragility modeling is outside the scope of the current effort.  However, a practical assessment of the integrity of any transportation network should contain specific fragility models for the key bridges in the network.  In the subsequent examples, the author uses plausible estimates of failure probabilities for generic bridge types.

\section{Formulation of optimal risk mitigation strategies}
\label{sec_mitigation}

The quantitative resilience-based performance metric, WIPW, as formulated in Section \ref{sec_WIPW}, can be employed to evaluate the transportation network performance, and it can also be incorporated in a decision framework to provide a common basis for evaluating and comparing alternative risk mitigation strategies on a common and rational basis.  Possible pre-disaster risk mitigation strategies to improve road network resilience include retrofitting existing bridges (links) or building new bridges. In either case, the formulation of the decision process is the same:  to make selections from a set of candidate links, representing either existing bridges and road segments or potential new construction, to maximize the network performance WIPW and, simultaneously, to minimize the associated cost. 

Suppose the set of candidates is represented by $S=\{s_1,s_2,...,s_s\}$ and the corresponding cost of each is $c=\{c_1,c_2,...,c_s\}$. Let $x_{t} \in X$, where $t=1,2,...,s$, denote the decision variables as below,

\begin{align}
x_{t} = \begin{cases}
1, s_{t}\in S \text{ is selected}\\
0, otherwise
\end{cases}. \label{decision_var}
\end{align}

Since the risk mitigation strategy (or decision) represented by $X$ will upgrade the existing network, all the parameters of the upgraded network are rewrote in the form of argument $X$, e.g., $WIPW(G(X))$, $w_{i}(X)$, $r_{i}$ and $P_{k}(i,j|X)$. Furthermore, some network parameters are uncertain and can be treated as random variables in the problem formulation; these variables will be represented with argument $\xi$. Thus, the first objective of the decision process is to maximize the network performance metric WIPW:

\begin{align}
\text{max } WIPW(G(X,\xi)) = \sum_{i=1}^{n}w_{i}(X)r_{i}(X,\xi) \label{max_wipw}
\end{align}

Let $\theta$ denote the cost associated with decision $X$; the second objective function is to minimize the total cost:

\begin{align}
\text{min } \theta(X,\xi) = \sum_{t=1}^{s}c_{t}(\xi)x_{t} \label{min_costs}
\end{align}

Eqs. \eqref{max_wipw} and \eqref{min_costs} pose a nontrivial multi-objective optimization problem, and a single solution that simultaneously optimizes these two competing objectives, i.e., maximizes WIPW and minimizes cost associated with mitigation strategies, does not exist.  However, a (possibly infinite) number of Pareto-optimal solutions do exist, which allows the tradeoff between the competing objectives and the subjective preferences of a decision maker to be factored into the decision process.  To compute the objective function of Eq. \eqref{max_wipw} requires the algorithm described in Figure \ref{evaluation_wipw}, a non-closed form formulation, which requires metaheuristic techniques to search for near-optimal solutions.  Accordingly, this research uses a Non-dominated Sorting Genetic Algorithm II (NSGA-II) \citep{deb2002fast} to search for the Pareto frontier; NSGA-II have been successfully applied to search for near-optimal solutions of similar network problems \citep{pishvaee2010memetic,mesbah2011optimization,keliy1996meta}.   GA is couped with Monte Carlo Simulation (MCS) to take uncertainties into consideration in the optimization process in the subsequent case study.  Since the parameter values have significant effects on the performance of the NSGA-II \citep{eiben2011evolutionary}, rigorous tests on parameter tuning are performed, resulting in the mutation parameter, crossover rate and population size in the GA to be set to 0.1, 0.7 and 100, respectively. The maximum number of iterations is 1,000, and the early termination criterion is 50, which means the program will stop if no better solution is found in consecutive 50 iterations.

\section{Illustration - risk-based mitigation decisions of
transportation network exposed to seismic hazards}
\label{sec_mitigation_illustration}

In this section, the role of the network resilience-based performance metric and the application of the decision methodology for risk mitigation are illustrated with a hypothetical road network exposed to a severe earthquake.  Two scenarios are discussed: (1) establishing priorities for pre-disaster retrofitting bridges that are critical for transportation network resilience; and (2) selecting among possible alternatives of new constructions which offer opportunities to improve the resilience of the network by altering its existing topology.  

\subsection{Road/bridge network}
\label{subsec_network}

Figure \ref{hypothetical_network} illustrates a hypothetical community road system, with 37 links representing the roads and 30 nodes representing the major road intersections and economic hubs. The community emergency response facilities (e.g. fire stations, hospitals, police, etc.) are located at Nodes 9 and 17.  For simplicity, every road is assumed to contain exactly one bridge. This assumption can be easily relaxed, if necessary.  If all the bridges in the network are in ``as new'' condition and reliabilities under service loads are assumed equal to 0.999, the upper bound of WIPW is 1.70, which means that on average there are 1.70 reliable IPW between any O-D pair in the road network under normal operational conditions. 

\myfigure{
	\centering
	\includegraphics[width=\textwidth]{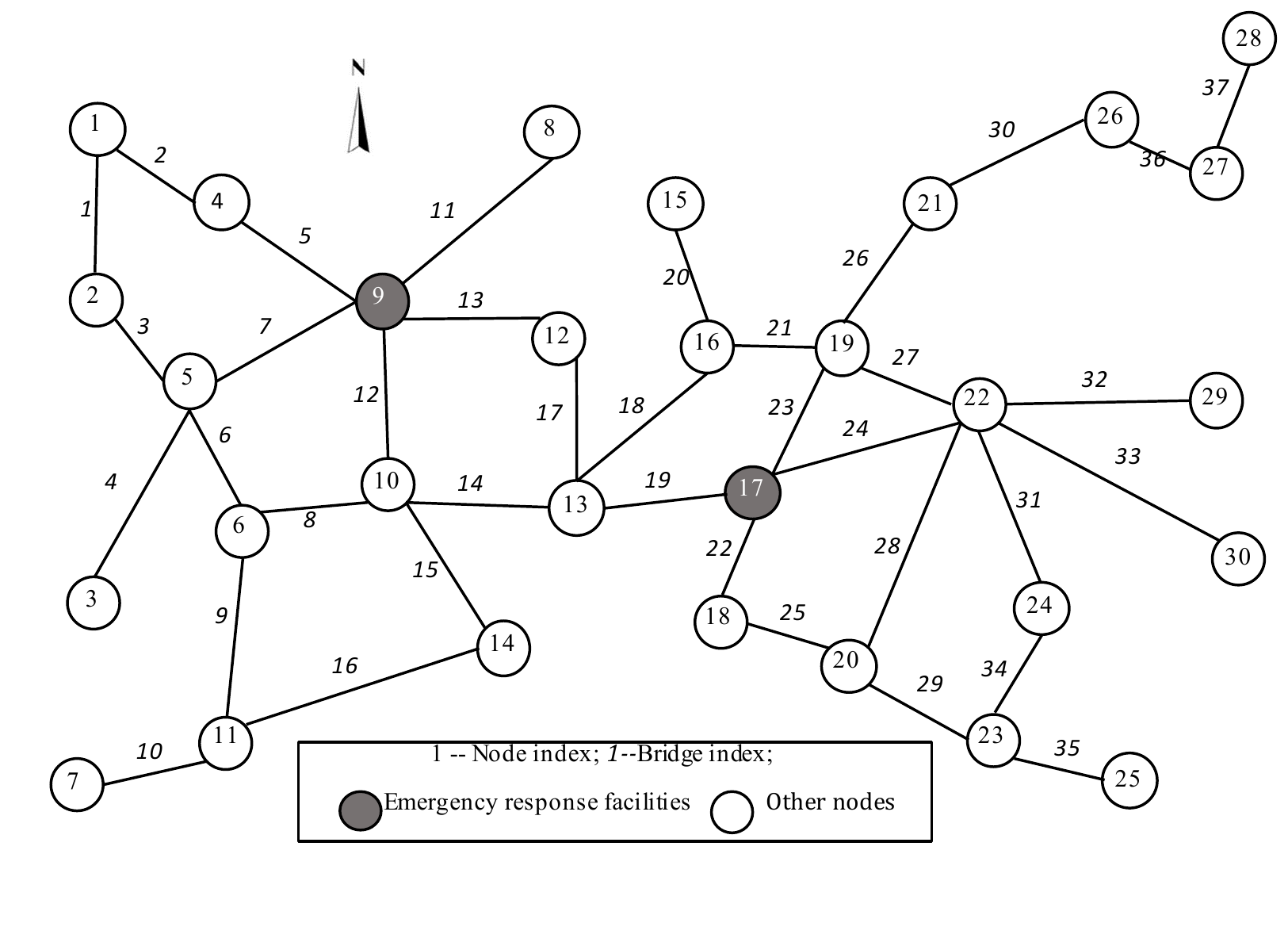}
}{Hypothetical bridge network}{hypothetical_network}

It is assumed that a  severe earthquake occurs, with magnitude MW equal to 7 and epicentral distance approximately 40 km from the centroid of the network (close to Node 13).  For illustrative purposes, it is assumed that out of the 37 bridges, 19 are steel (S) bridges and 18 are reinforced concrete (RC) bridges.   To reflect the diversity of bridge construction (including bridge configuration, material and area) within the network, it is assumed that the mean reliabilities of the 19 steel bridges and 17 RC bridges under the considered seismic action are 0.75 and 0.65, respectively.   Furthermore, to reflect the epistemic uncertainty associated with the estimate of bridge reliability, the reliability for each bridge is described by a normal distribution, with a mean value as tabulated in Table \ref{mean_network_parameter} and a coefficient of variation of 0.07 for all bridges; the implication of this assumption, often made in probabilistic risk analysis, is that the reliability of each bridge in this system can be estimated to be within $\pm 15\%$ with more than $95\%$ confidence.  This modeling approach results in a bridge portfolio that represents a diversity of construction practices. Correlations exist in the demand and capacity of these bridges as discussed in Section \ref{sec_bridge_reliability} due to common hazard and similar construction code and practices. These correlations are modeled as exponentially decreasing with the increase in the separation distance between bridges, i.e. $\rho_{ij}=\exp(-a_{ij}/L_{c})$, where $a_{ij}$ is the separation distance between bridges $i$ and $j$, and $L_{c}$ is the correlation length which is set to be the largest distance between two bridges in the network.   This simple correlation model reflects the fact that correlation in bridge performance due  to both seismic demand and common construction practices is likely to diminish as their separation distance increases.  In addition, the retrofit cost for each bridge is also modeled with a normal distribution, assuming that the mean cost is a function of the bridge deck area and its structural reliability \citep{fragkakis2004quantity}, and coefficient of variation is $8\%$.   The mean reliability, mean retrofit cost and mean ADT of each bridge are tabulated in Table \ref{mean_network_parameter}.  The distributions used to model the uncertainty associated with these parameters are summarized in Table \ref{statistics_network_parameter}.

\mytable{
	\begin{tabular}{c  c c  c c}
		\hline
	Bridge ID	&	Construction Type	&	Reliability	&	ADT (Vehicle/Hour)	&	Cost (Unit)	\\
		\hline
		1	&	RC	&	0.66	&	2200	&	3.57	\\
2	&	RC	&	0.76	&	1900	&	3.82	\\
3	&	S	&	0.82	&	2000	&	4.34	\\
4	&	S	&	0.88	&	1500	&	4.17	\\
5	&	RC	&	0.55	&	1900	&	4.87	\\
6	&	S	&	0.84	&	2200	&	3.49	\\
7	&	S	&	0.77	&	700	&	4.41	\\
8	&	S	&	0.82	&	2400	&	3.74	\\
9	&	S	&	0.77	&	2600	&	2.61	\\
10	&	S	&	0.85	&	300	&	3.55	\\
11	&	S	&	0.84	&	800	&	2.53	\\
12	&	RC	&	0.71	&	900	&	5.29	\\
13	&	S	&	0.89	&	2500	&	4.78	\\
14	&	S	&	0.78	&	600	&	3.25	\\
15	&	RC	&	0.77	&	2000	&	3.43	\\
16	&	S	&	0.78	&	500	&	4.33	\\
17	&	RC	&	0.61	&	2500	&	3.14	\\
18	&	RC	&	0.79	&	2800	&	2.98	\\
19	&	S	&	0.8	&	1300	&	2.88	\\
20	&	S	&	0.75	&	1700	&	3.28	\\
21	&	S	&	0.89	&	1500	&	3.98	\\
22	&	S	&	0.81	&	1200	&	4.82	\\
23	&	RC	&	0.76	&	1500	&	3.24	\\
24	&	S	&	0.75	&	700	&	4.8	\\
25	&	S	&	0.78	&	1800	&	3.8	\\
26	&	S	&	0.75	&	900	&	4.66	\\
27	&	S	&	0.8	&	600	&	4.46	\\
28	&	RC	&	0.71	&	800	&	3.33	\\
29	&	RC	&	0.65	&	1400	&	4.86	\\
30	&	RC	&	0.67	&	2800	&	3.45	\\
31	&	RC	&	0.69	&	1900	&	3.08	\\
32	&	RC	&	0.75	&	2900	&	3.74	\\
33	&	RC	&	0.79	&	1300	&	4.5	\\
34	&	RC	&	0.69	&	900	&	4.47	\\
35	&	RC	&	0.72	&	2200	&	3.36	\\
36	&	RC	&	0.83	&	700	&	4.46	\\
37	&	RC	&	0.73	&	3000	&	5.15	\\
		\hline		
	\end{tabular}
}{Mean values of network parameters } {mean_network_parameter}

\mytable{
	\begin{tabular}{c  c c  c c}
		\hline
		Parameters	&	Notation	&	Distribution	&	Mean	&	COV	\\
			\hline		
		Individual Bridge Reliability	&	$q$	&	Normal	& In Table \ref{mean_network_parameter}		&	0.07	\\
	Average Daily Traffic (ADT)	&	$T$	&	Uniform	&	In Table \ref{mean_network_parameter}	&	0.05	\\
	Retrofit Cost	&	$c$	&	Normal	&	In Table \ref{mean_network_parameter}	&	0.08	\\
\hline
		\end{tabular}
	}{Statistics of network parameters } {statistics_network_parameter}

Figure \ref{hist_wipw_no_retrofit} illustrates the histogram of the network WIPW under the considered hazard scenario, obtained from 1,000 Monte Carlo simulations of the network performance, which was found through experimentation to yield a stable estimate of the mean WIPW.  The mean residual WIPW of the current network without any risk mitigation actions is 0.61, and its coefficient of variation (COV) is 0.22.   That is, if the scenario earthquake were to occur, the average number of weighted independent pathways between all O-D pairs is less than 1, meaning that some areas in the community will become isolated from one another following the scenario earthquake.  This analysis provides a baseline for comparing the effectiveness of the alternative risk mitigation strategies to be evaluated in the subsequent section.

\myfigure{
	\centering
	\includegraphics[width=\textwidth]{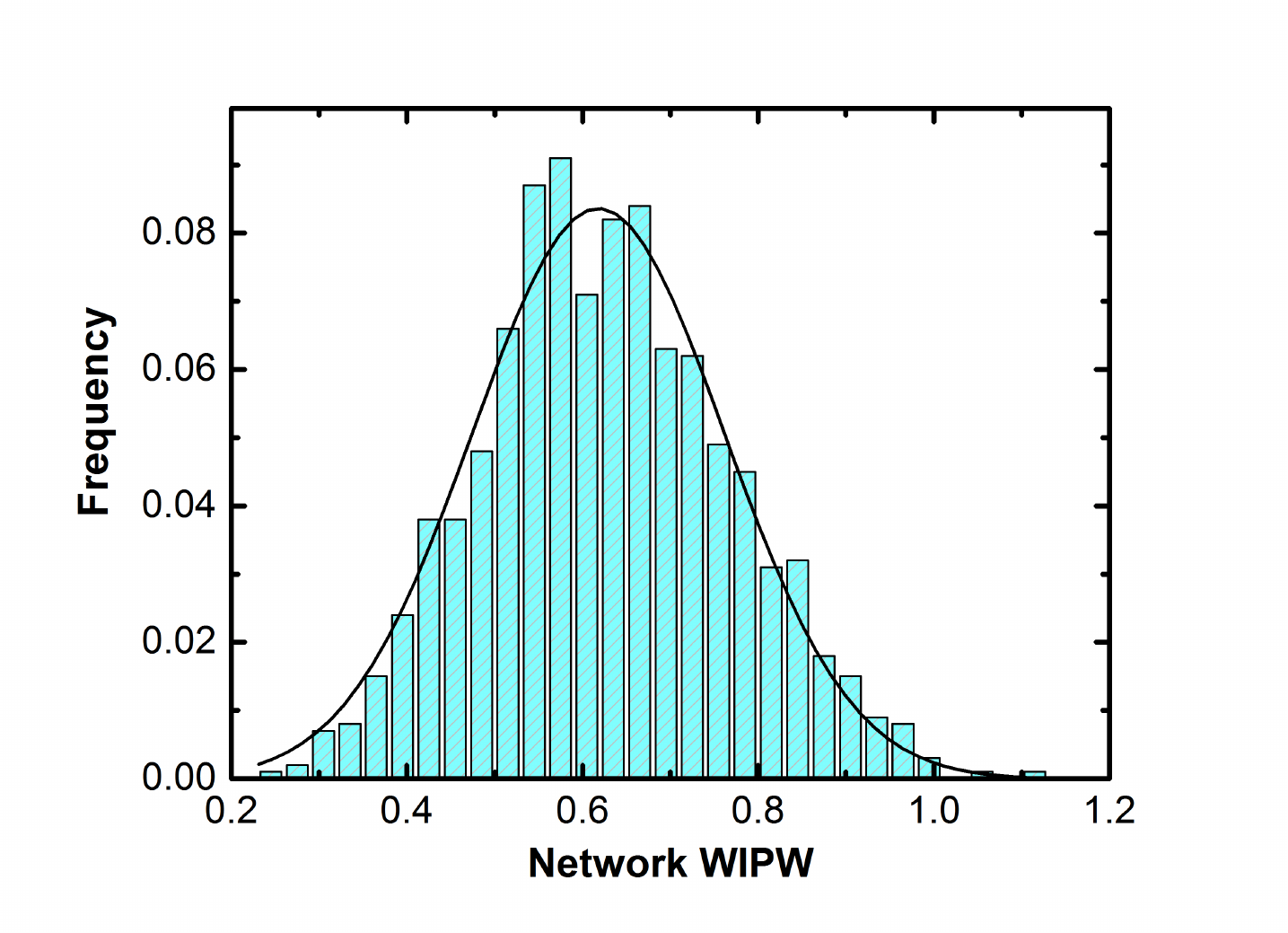}
}{Histogram of network WIPW before Retrofit}{hist_wipw_no_retrofit}

\subsection{Improving network resilience by strengthening critical bridges}
\label{subsec_improve_network_resilience}

When only limited resources are available for risk mitigation aimed at achieving a more resilient network through bridge retrofitting, it is critical to distribute those resources within the network in such a way that the overall network performance is maximized.    Since there are 37 bridges in the network and each can be either selected or not selected for retrofit, the solution space has  different combinations. As discussed previously, the GA is used to solve this multi-objective optimization problem as formulated in Section \ref{sec_mitigation}.  It is assumed that a retrofit intervention will bring the reliability of a bridge back to 0.999, the ``as new'' condition. 

Figure \ref{mitigation_tradeoff} shows the tradeoff between network performance and financial investment in risk mitigation, with the triangular markers representing the collection of optimal solutions (Pareto Frontier) when WIPW and retrofitting cost (measured in non-dimensional cost units) are considered as competing objectives.  With a total of 150 cost units, the alternative shown at the top right corner of the Figure 5 selects all 37 bridges for retrofit and the corresponding network WIPW increases to the pre-earthquake value of 1.7, as all bridges are upgraded to the ``near-perfect'' hazard-resilient condition.   Four solutions on the Pareto Frontier, marked as I, II, III and IV in Figure \ref{mitigation_tradeoff}, are identified in Figure \ref{details_of_optimal_solutions} for further examination.  Generally speaking, a larger financial investment will lead to more bridges being selected for retrofit and, consequently, a more resilient network. For instance, Solution I identifies seven bridges in the network as having retrofit priority, with a budget of 25 cost units, resulting a network WIPW of 0.76.   In contrast, Solution III indicates that with a budget of 86 cost units, 22 bridges can be selected for retrofit, leading to a WIPW of 1.23, which is a $101\%$ increase over the pre-retrofit WIPW of 0.61. Figure \ref{wipw_comparsion} compares the network WIPW before and after the retrofit associated with Solution III when uncertainties are considered in the evaluation.

Table \ref{details_of_optimal_solutions} reveals that the increase in the number of selected bridges from Solution I to Solution IV is not simply due to adding more bridges to the selected group associated with a lower budget. The likelihood of some bridges, e.g. bridges 11 and 13, being selected increases as the budget increases.   Other bridges, e.g. bridges 4 and 33, are selected initially when the budget is low; however, they are deselected as the budget increases because a neighboring bridge on an alternative path may become a more cost-effective candidate for improving the overall network resilience.  Finally, bridges 5 and 17 are always selected in optimal retrofit solutions because, when compared with other bridges, they both: 1) have much lower than average structural reliabilities (0.55 and 0.61); 2) carry heavy ADT (1,900 and 2,500); 3) are shared by IPWs between multiple node-pairs; and 4) are in close proximity to emergency response facilities. This dynamic prioritization mechanism holistically integrates the network characteristics and individual bridge properties in resilience-based decision for the road network.

\myfigure{
	\centering
	\includegraphics[width=\textwidth]{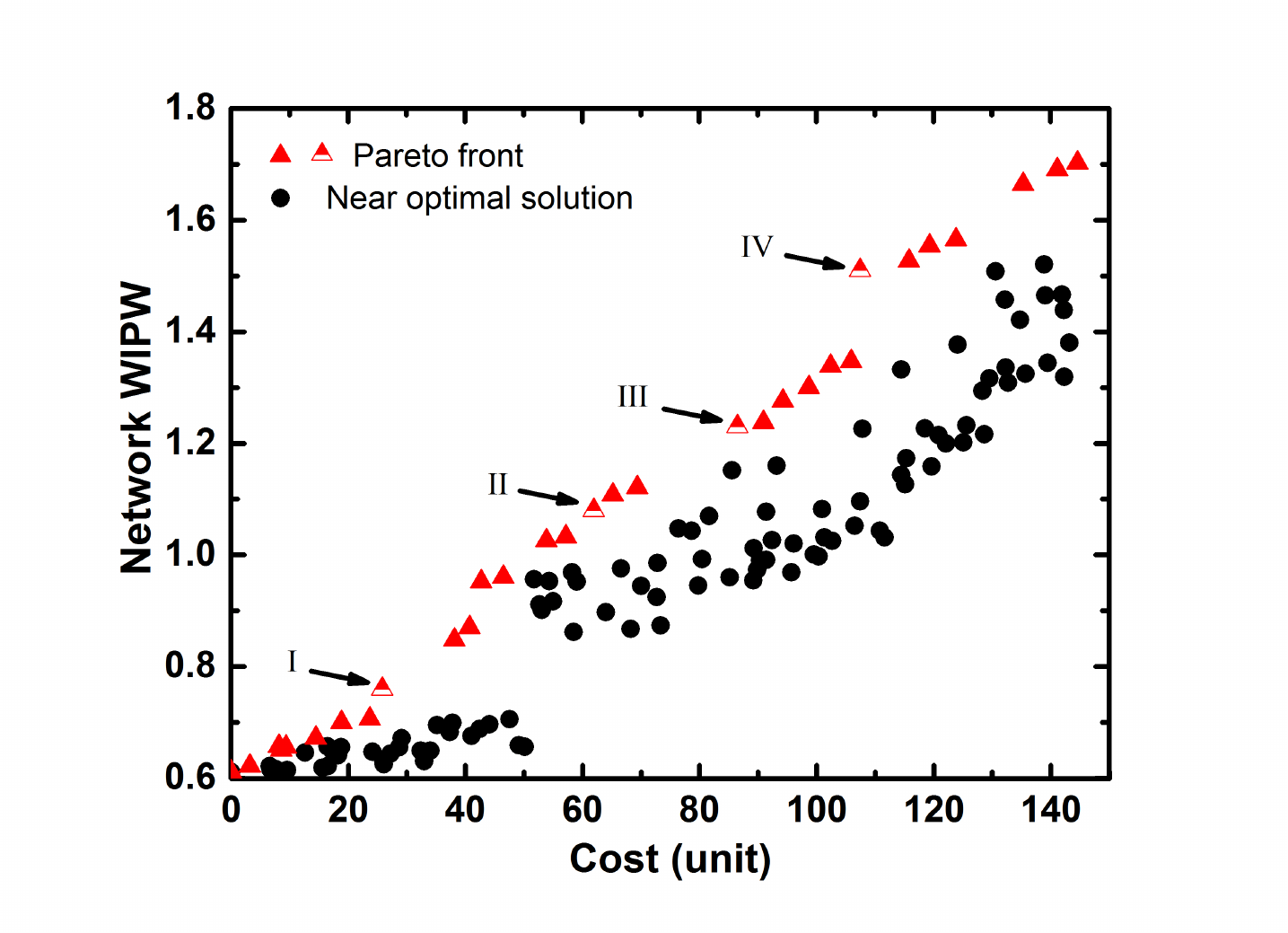}
}{Pareto Frontier of Optimal Solutions}{mitigation_tradeoff}

\mytable{
	\begin{tabular}{c  c c c l}
		\toprule
		Solution ID	&	Cost	&	Mean WIPW	&	$\#$ of bridges & Bridges selected	\\
		\hline
		I & 25 & 0.76 & 7 & 5,17,8,32,10,4,33\\
		\multirow{2}{*}{II} & 62 & 1.07 & 14 & 22, 5, 28, 17, 20, 14, 12, 19, 27, \\
		&  &  & & 15, 24, 16, 36, 21, 37, 31 \\ 
		\multirow{3}{*}{III} & 86 & 1.24 & 22 & 27,  11,  19,  16,  10,  31,  5,  17,  18,   \\
		&  &  & & 37,  23,  30,  28,  14,  12,  24,  21,  \\ 
		& & & &  22,  29,  36,  15,  20\\
		\multirow{3}{*}{IV} & 107 & 1.51 & 27 & 7,  6,  11,  19,  27,  31,  1,  5,  17,  8,    \\
		&  &  & & 18,  37,  25,  26,  23,  30,  28,  14,  2,   \\ 
		& & & &  24,  21,  35,  22,  29,  36,  20,  13\\
		\bottomrule		
	\end{tabular}
}{Details of four optimal solutions in Pareto frontier} {details_of_optimal_solutions}

\myfigure{
	\centering
	\includegraphics[width=\textwidth]{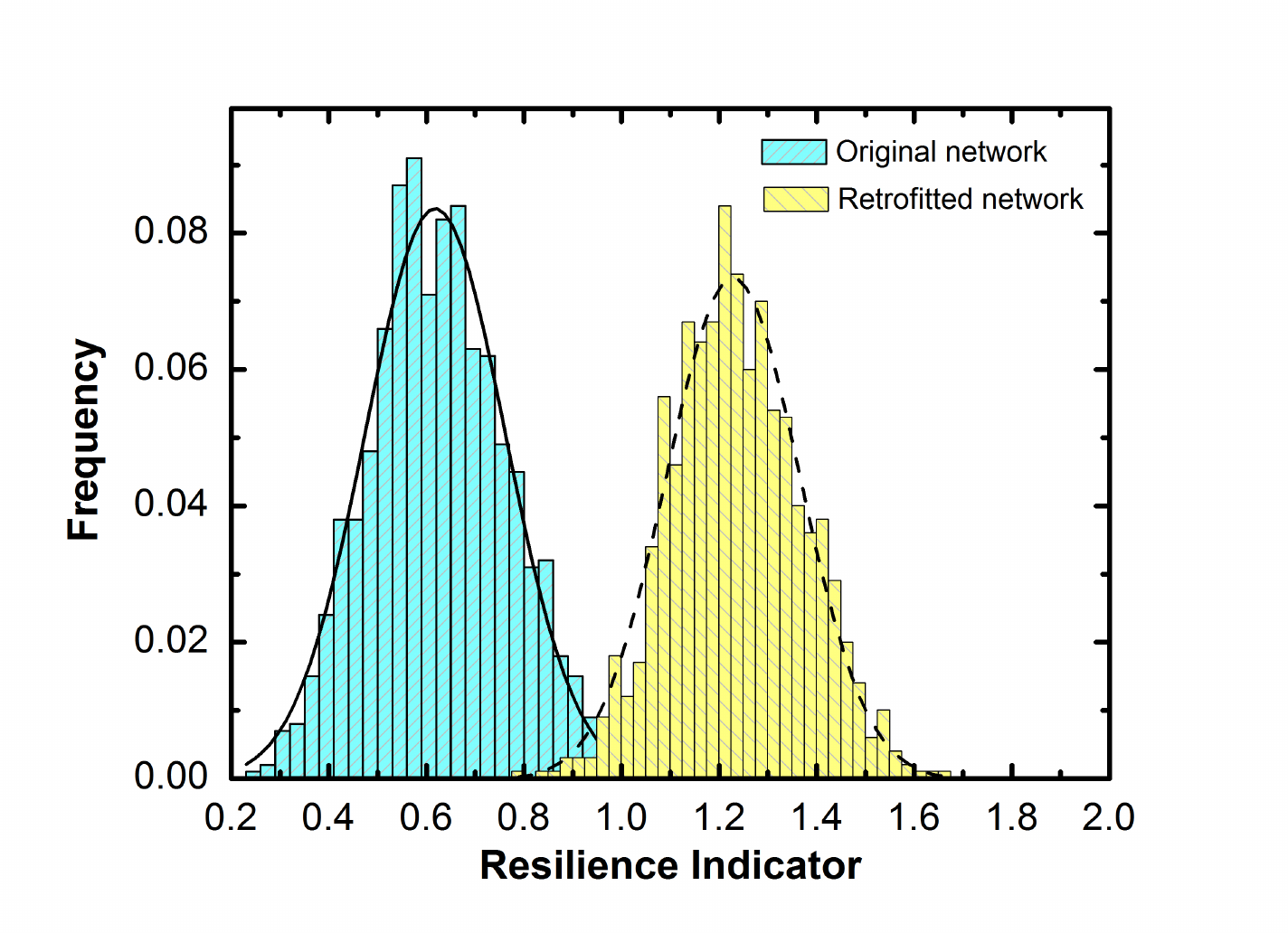}
}{Comparison of Network Resilience before and after Retrofit (Solution III)}{wipw_comparsion}

\subsection{Improving network resilience by changing network topology through new construction}
\label{subsec_new_construction}

Opportunities may exist in a developing community to improve the resilience of the existing transportation network by altering its topology through new construction.  For example, the hypothetical road network considered herein has two obvious areas of potential weakness as far as resilience is concerned. The first weakness is the branch from node 19 to node 28, highlighted with thickened lines in Figure \ref{new_construction}, where four bridges - 26, 30, 36, and 37- are connected in series to the rest of the network through node 19.  If any of the four bridges fail due to the earthquake, the area to the east of that bridge will be isolated from the rest of the community because an alternative path does not exist.   The second weakness is if both bridges 18 and 19 fail, the entire community road system will be divided into two unconnected subsystems, and the resilience of the road system will instantly drop to 0.39.  Relieving the burden on these two bridges by creating bypasses will increase the overall resilience of the network. 

Suppose that as part of community development, Links S1-S5, shown by dashed lines on Figure \ref{new_construction}, are possible candidates for construction to mitigate the above-mention potential risks.  Suppose, further, that resources are available to construct only two additional roads out of the five candidates: one road to be selected from S1, S2, or S3 is to increase the connectivity of the branch from node 19 to node 28, and one road to be selected from S4 or S5 to relief the burden on roads 18 and 19.   The decision method outlined in Section 5 is applied to select alternatives that will most enhance the overall resilience of the road network. Table \ref{impact_new_constructions} lists the mean and percentage increase of the network resilience for each combination of possible selections. While all possible solutions improve the overall network resilience significantly, the combination of S2+S4 is most effective, with a 72.3\% increase in network resilience from 0.61 to 1.05.

\myfigure{
	\centering
	\includegraphics[width=\textwidth]{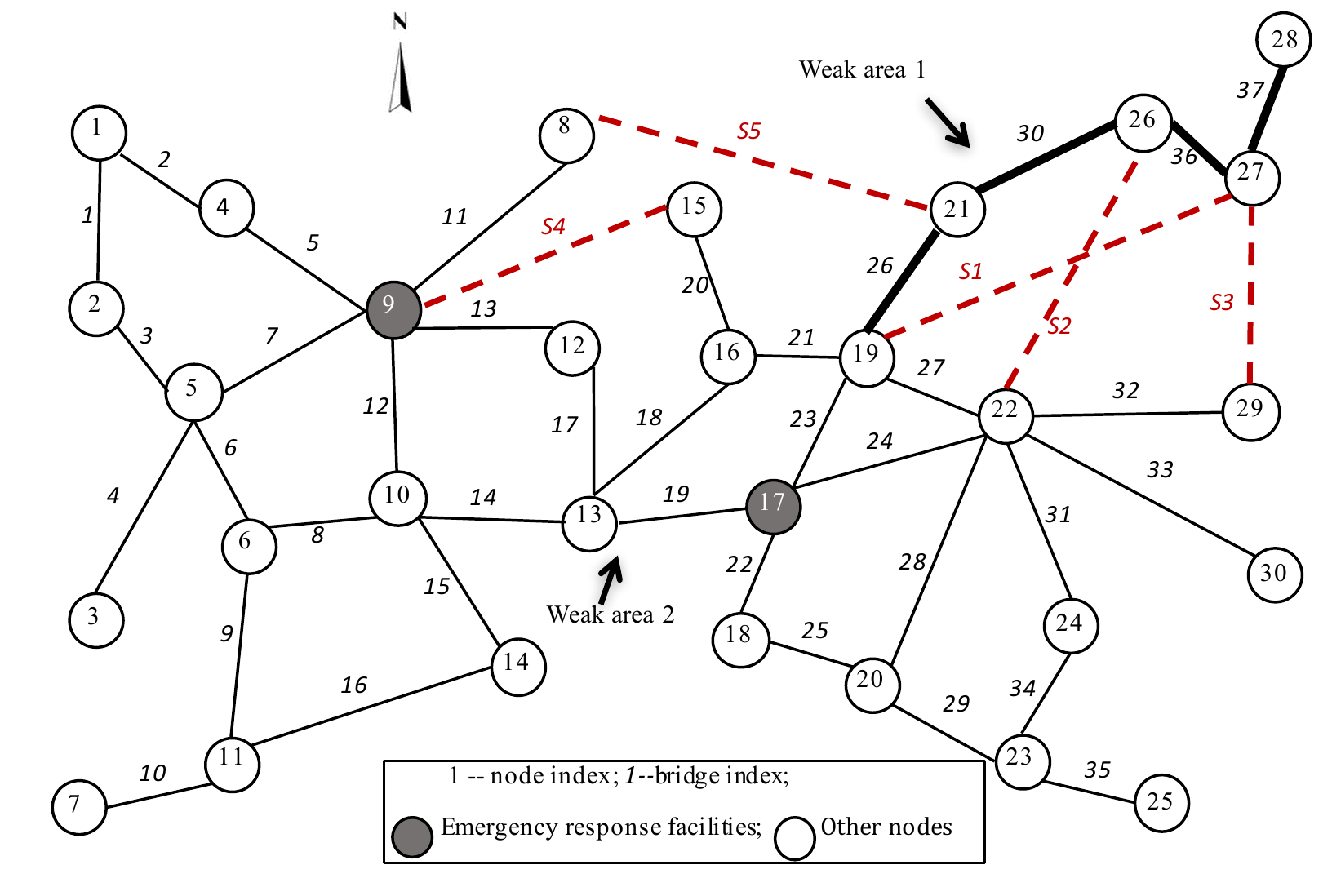}
}{Hypothetical Bridge Network with Candidates for New Construction }{new_construction}

\mytable{
	\begin{tabular}{c  r r}
		\hline
		Possible combinations 	&	Mean of resilience	&	Increase in resilience	\\
		\hline		
		S1+ S4	&	1.03	&	68.80\%	\\
		S1+ S5	&	0.88	&	44.30\%	\\
		S2+ S4	&	1.05	&	72.30\%	\\
		S2+ S5	&	0.89	&	45.90\%	\\
		S3+ S4	&	1.01	&	65.60\%	\\
		S3+ S5	&	0.85	&	39.30\%	\\
		\hline
	\end{tabular}
}{Impact of New Construction on Network Resilience} {impact_new_constructions}

\section{Summary}
\label{sec_mitigation_summary}

This chapter has introduced a quantitative metric based on graph theory to measure resilience of a road/bridge network which permits risk mitigation alternatives for improving the resilience of transportation system to be evaluated and compared on a common basis.  The method systematically integrates the network topology, redundancy level, traffic patterns, location of community emergency response facilities as well as failure probability of individual bridges, into the resilience indicator. Resilience analysis suggests that network resilience can be improved by increasing reliabilities of critical bridges through appropriate retrofitting, optimizing network topology through new construction, by altering traffic flow patterns through appropriate routing policies, and by strategically allocating emergency response facilities. However, the role that the transportation system plays before, during, and after a disaster will be unique for each community. Stakeholders of communities of different sizes, populations and social-economic vulnerabilities and support systems are likely to reveal different values and preferences in evaluating resilience of their transportation systems. The resilience indicator proposed in this chapter provides a transparent framework for incorporating other attributes in addition to those discussed herein by adjusting the weights in the network resilience quantification. The proposed decision framework for effective risk mitigation through bridge retrofitting or new construction is formulated as a multi-objective optimization problem, which allows tradeoffs to be made between competing resilience and cost objectives and the subjective preferences of a decision maker to be factored in decision process.

\chapter{Resilience-based post-disaster recovery strategies for road-bridge networks}
\label{ch_recovery}

\section{Introduction}
\label{sec_recovery_introduction}

A well-accepted definition of infrastructure system resilience is presented in \citep{bruneau2003framework}, as illustrated in Figure \ref{concept_of_resilience}, Several approaches to quantify resilience can be found in \citet{bruneau2003framework}, \citet{chang2004measuring}, \citet{cimellaro2006quantification, cimellaro2010framework, cimellaro2010seismic}, \citet{zobel2011representing},  \citet{bocchini2013resilience}.   Numerous studies have been focused on the post-hazard restoration of physical networks. For example, \citet{chang2001measuring} utilized network coverage and transport accessibility to quantify the post-disaster performance of a transportation network and applied these concepts to a rail and highway transportation system in Kobe, Japan.   \citet{shinozuka2004resilience} discussed the restoration curve in terms of robustness and rapidity in which the recovery indicator is associated with the service level (e.g. power supply for electric networks or water supply for water systems) and the rapidity is measured by the average recovery rate, expressed in percentage recovery/time.   \citet{wang2004resource} constructed a depot location model to minimize the total inter-cell transportation cost for the electric power restoration process.  \citet{ccaugnan2006post} compared the alternative restoration strategies for the electric power transmission systems with respect to the expected duration of power outages in Los Angeles using a discrete-event simulation model. The study by \citet{xu2007optimizing} of an electric system was aimed at minimizing the area above the restoration curve; this area, shown in Figure \ref{concept_of_resilience} as the light-shaded triangular area, is named the resilience triangle in \citet{bruneau2003framework}. \citet{miles2006modeling} proposed one of the first comprehensive concept models of post-disaster community recovery using an object-oriented design technique, in which the variables and relationships between different sectors were clearly defined. Their model provides a common basis for developing computer models of socioeconomic recovery from disasters and its flexibility allows for the incorporation of various indicators and algorithms within the framework, and was implemented successfully in a prototype computer simulation with a graphical user interface.  \citet{karlaftis2007fund} developed a three-stage approach to allocate available resources to the restoration of a transportation system in terms of the contribution of each bridge to the operation of the network.  \citet{frangopol2011resilience} optimized the post-disaster restoration schedule for a transportation system with respect to total cost and resilience, the definition of which is the area below the recovery trajectory (the dark-shaded trapezoid shown in Figure \ref{concept_of_resilience}) using total travel time and total travel distance as the network performance metrics.  In a later study \citep{bocchini2012restoration}, the authors added the time required to recover a certain level of network functionality (less than complete recovery) as another objective in optimizing the network restoration schedule. More recently, \citet{karamlou2014optimal} proposed a multi-objective optimization model to maximize the network resilience and minimize the required time to connect critical nodes (e.g., healthcare facilities and operations centers) in the network. 

The network performance metrics used in post-disaster recovery in the studies as reviewed above include rapidity (recovery time), monetary loss (user cost), service performance (travel time and travel distance), and area under the recovery curve; however, no methods could be located which focus on the shape of the recovery trajectory.  The shape, however, provides additional information and a novel perspective on the efficiency of the network restoration process. For example, as shown in Figure \ref{recovery_trajectories}, the four recovery trajectories share the same rapidity, i.e. recovery time.  Assuming equal levels of community investment, curve 1 represents the best recovery strategy and curve 4 is the worst among all the alternatives. Furthermore, recovery curve 2 is better than curve 3 due to its more efficient early-stage recovery which reduces early losses caused by network service disruption.  This efficiency in early phase of recovery could also facilitate the recovery of other infrastructure systems whose service capabilities highly depend on the functionality of transportation network (e.g. emergency response and rescue immediately following hazard event).   To the best of our knowledge, the existing network recovery scheduling frameworks are generally not capable of distinguishing recovery trajectories 2 and 3. While \citet{bocchini2012restoration} proposed to use the time required for network to recovery to a pre-defined level of functionality as an additional measure to evaluate the efficiency of the recovery process, that proposal still leaves curves 2 and 3 indistinguishable if the prescribed function level is close to the intersection point of the two curves. 

\myfigure{
	\centering
	\includegraphics[width=\textwidth]{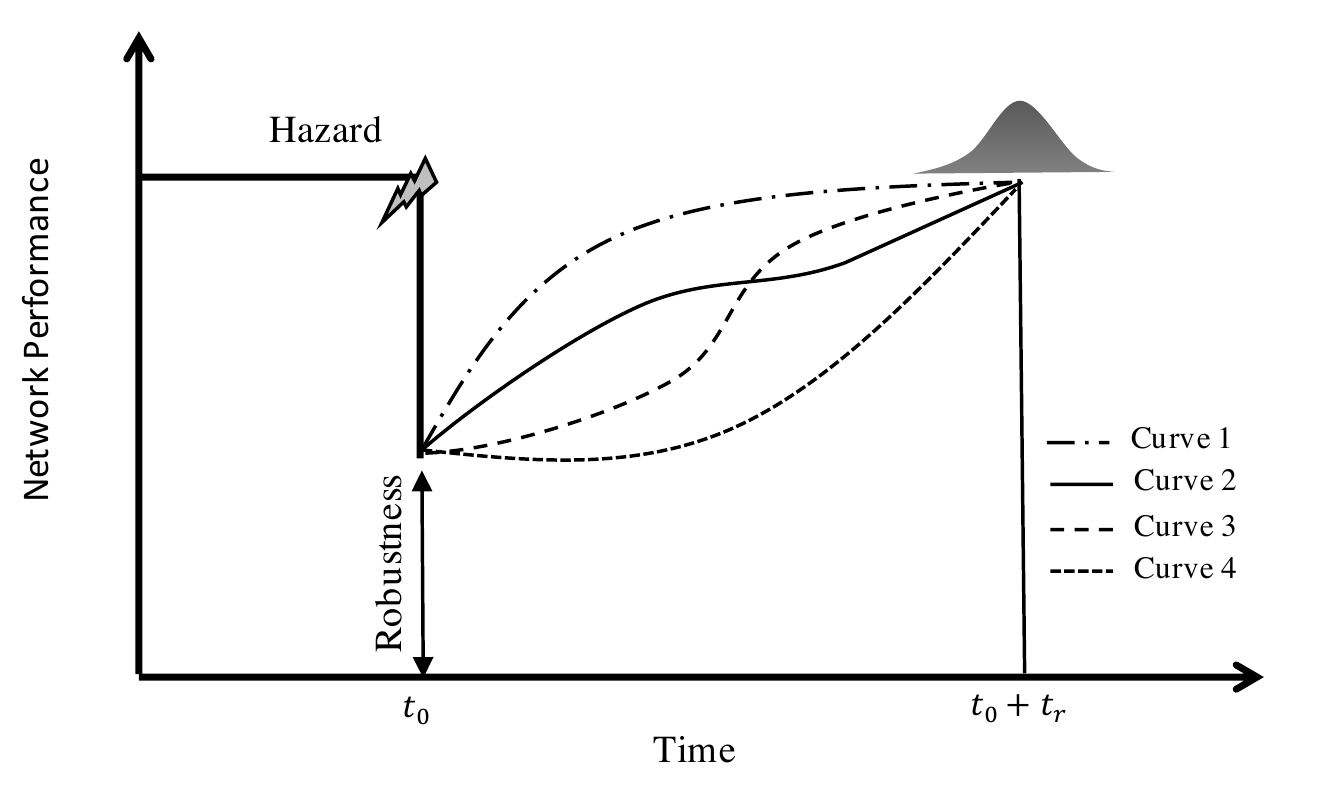}
}{Recovery trajectories}{recovery_trajectories}

\section{Highlight contributions}
\label{sec_highlight}

In this chapter, the author first introduces a novel metric for evaluating the relative efficiency of alternative network recovery strategies.   This chapter then develops a restoration scheduling methodology for network post-disaster recovery that minimizes the overall network recovery time and optimizes the recovery trajectory, which ultimately will reduce economic losses due to network service disruption.  In the proposed method, the number of simultaneous repair interventions (actions) is constrained by the available resources in the community throughout the recovery period.  In addition, this model dynamically updates the damage level of each individual bridge and the corresponding overall performance of the network during the recovery process until all the damaged bridges are restored.  The restoration scheduling model is stochastic in nature because the uncertainties associated with parameters that are critical for network recovery, e.g. the restoration intervention duration for each damaged bridge and traffic flow on roads and bridges, are propagated throughout the analysis.  The optimization problem in this study is formulated as a version of the dynamic job shop problem, which is known to be NP-hard \citep{gonccalves2005hybrid}.  Genetic algorithm (GA) implementation is used to search for near-optimal solutions efficiently. Monte Carlo Simulation (MCS) is employed to sample the stochastic parameters to quantify the uncertainties associated with network recovery process.

The remainder of the chapter is organized as follows. Section \ref{sec_recovery_metric} introduces the resilience-based transportation network performance metric used in this study. In section \ref{sec_recovery}, the author defines the metrics for measuring the efficiency of the network recovery process, and then develop the mathematical formulation for resilience-based network recovery optimization.  In Section \ref{sec_numerical_application}, a hypothetical bridge network comprised of 30 nodes and 37 bridges is generated to illustrate the implementation of the developed methodology in a context of a considered scenario earthquake. A sensitivity study using this network illustrates the impact of the resourcefulness of a community and time-dependent commitment of resources on the network recovery time and trajectory.   Conclusions and future work are summarized in Section \ref{sec_recovery_summary}.

\section{Resilience-based performance metric of road networks}
\label{sec_recovery_metric}

Many performance metrics for transportation networks, such as maximum traffic flow capacity and minimum travel time or distance, can be used to measure network performance under normal operational conditions, but are not directly applicable to the analysis of post-disaster recovery following severe natural hazards. Immediately following an extreme event, people typically are more concerned about whether they are able to travel from one place to another than the distance or the speed at which they can travel.  Connectivity reliability, as a network performance measure, could address the above-mentioned deficiencies, but does not reflect the different levels of importance in the roles that different roads and bridges play in the functionality of the network.  Thus, connectivity reliability does not fully support resilience-based decisions on engineering interventions that are directly implemented at the network component (roads and bridges) level.  A network performance metric recently introduced by
\citet{zhang2016resilience} and formulated in the context of community resilience to natural hazards, uses the weighted average number of reliable independent pathways between all origin-destination (O-D) pairs, denoted WIPW, as a performance measure of transportation networks.  This study adopts the WIPW as the network performance metric (the vertical coordinate of the resilience curves illustrated in Figures \ref{recovery_trajectories}) for optimizing post-event network recovery scheduling.

Let $G=(V,A)$ denote the road network, where $V$ is the set of nodes, which is partitioned into a set  of $E=\{1,2,...,e\}$ emergency nodes (including critical emergency response facilities, e.g. fire stations and hospitals) and a set  of $N=\{e+1,e+2,...,n\}$ normal nodes (representing major destinations, e.g. residential areas, economic hubs, and major road intersections); and  $A=\{1,2,...,m\}$ is the set of arcs (links) that represent roads without or with a maximum of one bridge. The network performance metric, WIPW, is written as,

\begin{align}
WIPW(G) =\sum_{i=1}^{n}w_{i}\frac{1}{n-1}\sum_{j=1,j\neq i}^{n}\sum_{k=0}^{K_{ij}}w_{k}(i,j)R_{k}(i,j) \label{wipw}
\end{align}

where $K_{(i,j)}$ is the total number of independent pathways (IPW) between node  $i\in V$ and node $j\neq i \in V$ ; $P_{k}(i,j)$ represents the $k$th IPW between node $i\in V$ and node $j \neq i \in V$, and  $R_{k}(i,j)$ represents the reliability (probability of surviving a hazard) of $P_{k}(i,j))$.The weighting factor $w_{k}(i,j)$ is applied to $P_{k}(i,j))$, which is a function of the average daily traffic (ADT, denoted as $F_{ij}$ in the subsequent sections) and the length (denoted $L_{ij}$ as in the subsequent sections) of each arc that is a portion of the $P_{k}(i,j)$; weighting factor $w_{i}$ is applied to node $i$, which is a function of the distance from node $i$ to its nearest emergency response facilities represented by emergency node $e\in E$. The detailed formulation of each item in Eq. (\ref{wipw}) and the complete algorithm to evaluate WIPW can be found in Chapter \ref{ch_WIPW}.  

This resilience-based network performance measure encompasses the following four important characteristics of network resilience:  (1) the network redundancy, encapsulated in the term $K_{(i,j)}$ in Eq. (\ref{wipw}), reflects the number of alternative or back-up independent paths between all possible O-D pairs.  (2) the network component reliability, encapsulated in the term $R_{k}(i,j)$, relates to the probability of bridges (and roads) being functional (fully or partially) after a given hazard event.  For example, bridges with a higher reliability are likely to have less damage and require less time and resources to recover following a disaster; such attributes should be considered in effective network risk mitigation and recovery strategies. (3) the importance levels of network components, encapsulated in the term  $w_{k}(i,j)$ (as a function of arc ADT and length) in Eq. (\ref{wipw}), reflect the different service levels of bridges and roads in terms of the roles that they play in supporting the overall network functionality.  The ADT describes the historical traffic flow on each roadway.  It  is more accurate as a roadway importance measure than the traffic pattern predicted by existing traffic assignment models \citep{davis1997accuracy} because the user pattern of the road system is affected by many factors, including distribution of origins and destinations, distance and travel time of paths, congestion, facilities along the path, user preferences, etc..  (4) the role of the transportation network in post-disaster emergency response is considered in the performance metric, through the term  in Eq. (\ref{wipw}), by applying heavier weights on roads and bridges that are topologically in close proximity to emergency facilities, as they likely play important roles in community emergency response and rescue immediately following the hazard event. 

The WIPW was originally introduced as a network performance metric for pre-event risk mitigation decisions.  In adapting the WIPW in this chapter for network post-disaster recovery scheduling optimization, its formulation in Eq. (\ref{wipw}) needs to be modified. Specifically, the pre-event reliability of IPW, $R_{k}(i,j)$, must be replaced by its post-event serviceability.  The post-event service level of an arc (a road segment or a bridge) is a function of its damage level denoted as $q_a$ for $a \in A$, which can be measured on a 0 to 4 scale, corresponding to the damage levels of none, slight, moderate, extensive and complete \citep{scawthorn2006hazus}.  The author idealizes the service level of each arc  $a\in A$ to be $\left(1-q_{a}/{4}\right)$ \citep{bocchini2012restoration}; i.e, if an arc is fully damaged (``complete'' damage state), $q_{a}$ is set to be 4, and the corresponding service level is 0; otherwise, the arc service level ranges from 0 to 1. The author further approximates the service level of the IPW $P_{k}(i,j)$ as the product of the service levels of all arcs $a \in P_{k}(i,j)$. Accordingly, the WIPW for a damaged transportation network can be computed as,

\begin{align}
WIPW(G) = \sum_{i=1}^{n}w_{i}\frac{1}{n-1}\sum_{j=1,j\neq i}^{n}\sum_{k=0}^{K_{ij}}w_{k}(i,j)\min_{\forall a \in P_{k}(i,j)}\left(1-\frac{q_{a}}{4} \right) \label{wipw_recovery}
\end{align}

For post-event recovery scheduling investigated in this chapter, a damaged transportation network is considered as fully recovered if the network WIPW computed using Eq. (\ref{wipw_recovery}) returns to its pre-disaster level (without damaged network components).    

\section{Optimization of network recovery scheduling}
\label{sec_recovery}

This research introduces two metrics for resilience-based network recovery planning. The first metric is the total recovery time (TRT),  $t_{r}$, after which the network is restored to its ``undamaged'' condition (damage level for all arcs equals to 0, and network WIPW resumes to its pre-disaster value).  As discussed previously, the TRT alone is not sufficient to evaluate the efficiency of network recovery strategies which is partially encapsulated in the shape of the recovery trajectory. For example, Figure \ref{recoverySRT} shows two recovery trajectories as a function of time resulting from two different network restoration schedules, where the network performance (vertical axis) is measured by WIPW;  $R_{0}$  and $R_{h}$ denote , respectively, the network WIPW before and immediately after the extreme event.  It is obvious that while restoration schedules 1 and 2 lead to approximately the same network recovery time, schedule 1 is notably more efficient than schedule 2 with respect to the economic losses incurred due to interrupted network service during recovery. The author therefore introduces a second metric for evaluating the effectiveness of network restoration schedules - the skew of the recovery trajectory (SRT), defined as the centroid of the area below the recovery trajectory (from $t_{0}$ to $t_{k}$) with respect to $t=t_{0}$.  The SRT associated with schedules 1 and 2 are marked in Figure \ref{recoverySRT} as $s_{1}$  and $s_{2}$, respectively. If the recovery was instantaneous,  $s$ would equal to 0.  This two-dimensional recovery metric, i.e. TRT and SRT, defines the objective functions in finding the optimal scheduling for the network recovery. 

\myfigure{
	\centering
	\includegraphics[width=\textwidth]{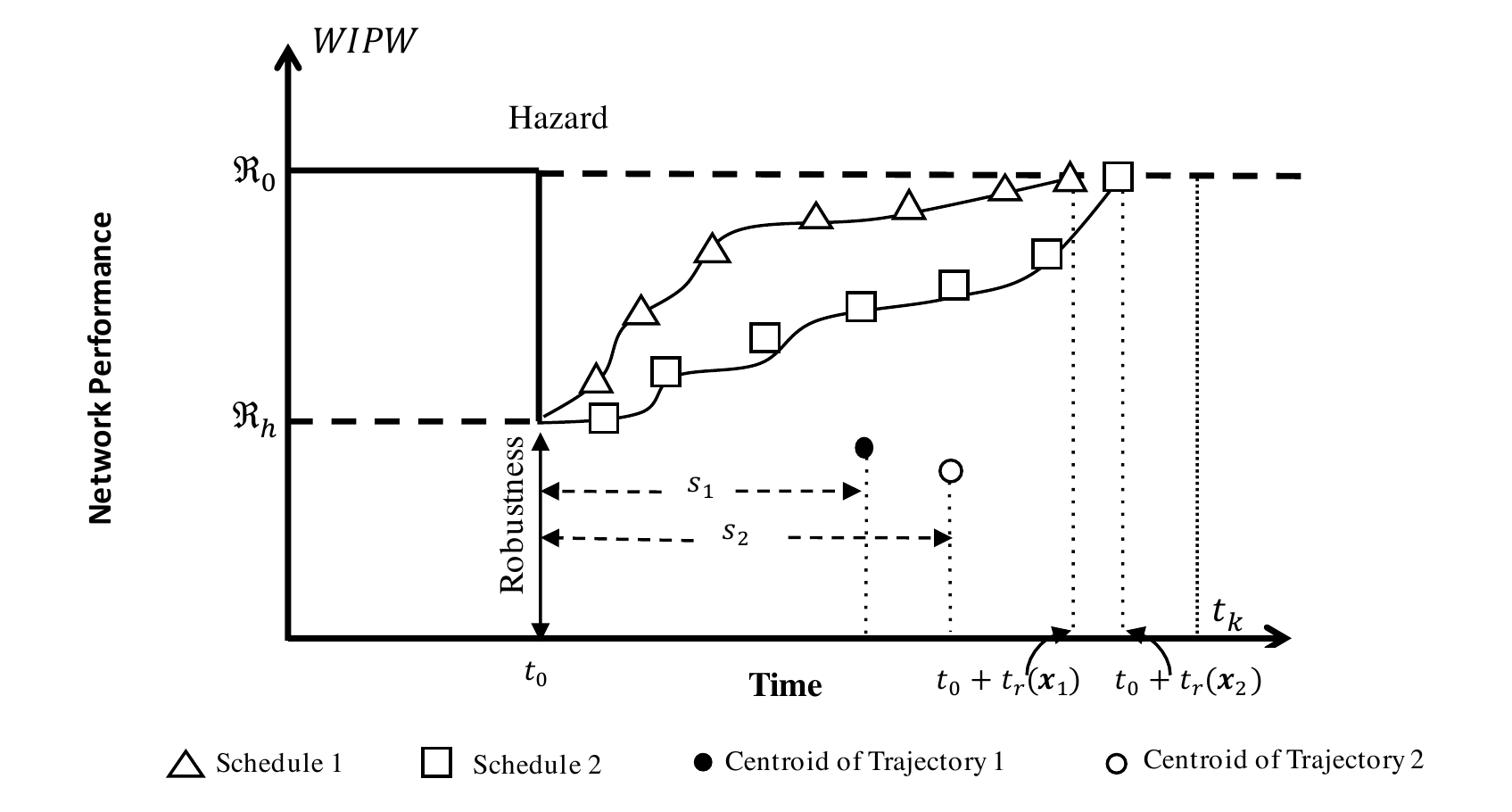}
}{Recovery trajectories}{recoverySRT}

Although the scheduling framework introduced in this chapter applies to any arc within the network, i.e. both bridges and road segments, the subsequent discussion is focused on bridges as they are the most vulnerable arcs in the transportation network.  Let $B =\{1,2,...,m\}$ denote the set of network bridges.  The recovery scheduling is to determine an optimal schedule $x=\{x_1,x_2,...,x_d\}$ for the repair of all $d \leq m$ damaged bridges, where $x$ is the time at which restoration is initiated for bridges $b=1,...,d$, such that both the network TRT and SRT are minimized. Let $D_b$ denote the duration of restoration intervention for each bridge $b=1,...,d.$ The network TRT associated with the schedule $x$, $t_r(x)$, is then:

\begin{align}
t_{r}(x) = \max_{b=1,2,...,d}(x_b + D_b) - t_0 \label{trt}
\end{align}

The network SRT associated with the scheduling plan x, s(x), the centroid of the area under the recovery trajectory as shown in Figure \ref{recoverySRT}, can be calculated by Eq. (\ref{SRT}), which requires the integrals of WIPW, i.e., $R_{t}$, as a function of time.  As discussed previously computing WIPW involves using Dijkstra's algorithm \citep{skiena1990dijkstra}  to search for IPWs for all O-D pairs, which cannot be performed in closed-form.  This study therefore estimates WIPW at discrete points in time; consequently, the recovery trajectory (expressed in terms of WIPW) is discretized into step functions.  The author sets $T=\{t_0,t_1, ..., t_k\} $ such that $t_0 \leq t_1 \leq ... \leq t_k$, in which the difference between any adjacent time points is a constant time increment $\Delta t$ . The SRT can then be approximated by:

\begin{align}
s(x) = \frac{\int_{t_0}^{t_0 + t_k}R_{t}(t-t_0)dt}{\int_{t_0}^{t_0 + t_k}R_{t}dt} \approx \frac{\sum_{i=0}^{k}t_{i}R(t_i)\Delta t}{\sum_{i=0}^{k}R(t_i)\Delta t} \label{SRT}
\end{align}

where $t_k$, set to be larger than any possible TRT, represents a common reference timeframe for computing SRT for different strategies. During the recovery phase, the damage level of each bridge $b$  at any time $t\in T$  is:

\begin{align}
q_b^t(x_b)=q_b[t < x_b + D_b] \label{damage_level_recovery}
\end{align}

where $[P]$ is the Iverson bracket, which returns 1 if $P$ is true, and 0 otherwise. Eq. \eqref{damage_level_recovery} ensures that the bridge $b$ remains at its initial damage level $q_b$ until its scheduled restoration intervention is completed; its damage level then becomes 0 after the completion of the intervention (i.e. when $t \geq x_b + D_b$). This assumption implies that selected lane closure during a bridge restoration project is not considered. That is, a bridges is not treated as a feasible link in searching for IPWs in the estimation of WIPW (i.e. $R(t)$ in Eq.(\ref{SRT})).  At any given time $t \in T$, the number of simultaneous restoration interventions within the entire network, denoted by $N_{SI}^{t}$ , can be computed as:

\begin{align}
N_{SI}^{t} = \sum_{b=1}^{m}[x \geq x_b][t\leq x_b + D_b] \leq N_{SI}^{max} \label{simultanous_interventions}
\end{align}

where $[P]$ is the Iverson bracket as introduced above; $N_{SI}^{max}$ denotes the maximum number of simultaneous restoration interventions in the network allowed by the human and financial resources available in the community for the recovery of the road network following the hazard event. Accordingly, $N_{SI}^{max}$, imposes a constraint to the restoration scheduling, and thereby can impact the overall network recovery characteristics expressed in terms of TRT, $t_{r}(x)$, and SRT, $s(x)$.

The optimal restoration sequence for all damaged bridges and the time at which restoration is initiated for each bridge are obtained by minimizing the network TRT [as defined by Eq. (\ref{trt})] and SRT [as defined by Eq. (\ref{SRT})], under the constraint that only a prescribed maximum number of simultaneous restoration actions are possible at any given time [as expressed by Eq. (\ref{simultanous_interventions})].   The complete optimization model is summarized in Figure \ref{recoveryModel}. The ADT of each roadway [used in  calculating the term $w_k(i,j)$ in estimating the WIPW for the damaged network as expressed in Eq. (\ref{wipw_recovery})] and the duration of restoration intervention for each bridge are treated as random variables denoted, respectively, as $F_{ij}(\widetilde{\xi})$ for all $(i,j) \in A$  and $D_b(\widetilde{\xi})$ for all $b \in B$, where $\xi$ is stochastic variables. The distribution of $F_{ij}(\widetilde{\xi})$ can be derived from historical ADT measurements which are readily available from federal, state or local bridge owners.  $D_b(\widetilde{\xi})$ is assumed to have a normal distribution with a mean that is a function of both the damage level ($q_b$) and deck area of the bridge ($A_b$) \citep{fragkakis2004quantity}. Realizations of these random variables are denoted as $F_{ij}(\xi)$ and $D_b(\xi)$.

The decision problem under investigation, as formulated in Figure \ref{recoveryModel}, is closely related to the NP-hard parallel machine scheduling problem \citep{lenstra1977complexity, ullman1975np, cheng1990state}.  It is assumed that bridge repair scheduling is non preemptive, that is, once a crew has begun repair on a given bridge, they must complete their work before moving to another bridge.  Additionally, the problem under investigation is further complicated by the necessary estimation of WIPW which, as discussed, is computed iteratively through a series of weighted shortest path problems.   Approximation approaches are commonly utilized for addressing complex scheduling problems (e.g., \citet{cheng1996parallel}).  Accordingly, the genetic algorithm (GA) is modified for the scheduling problem described in \citet{gonccalves2005hybrid} to be applicable in this study to identify the near-optimal solutions.  The classical weighted-sum method \citep{deb2001multi, kim2005adaptive} is applied to the two objectives, TRT and SRT, to form a fitness function as below:
\begin{align}
\text{min}[ct_{r}(x) + (1-c)s(x)] \label{recovery_fitness_function}
\end{align}

where the author introduces $c \in [0,1]$ as a weighting factor to impose the relative importance between the two objectives.  A community (or government decision makers) can use different weighting factors based on their preferences, value and tolerances to different risks in order to obtain the ``best'' strategy to their specific situation.  For the example presented in the subsequent section, the work simply applies equal weights to the two objectives for illustration.  It is worth to note that even though the GA is implemented with one fitness function in the way the author chooses to perform the optimization, there are alternative approaches to keep the objectives separate, e.g., non-sorting genetic algorithm II \citep{deb2002fast}. The optimization process is summarized in Figure \ref{flow_chart_ga}. The stochastic variables (i.e. $F_{ij}(\widetilde{\xi})$ and $D_b(\widetilde{\xi})$ ) are realized using Monte Carlo Simulation (MCS).

 \myfigure{
 	\centering
 	\includegraphics[width=1.15\textwidth]{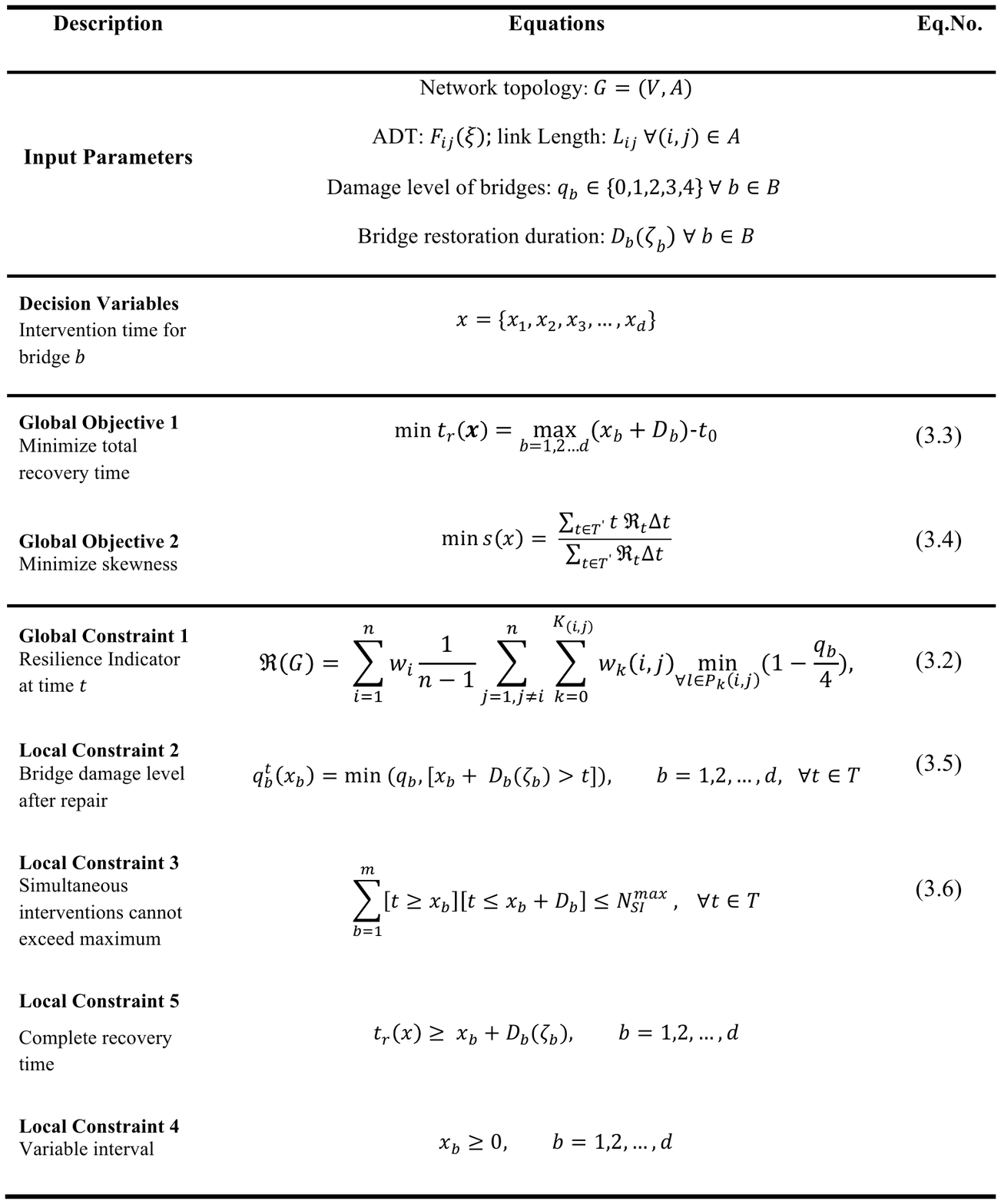}
 }{Summary of the realized optimization formulation}{recoveryModel}

\myfigure{
  	\centering
  	\includegraphics[width=1.4\textwidth]{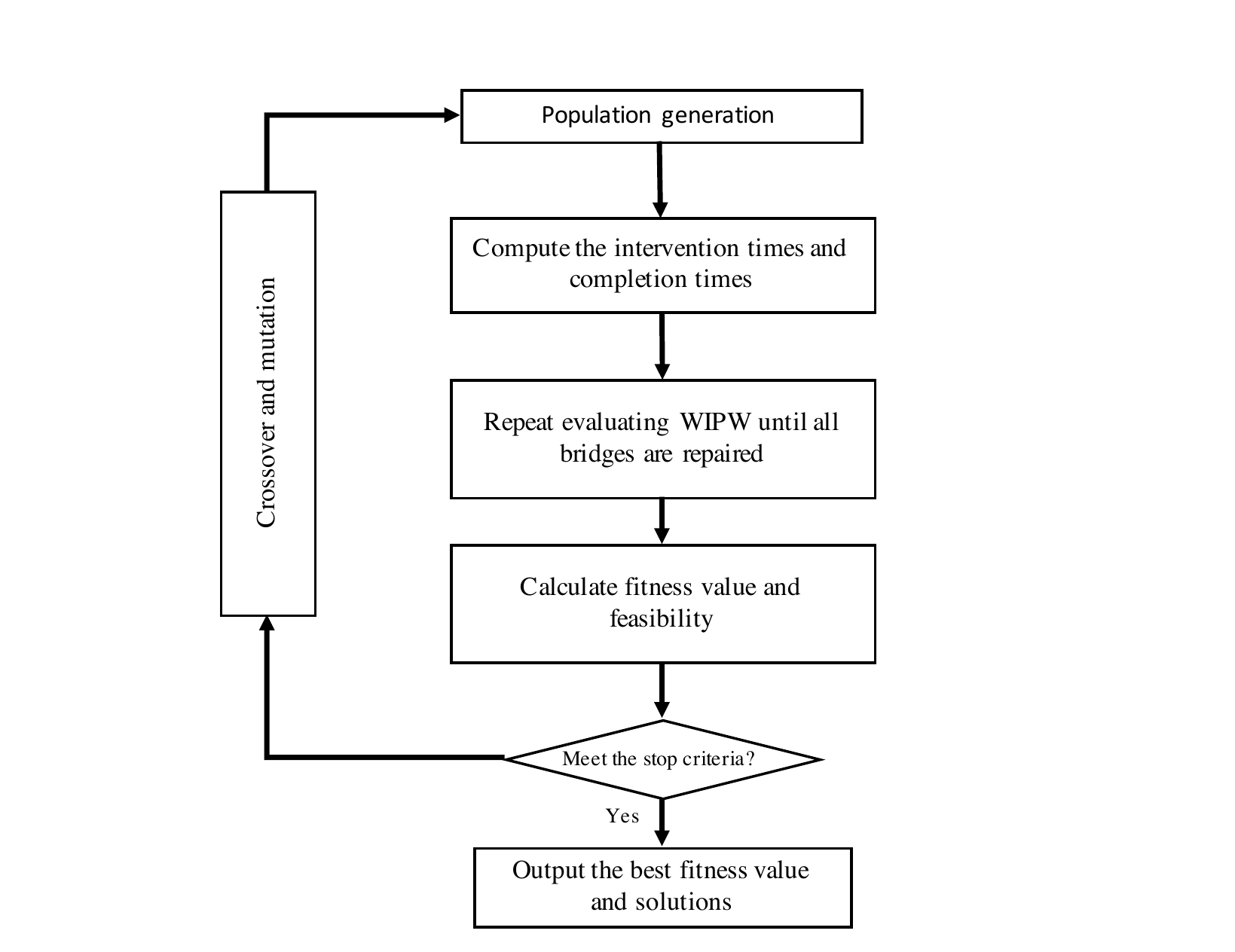}
}{Flow chart of genetic algorithm (GA)}{flow_chart_ga}
 
Note that the two objectives, TRT and SRT, represent two very distinct characteristics of the network recovery.  Summing the two objectives to form a fitness function is simply the way the author chooses to solve the optimization problem efficiently, which is independent of the multi-objective optimization formulation for the recovery as shown in Figure \ref{recoveryModel}.   There are alternative approaches to keep the objectives separately, e.g., non-sorting genetic algorithm II, if preferred.

\section{Numerical Application}
\label{sec_numerical_application}

\subsection{Bridge network characteristics}
\label{subsec_network_characteristics}
The proposed methodology is illustrated using a hypothetical road transportation network, shown in Figure \ref{hypothetical_network_recovery}, in which there are 37 arcs and 30 nodes. The gray nodes (Node 9 and 17) represent emergency nodes and other nodes represent normal nodes. For simplicity, each arc is assumed to be associated with exactly one bridge; out of the 37 bridges, 19 are steel (S) bridges and 18 are reinforced concrete (RC) bridges.  A scenario earthquake with a magnitude 7.0 and an epicenter distance of 40 km from the centroid of the network is considered for this illustration.  The network parameters used in this illustration, including bridge type and ADT, are tabulated in Table \ref{mean_network_parameter_recovery}.  The initial damage levels for all bridges for the stipulated hazard event are tabulated in Table \ref{mean_network_parameter_recovery}.  Among the 37 bridges, 15 bridges sustained negligible damages, while the other 22 suffer damage at different levels: 6 have major (complete or extensive) damages, 5 are moderately damaged and 11 have slight damages. These initial damage levels are assigned inversely proportional to the bridge reliabilities \citep{zhang2016resilience}, and are used as a starting point for recovery scheduling. To simplify the computation, the damage level is converted to binary with moderate damage as the threshold. That is, if damage level is less or equal to 2, the bridge is functional;  otherwise, it is not. The restoration duration of each damaged bridge is also presented in Table \ref{mean_network_parameter_recovery}. The uncertainty models that it is assumed for the parameters  $F_{ij}(\widetilde{\xi})$ and $D_b(\widetilde{\xi})$ are summarized in Table \ref{statistics_network_parameter_recovery}.

The network WIPW prior to and immediately following the scenario event are 1.74 and 0.62, respectively, representing a 64.4\% sudden drop in network performance (the vertical coordinates of the resilience curve).  In the subsequent network recovery optimization, it is assumed that the bridges with a damage levels 2-4 cannot carry traffic until the completion of their scheduled repair; furthermore, the limited available resources for recovery, representing a mid-income (average) community, only allow a maximum number of 4 bridges to be repaired and restored simultaneously, i.e. $N_{SI}^{t} \leq 4$.

\myfigure{
	\centering
	\includegraphics[width=\textwidth]{network.pdf}
}{Hypothetical bridge network}{hypothetical_network_recovery}

\mytable{
	\begin{tabular}{c | r r r}
		\toprule
		Bridge ID	&	Construction Type	&	ADT (Vehicle/Hour)	&	Month of Restoration	\\
\hline							
1	&	RC	&	2200	&	4.1	\\
2	&	RC	&	1900	&	1.71	\\
3	&	S	&	2000	&	10.21	\\
4	&	S	&	1500	&	0	\\
5	&	RC	&	1900	&	6.52	\\
6	&	S	&	2200	&	0	\\
7	&	S	&	700	&	0	\\
8	&	S	&	2400	&	0	\\
9	&	S	&	2600	&	3.99	\\
10	&	S	&	300	&	4.75	\\
11	&	S	&	800	&	1.44	\\
12	&	RC	&	900	&	2.38	\\
13	&	S	&	2500	&	0	\\
14	&	S	&	600	&	2.42	\\
15	&	RC	&	2000	&	1.4	\\
16	&	S	&	500	&	2.11	\\
17	&	RC	&	2500	&	3.32	\\
18	&	RC	&	2800	&	0	\\
19	&	S	&	1300	&	2.49	\\
20	&	S	&	1700	&	0	\\
21	&	S	&	1500	&	9.04	\\
22	&	S	&	1200	&	3.34	\\
23	&	RC	&	1500	&	0	\\
24	&	S	&	700	&	1.28	\\
25	&	S	&	1800	&	0	\\
26	&	S	&	900	&	5.02	\\
27	&	S	&	600	&	5.25	\\
28	&	RC	&	800	&	6.65	\\
29	&	RC	&	1400	&	0	\\
30	&	RC	&	2800	&	2.35	\\
31	&	RC	&	1900	&	2.46	\\
32	&	RC	&	2900	&	0	\\
33	&	RC	&	1300	&	1.65	\\
34	&	RC	&	900	&	0	\\
35	&	RC	&	2200	&	0	\\
36	&	RC	&	700	&	0	\\
37	&	RC	&	3000	&	0	\\
	\bottomrule		
	\end{tabular}
}{Mean values of network parameters for recovery} {mean_network_parameter_recovery}

\mytable{
	\begin{tabular}{c | r r l}
		\toprule
		Damage Level	&	Condition	&	$\#$ of bridges	&	Bridge ID	\\
		\hline
		 \multirow{2}{*}{0} & No damage & 15 & 4, 6, 7, 8, 13, 18, 20, 23 \\
		 &  &  & 25, 29, 32, 34, 35, 36, 37 \\ 
		\multirow{2}{*}{0-1}	&	Slight damage	&	11	&	2, 11, 12, 14, 15, 16 	\\
		&  &  & 19, 24, 30, 31, 33 \\
		1-2	&	Moderate damage	&	5	&	1, 9, 17, 22, 27	\\
		2-3	&	Extensive damage	&	3	&	5, 10, 26	\\
		3-4	&	Collapsed	&	3	&	3, 21, 28	\\
		\bottomrule		
	\end{tabular}
}{Mean values of network parameters for recovery} {mean_network_parameter_recovery}

\mytable{
	\begin{tabular}{c | c c  c c}
		\hline
		Parameters	&	Notation	&	Distribution	&	Mean	&	COV	\\
		\hline		
		Average Daily Traffic (ADT)	&	$F_{ij}(\xi)$	&	Uniform	&	In Table \ref{mean_network_parameter_recovery}	&	0.05	\\
		Month of restoration	&	$D_d(\xi)$	&	Normal	&	In Table \ref{mean_network_parameter_recovery}	&	0.05	\\
		\hline
	\end{tabular}
}{Statistics of network parameters for recovery} {statistics_network_parameter_recovery}

\subsection{Optimal schedule for network restoration}
\label{subsec_optimal_schedule}

The optimal bridge restoration schedule is determined using the GA summarized in Figure \ref{flow_chart_ga}.  The specific tuning parameters and stopping criterion play a critical role in the efficacy of the algorithm. These parameters include population size, crossover rate, mutation rate, and elitist mechanisms (see, e.g., \citep{davis1997accuracy}).  A generation in the GA refers to one complete cycle as depicted in Figure \ref{flow_chart_ga}.   Table \ref{parameters_setting_ga}	 summarizes the GA parameters used in this illustration, which were determined after extensive experimentation. 

\mytable{
	\begin{tabular}{c  c}
		\hline
		Parameters	&	Value	\\
		\hline		
	Population	&	50	\\
Cross-over rate	&	0.9	\\
Mutation rate	&	0.3	\\
Elitist selection	&	20	\\
Maximum generations	&	1,000	\\
Maximum time	&	1,800 seconds	\\
		\hline
	\end{tabular}
}{Parameters setting in the GA} {parameters_setting_ga} 

This experiment generates 500 random instances using MCS and optimize the restoration of each of the 500 instances using the aforementioned GA. Figure \ref{probability_density} illustrates the distribution of optimal network recovery time under stochastic conditions.  Considering the uncertainties associated with restoration duration and ADT for each bridge, the mean network recovery time is 21.3 months and the standard deviation is approximately 20 days.  The $\Delta t$ and $t_{k}$  in Eq. (\ref{SRT}) are set to be 1 day and 50 months, respectively. In the remainder of this section, the author focuses our discussion on the optimal solution for a single instance of the simulation, $\widetilde{x}$ , where all the random variables are taken as their mean values.  This discussion is applicable for all other simulated scenarios.   

\myfigure{
	\centering
	\includegraphics[width=\textwidth]{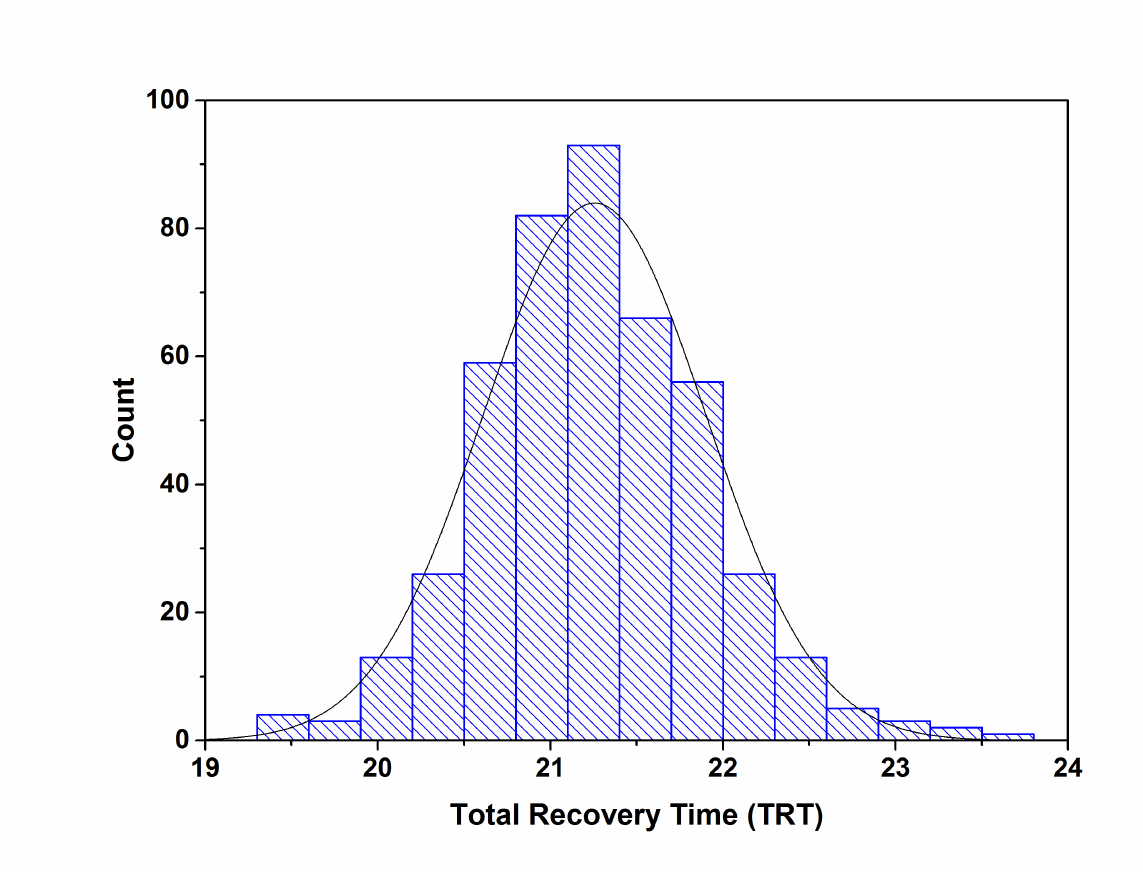}
}{Probability density of the total recovery time (TRT)
}{probability_density}

Figure \ref{recovery_fitness_function} shows the fitness values defined by Eq. (\ref{recovery_fitness_function}) over the CPU time with three random initial populations in genetic algorithm.  GA is allowed to run up to 60 minutes and it is observed that GA converges in 30 minutes. Therefore, the maximum allowable running time is set as 30 minutes for all experiments in this section. The best fitness value in the initial GA population is 78.14 and decreases to 49.95 in 30 minutes (600 generations).  The fitness value of 49.95 is the sum of the network TRT (21.42 months) and SRT (28.53 months). Figure \ref{trt_srt} illustrates the quality of the near-optimal solution from each generation of GA with respect to the fitness function, which shows that TRT and SRT are highly positively correlated. Note that this correlation will be much less if all feasible solutions are included in the Figure \ref{trt_srt}.  Even with only the near-optimal solutions, Figure \ref{trt_srt} indicates that for a given TRT there exist many alternative strategies with different SRT representing different recovery trajectories. This observation confirms that TRT alone, if used as the sole objective for recovery, is not sufficient to ensure the optimal recovery schedule with the best trajectory.  The SRT ensures the recovery efficiency among the alternative schedules that represent the same recovery time.

\myfigure{
	\centering
	\includegraphics[width=\textwidth]{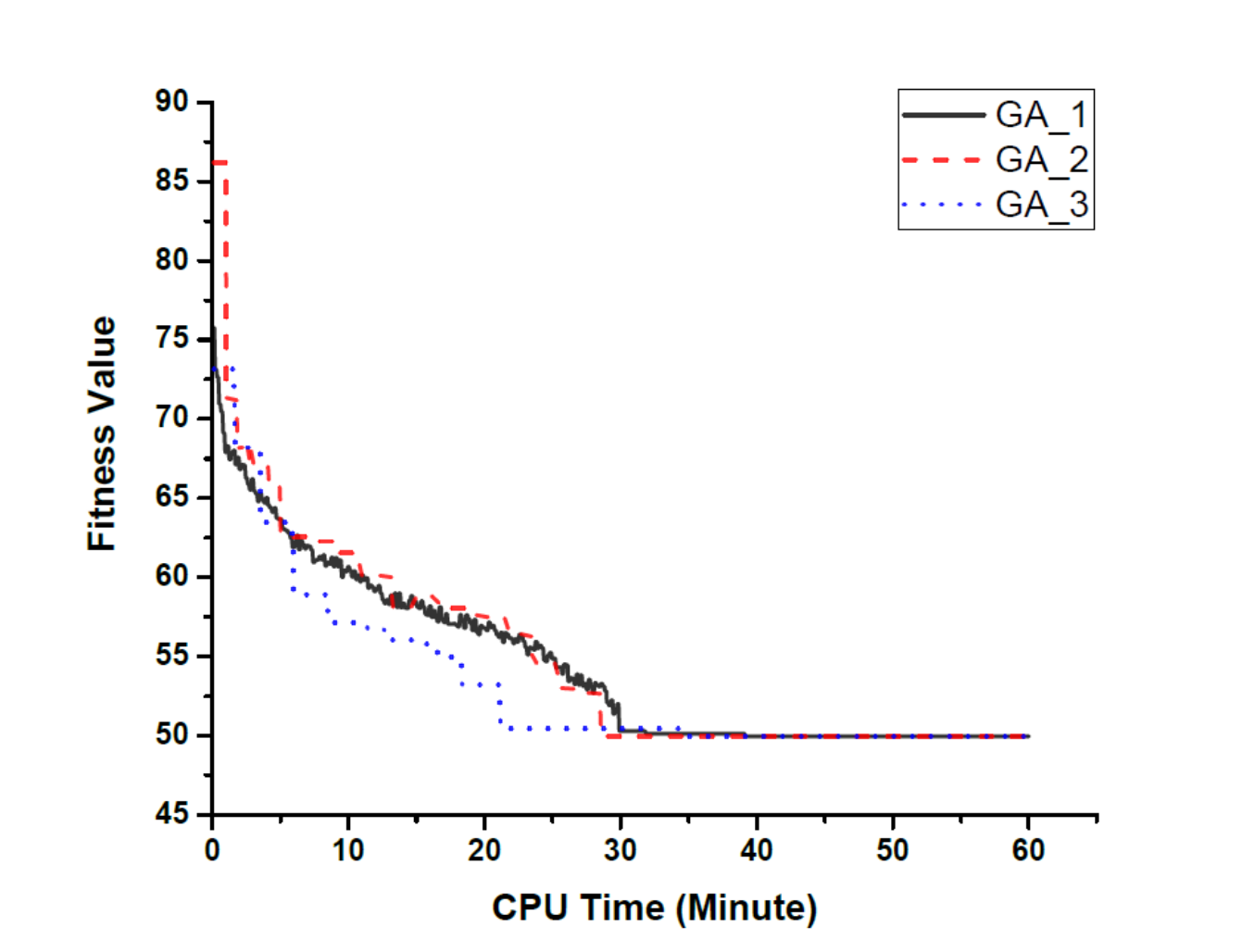}
}{Fitness values over time with three random initial populations in GA 
}{recovery_fitness}

\myfigure{
	\centering
	\includegraphics[width=\textwidth]{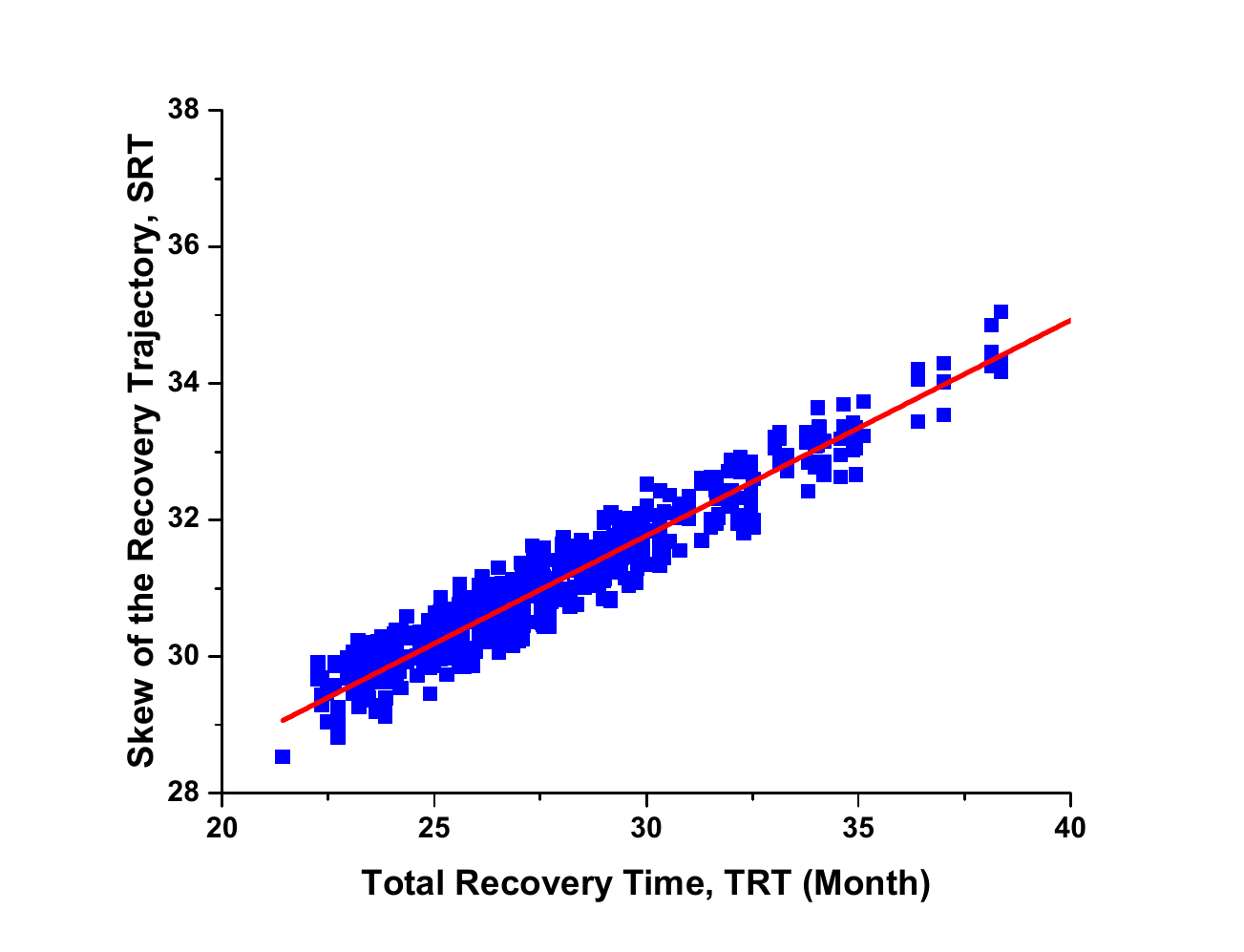}
}{TRT and SRT of the optimal solutions
}{trt_srt}

The optimal network restoration schedule is illustrated in Figure \ref{optimal_schedule}, which shows the times at which restoration intervention is initiated and completed for each damaged bridge. The length of the bar associated with each bridge is the duration of the intervention.  This optimum schedule allows all damaged bridges in the network to be restored in less than 21 months, given that the maximum of 4 bridges can be repaired simultaneously.  To evaluate the efficiency of this scheduling, the work compares this optimal solution with a naïve network restoration schedule in which all the damaged bridges tabulated in Table \ref{statistics_network_parameter_recovery} are repaired in a randomly selected sequence as shown in Figure \ref{random_schedule}.  In this case, 23.5 months are required to completely restore the network.    Figure \ref{recovery_trajectory} illustrates that the network recovery trajectory associated with the optimal recovery plan is superior to that of the random restoration sequence.  Note that these two recovery schedules employ the same amount of resources and the only difference between them is the sequence in which the damaged bridges are restored to their pre-event conditions. These trajectories based on the mean  restoration time. Because the variance of all work crews are identical, the GA solution and naive solution share the same standard deviation. However, if the COV varies across teams, variance comparison between optimal and random solutions should be added.
Although, the 2.5-month seems to be an insignificant improvement in this illustration, one that could be achieved by adding some common sense to the completely random strategy, the advantage of this scheduling algorithm would become more obvious when dealing with large, extensively-damaged networks where the decision variables, possible alternative strategies and constraints create a complex decision problem where intuition may not apply.    Furthermore, if a short-term network recovery objective is to ensure that there is at least one path, on average, between each O-D pair (i.e. the performance metric WIPW is equal or greater than 1), then the optimal recovery scheduling achieves this objective 8 months earlier than the random scheduling.  It is noted that the bridges selected in the early phase in the optimal solution, i.e. bridges 14, 24, 28, are the bridges with the highest impact on the overall network performance, as measured by WIPW, because they are shared by multiple O-D pairs and are close to the emergency facilities.  These results demonstrate that the optimal scheduling of bridge restoration can greatly improve the efficiency of the transportation network recovery. Bridge authorities can use this information to allocate their limited resources intelligently to both minimize total network recovery time and to maximize the recovery efficiency.

\myfigure{
	\centering
	\includegraphics[width=\textwidth]{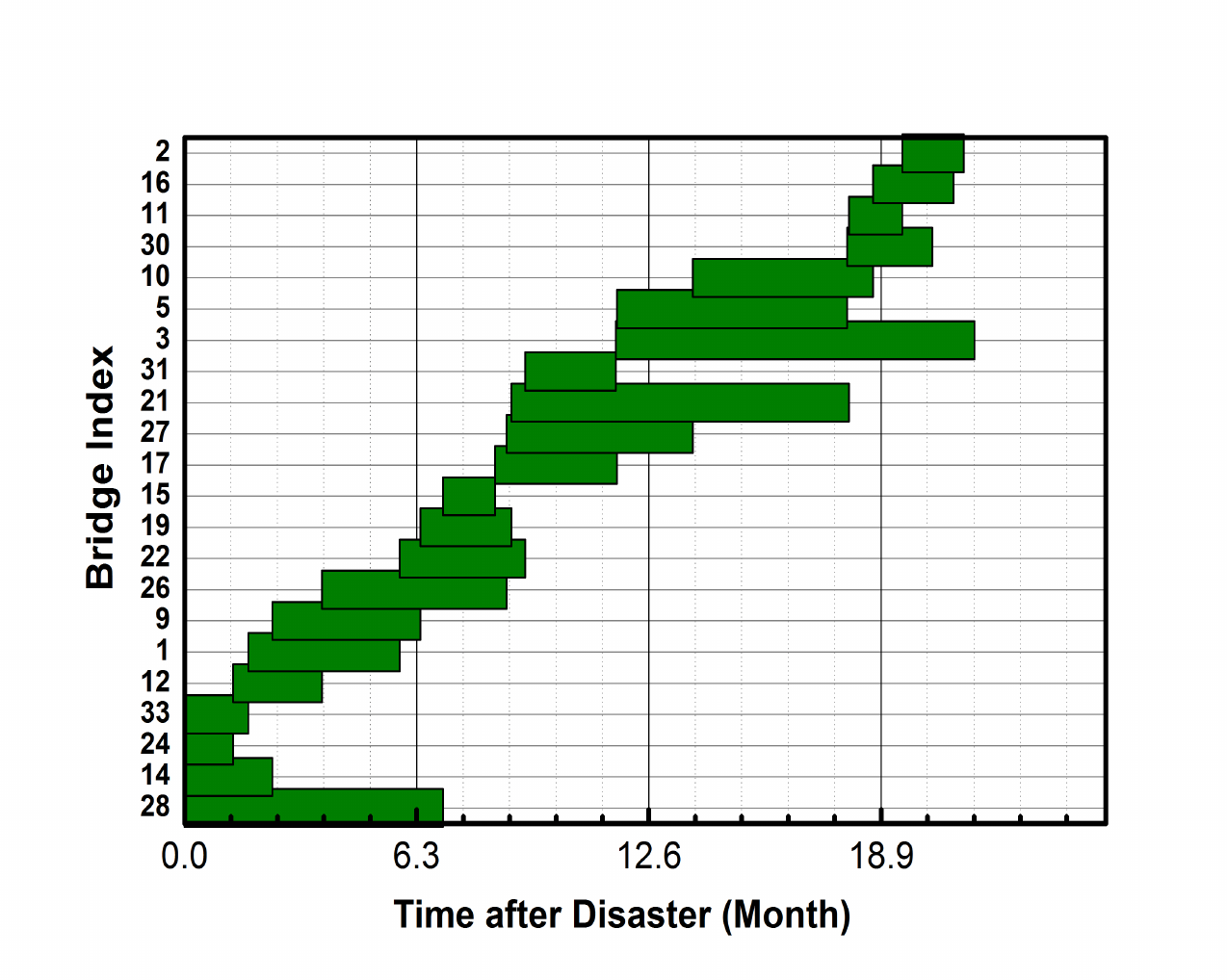}
}{Optimal scheduling with times for initiation and completion of restoration for each damaged bridge}{optimal_schedule}

\myfigure{
	\centering
	\includegraphics[width=\textwidth]{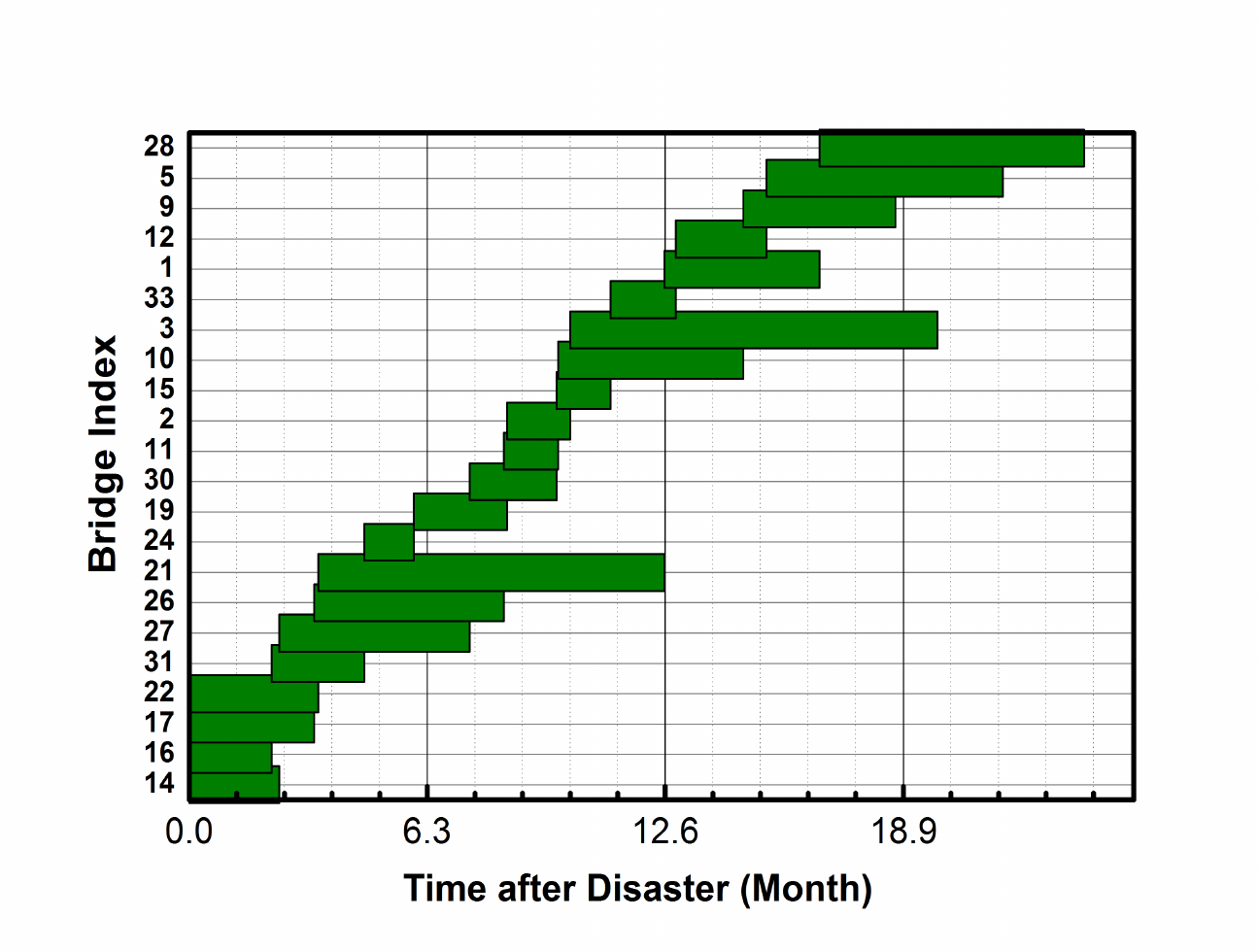}
}{Naïve (random) bridge restoration sequence}{random_schedule}

\myfigure{
	\centering
	\includegraphics[width=\textwidth]{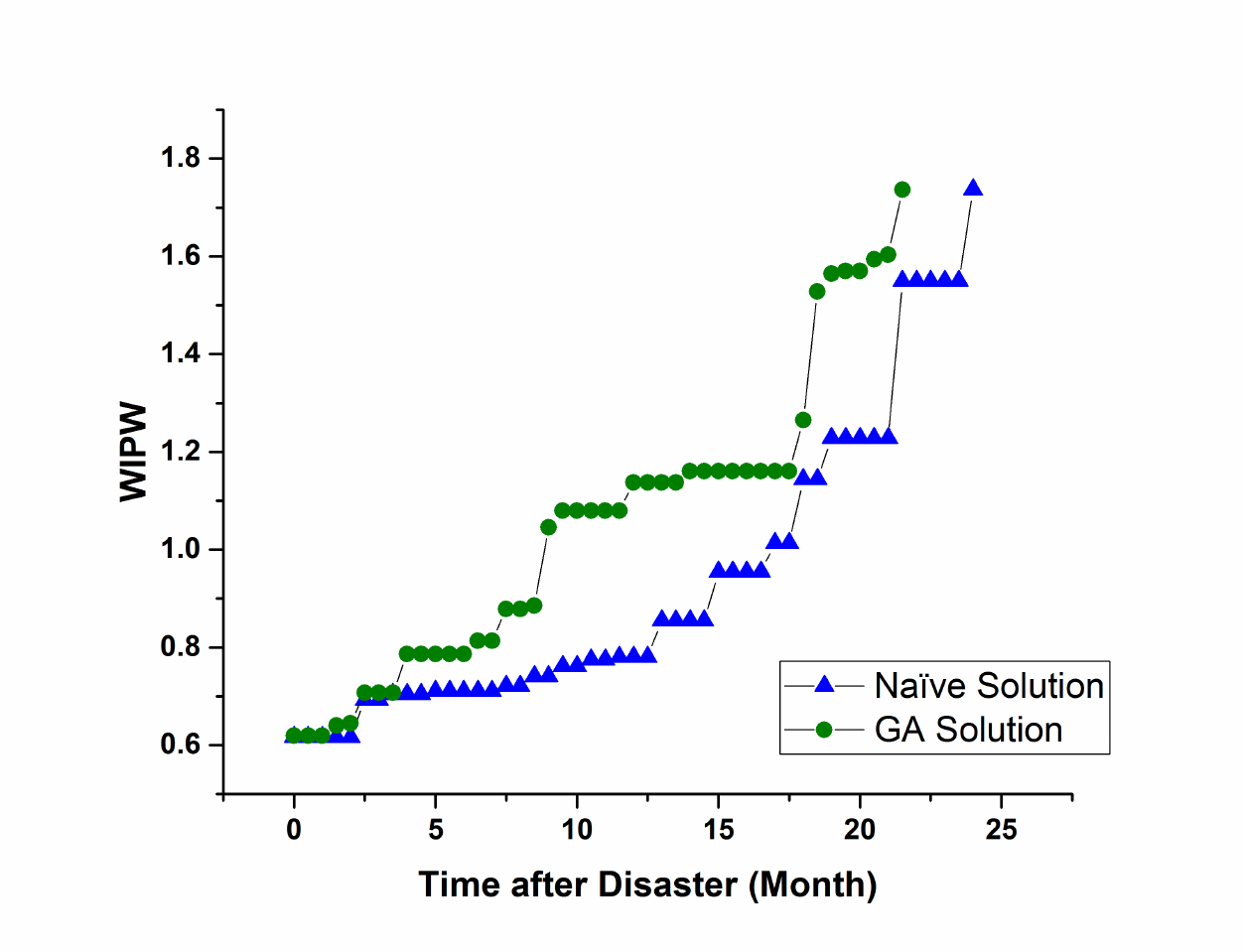}
}{Network recovery trajectories with different restoration schedules}{recovery_trajectory}

To investigate the sensitivity of the scheduling to the resourcefulness of the community (an important characteristic of community resilience), rather than assuming a constant amount of recovery resources (i.e. $N_{SI}^{max}=4$ ), it is now supposed that the maximum number of simultaneous restoration interventions,  $N_{SI}^{max}$, is a dynamic quantity, which changes the recovery process to reflect different community social-economic structures and investment patterns.  For instance, poorer communities may have fewer resources available immediately following the event and may depend on outside resources which take time to deploy.  To address this case, it is assumed that the value of $N_{SI}^{max}$ is 2 for the first seven months, 4 from the seventh to fourteenth months, and 6 after the fourteenth month.  Conversely, a wealthy community might have more resources on hand to address the initial impacts of a disaster, but may decrease the investment in recovery as the network resilience improves.  To model such a situations, it is assumed that $N_{SI}^{max}$ is dynamically decreasing from $N_{SI}^{max} = 6$ for the first seven months, 4 for the following seven months, and then 2 until all bridges are repaired.  Figure \ref{recovery_trajectoy_multi_capacity} illustrates the associated resilience trajectories for the three communities with different resource investment patterns: wealthy (6->4->2), middle-income as considered previously with constant recovery resources (4->4->4) and poor (2->4->6).  The recovery time of the wealthy and middle-income communities are about the same; however, the skew of the recovery trajectory of the wealthy community indicates a more efficient recovery schedule. The network performance metric WIPW of the wealthy community achieves the value of 1 at month seven which is two months earlier than is achieved by the average community. Furthermore, although the poor community has 6 simultaneous interventions in progress after 14 months, its total recovery time still is 3 months longer than that of both the wealthy and the average communities. It is evident that the resource allocation pattern as a function of time can significantly impact the efficiency of the network recovery.

\myfigure{
	\centering
	\includegraphics[width=\textwidth]{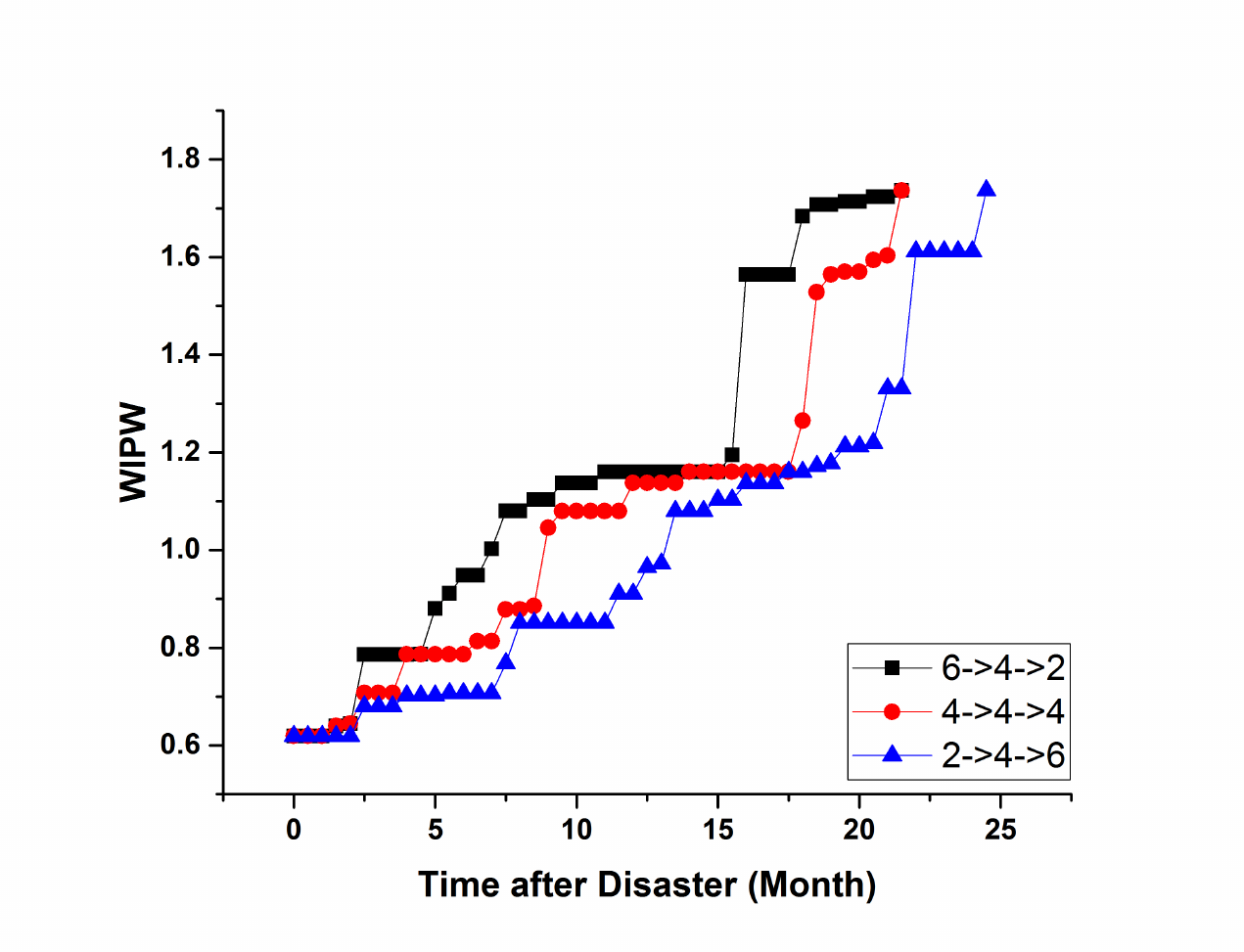}
}{Resilience as a function of time under different recovery capability}{recovery_trajectoy_multi_capacity}

\section{Summary}
\label{sec_recovery_summary}

This chapter has presented a novel and dynamic model to optimize the restoration schedules of transportation networks following extreme events.  The model incorporates a network resilience-based performance metric, recovery trajectory, community resourcefulness, and uncertainties relating to damage levels and restoration duration of damaged bridges in a quantitative resilience-based decision framework for road network recovery.  This work employs a powerful metaheuristic approach to find near-optimal solutions for recovery scheduling.  The following three aspects of the restoration process are addressed. 

First, the author introduced a two-dimensional metric - total recovery time (TRT) and the skew of recovery trajectory (SRT) - as measures for the network recovery planning, both of which are significantly affected by the restoration sequence when resources for post-disaster recovery are limited.  This two-dimensional metric can be used collectively as the objectives in identifying the optimal post-event scheduling for the network recovery.   

Second, according to Figure \ref{recovery_trajectory} and Figure \ref{recovery_trajectoy_multi_capacity}, simple functions (linear or nonlinear) for recovery trajectory apparently do not exist in even simple networks such as the one considered herein. Different prioritization schedules can lead to significant changes in both TRT and SRT.  The dynamic nature of available resources (immediately on-hand or delayed due to outsourcing) can have a notable impact on the network recovery time and trajectory.  Even under circumstances in which the total amount of resources utilized in the recovery period is the same, the recovery trajectory is sensitive to the timing in which such resources are made available.   

Third, both the scheduling framework and the genetic algorithm used to obtain the optimal solutions can be extended to handle large and complex networks. This problem is NP-Hard \citep{lenstra1977complexity}; accordingly, the problem solving time may increase significantly with the size of the network. However, there is no limitation regarding network size on the mathematical formulation itself presented herein. Moreover, many advanced techniques can be employed to reduce the solving time, such as parallelized computing.  Finally, the distribution of recovery time under stochastic conditions, the intelligent network restoration scheduling (intervention time and completion time) and the corresponding recovery trajectory can be visualized easily, which provides a wealth of information for bridge authorities to make decisions regarding the post-disaster recovery of transportation networks.

\chapter{Stage-wise Decision Framework for Transportation Network Resilience Planning }
\label{ch_stage_wise}

\section{Introduction}
\label{sec_stage_introduction}

In Chapter \ref{ch_WIPW} and \ref{ch_recovery}, the author has discussed the definition of a new metric to quantify the transportation network performance, i.e., WIPW, and proposed  two more metrics for recovery, i.e., total recovery time and the skew of recovery trajectory. The corresponding stochastic optimization models are developed to support resilience-based decision for both pre-disaster mitigation and post-disaster recovery. In this chapter, this research present a more flexible decision framework by adjusting the weights in WIPW according to the different stages with the nonoccurence  of extreme events to improve resilience of the transportation network. 
Resilience - defined as the ability of social units (e.g., organizations, communities) to mitigate hazards, contain the effects of disasters when they occur, and carry out recovery activities in ways that minimize social disruption and mitigate the effects of future hazards - can be measured, as shown in Figure \ref{concept_of_resilience}, by four major attributes: robustness (ability to withstand extreme events), rapidity (speed of recovering from impacts of extreme events), redundancy (alternative or backup components within the system), and resourcefulness (availability of emergency resources to respond to disasters) \citep{bruneau2003framework}. A notable growing body of literature has been published to evaluate the performance of transportation systems from the perspective of network resilience to natural hazards \citep{chang2001measuring, murray2006comparison,ta2009structuring,cox2011transportation, frangopol2011resilience,ip2011resilience,ccetinkaya2015multilevel, gillen2016measuring, zhang2016resilience, zhang2017resilience}.  Research studies have recently extended to decision formulations regarding risk mitigation, response and recovery of roadway networks at a community or regional scale.  \citet{bell2008attacker} used travel time reliability as the network performance indicator to study the transportation vulnerability under terrorist attacks, in which decisions regarding the path used in the road network were made to minimize the maximum of expected loss from a predefined disruption. \citet{bell2017investigating} proposed capacity weighted spectral partitioning to identify the bottleneck of transportation network, which does not require information of origin-destination matrix and path assignment.  \citet{chang2001measuring}  developed a decision method for post-disaster evacuation using a proposed transportation simulation model which was capable of capturing the change of traffic pattern following seismic events. \citet{miller2012measuring} proposed a two-stage stochastic programming model to maximize transportation network resilience, defined as the expected traffic demand supplied following a disastrous event, by enhancing pre-disaster preparedness and optimizing post-disaster scheduling.  This model was later extended in \citet{faturechi2014travel} to a three-stage, bi-level stochastic mathematical program, in which the upper-level is a sequence of actions for both pre-disaster mitigation and post-disaster restoration, and the lower-level reconstructs the routine for affected users using partial user equilibrium model.   \citet{sheu2014method} presented a survivor perception ``attitude'' resilience conceptual model with the focus on maximizing the survivor resilience after disaster.  \citet{sheu2014method} integrated three sub-networks (shelter network, medical network, and distribution network) to build a seamless centralized emergency supply network in response to natural disasters.  \citet{zhang2017resilience} optimized roadway network recovery from an earthquake event by incorporating network topology, redundancy, traffic flow, damage states and available resources into a stochastic decision processes, resulting in an optimal schedule for sequencing restoration interventions for all damaged bridges, which lead to the fastest (in terms of time) and most efficient (in terms of indirect loss) network recovery process.  

Reviewing these studies has revealed the following:  i) most research studies only investigated the decision formulation or process at one point in time with respect to the occurrence of a hazard, and ii) different studies selected different network performance metrics as the basis for their decisions at hand.  It has become apparent that the time at which a decision is made (with respect to the hazard occurrence) determines the situation (or constraints) and purpose (or objectives) of the decision process.  It is therefore necessary for a decision framework to reflect these dynamic perspectives across the different stages of resilience planning through carefully defined performance metrics, interventions, objectives and constraints. There is an increasing need for a consistent methodology to support stage-wise resilience planning of roadway networks as more and more communities are beginning to incorporate resilience as a concept in their day-to-day risk management practices.

\section{Stage-wise decision framework}
\label{sec_stage_framework}

The resilience concept, as illustrated in Figure \ref{concept_of_resilience}, clearly has a time dimension. On one hand, risk mitigation and disaster management interventions at different points in time with respect to the occurrence of hazard are very different, necessitating different decision support \citep{mcallister2015community}. On the other hand, these decisions, although made at different planning stages, should serve the same overarching goal of enhancing network resilience in a systematic fashion.  This section proposes a stage-wise decision framework, as illustrated in Figure \ref{stage_wise_framework}, to support resilience planning of the transportation network in three phases: Stage I - pre-disaster mitigation; Stage II - post-disaster emergency response; and Stage III - long-term recovery; i.e.: 

\begin{itemize}
	\item Stage I refers to the mitigation planning stage prior to a disaster. An often-used mitigation strategy for roadway networks is to retrofit network components (i.e. bridge and roads), with an objective of enhancing network robustness (cf. Figure  \ref{concept_of_resilience}) and subsequently reducing damage and functionality loss of the network when a hazard event occurs.  This pre-event mitigation decision problem concerns how many and which network components should be selected for retrofit subject to the available finances and other resource limitations. The author proposes, for Stage I, a prioritization mechanism for retrofit interventions that maximizes the network robustness under budget constraints.  
	\item Stage II refers to the chaotic emergency response stage during or shortly after the occurrence of a hazard event.  The primary functionality of a roadway network in Stage II is to connect affected population to critical facilities, e.g. hospitals, fire stations, police stations, etc., and to provide routes enabling repair crews to access damaged components of utility (e.g. power and water) networks.  This functionality requirement in Stage II is often resumed through rapid repairing or establishing temporary pathways under time pressure. This study therefore presents a decision method to identify the damaged network components for rapid repair that can resume, within the shortest time frame, the connectivity between the O-D pairs that are critical for life rescue.  
	\item Stage III refers to the long-term recovery phase, in which the impaired network is gradually restored to its pre-disaster functionality level through repair interventions.   The recovery time of the network is the time it takes to repair or reconstruct all the damaged roads and bridges, which is governed by the availability of construction resources (i.e. funding, contractors, construction materials, etc.) and how they are distributed spatially along the time steps of the recovery phase. For Stage III, the author proposes a scheduling approach to sequence the repair interventions constrained by the characteristic of resource availability in order to resume network functionality efficiency with the most desirable recovery trajectory and the shortest recovery time.  
\end{itemize}

The proposed stage-wise decision framework includes two major components: 1) a stage-wise network metric (SWM) system, that is formulated to be compatible with the unique objectives of the specific decisions in each of the planning stages; and 2) a stage-wise formulation of the decisions (SWD), using the SWM system as the network performance indicators in different planning stages as the basis for decision-making. The formulations of SWM and SWD are discussed in Sections \ref{sec_stage_metrics} and \ref{sec_stage_formulation}, respectively.

\myfigure{
	\centering
	\includegraphics[width=\textwidth]{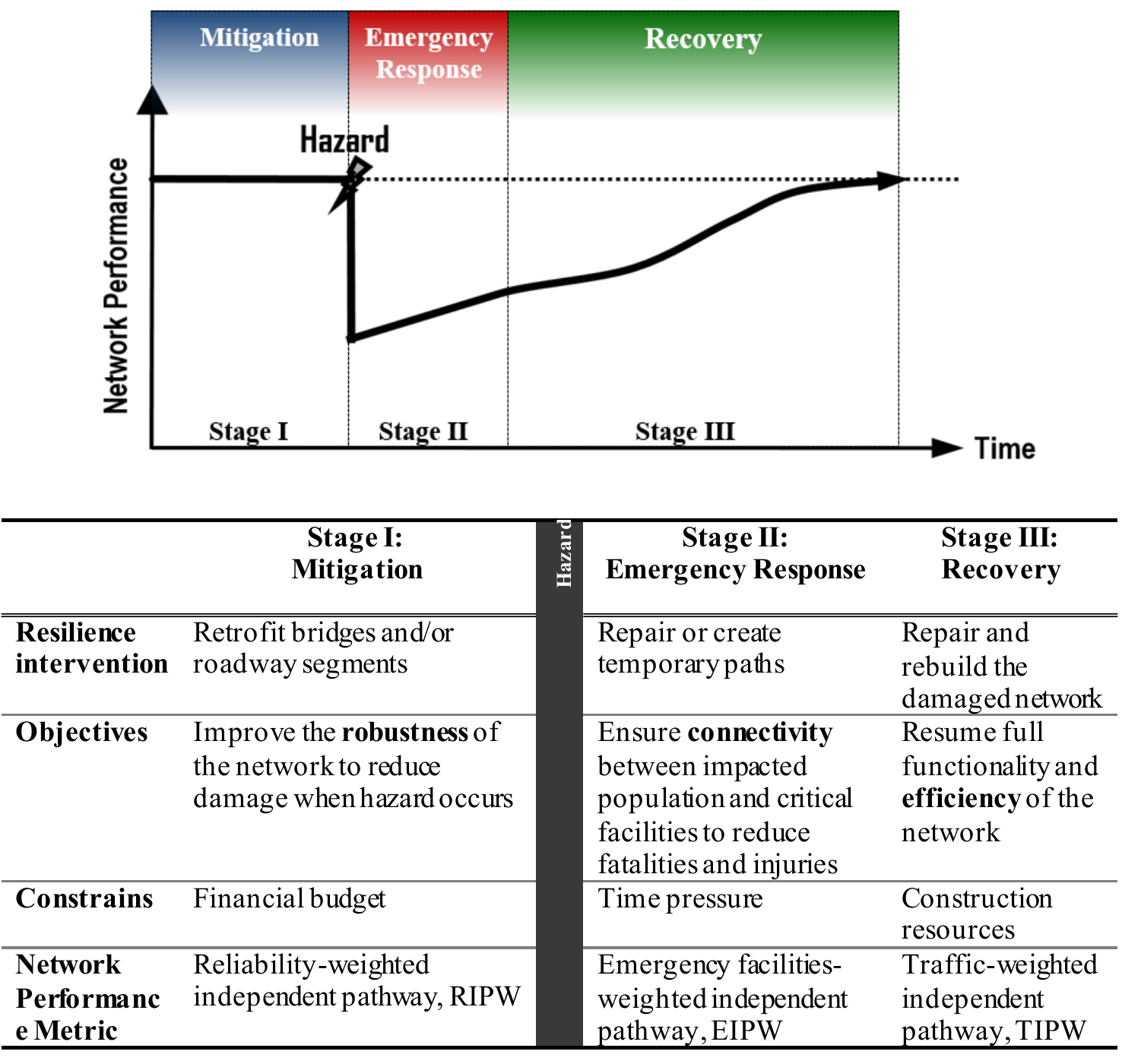}
}{Illustration of the stage-wise decision framework}{stage_wise_framework}

\section{Stage-wise network performance metric system}
\label{sec_stage_metrics}

Different objectives of distnctive planning phases require specific network performance metrics, and different choices of metrics used in a decision formulation will ultimately result in different decisions.  \citet{cutter2008place} emphasized that the most important criteria for metric selection is validity, which speaks to the question of whether the metric is representative of the resilience dimension or the decision of interest. General-purpose network performance metrics, either topology-based or functionality-based, usually cannot appropriately reflect the specific objectives of the decisions in each of the resilience planning stages. In particular, topology-based metrics (e.g. connectivity, density, etc.) alone do not reflect network functionality requirements which often are the focuses of resilience-based decisions; on the other hand, functionality-based metrics (e.g. travel cost, maximum traffic capacity) ordinarily require ``too much'' information (e.g. pre- and post-event network supply-demand at different spatial and temporal resolution), which may not be readily available and, more importantly, may not be directly relevant or representative of the specific objectives of the decisions to be made. That is, a metric that incorporates ``too much'' information could lose its sensitivity as a performance indicator to a specific dimension of network performance that is directly relevant to the current decision.  To best support the stage-wise resilience planning of roadway networks, the author first develops a stage-wise network performance metric (SWM) system that includes three metrics all based on the concept of independent pathways (IPW): reliability-weighted IPW (RIPW), emergency facility-weighted IPW (EIPW), and daily traffic-weighted IPW (TIPW), to support decisions in Stage I, II and III, respectively.  

Edge-independent pathways, i.e. IPW, referred to in the SWM system, is an important concept in network science.  The set of all IPWs in a network is defined as all paths between a two distinct nodes that do not share any common edges (e.g., road segments)\citep{newman2011structure}.  IPW provides a measure of network redundancy, which is an essential topological feature of roadway networks that strongly and positively correlates to network resiliency as observed in the 1989 Loma Prieta earthquake in California \citep{ardekani1992transportation}.  Recognizing that redundancy in transportation network  enhances resiliency, the author employs IPW is as a common substrate for the SWM system;  and depending on the specific objective of the decisions in each stage, the network functionality-specific and component-specific attributes are weighted into the formulation of the IPW-based SWM system, as presented next. 

The author describes a transportation network on a directed graph $G=(V,A)$ when $V=\{1,2,...,n\}$ is the set of nodes and $A=\{1,2,..,m\}$ is the set of edges that represent road segments without or with maximum of one bridge. The set $V$ is partitioned into two mutually exclusive sets $E=\{1,2,...,e\}$ comprised of emergency nodes (representing critical facilitates, e.g., fire stations and hospitals) and $N=\{e+1, e+2, ..., n\}$ of non-emergency nodes (representing major destinations, e.g., residential areas, economic hubs, and major road intersections). Let $K_{ij}$ and $P_{ij}^k$, respectively, denote the total number of IPWs and the $k^{th}$ IPW between node $i\in V$ and $j\in V$. Each IPW is a set of ordered edges connected in series. An algorithm for computing $K_{ij}$ and searching for $P_{ij}^{k}$ between all node pairs can be found in \citet{zhang2016resilience}. Let $IPW_i$ denote the average number of IPWs between node $i \in V$ and all the other $n-1$ nodes, i.e., 

\begin{align}
IPW_i = \frac{1}{n-1}\sum_{j=1}^{n-1}K_{ij} \label{ipw}
\end{align}

The IPW of network $G$ is defined as the overall average of all $IPW_{i}$ values for every node $i\in V$,
\begin{align}
IPW(G) = \frac{1}{n}\sum_{i=1}^{n} \label{network_ipw}
\end{align}

Note that IPW is independent of traveler behavior and origin-destination (O-D) demand values and can be computed efficiently using either Dijkstra's algorithm or the Ford-Fulkerson algorithm for max flows \citep{ahuja1993network}.

\subsection{Reliability weighted IPW, RIPW}
\label{ripw}

In Stage I, the purpose of pre-disaster mitigation through component retrofit is to enhance the network robustness as discussed in Section \ref{sec_stage_framework}, i.e. to reduce the network damage when hazard occurs.  Increasing reliability (or reducing failure probability) of selected network components to minimize the possible damages to the IPWs between all O-D pairs is the objective of retrofit interventions. Accordingly, the author defines reliability-weighted IPW, RIPW, as the network performance metric to support the Stage I pre-event mitigation decisions.  Assuming conservatively that reliabilities of edges are statistically independent, the reliability of path $p_{ij}^{k}$, denoted as $R_{ij}^{k}$, is the product of the reliabilities of all edges included in the path $P_{ij}^{k}$,
\begin{align}
R_{ij}^{k} = \prod_{l\in P_{ij}^{k}} q_{l} \label{path_reliability}
\end{align}

where $l$ and $q_{l}$ are the edge index and the corresponding reliability. The average number of reliable IPWs for node $i \in V$ is denoted as $Q_i^R$ and calculated as shown in Eq (\ref{node_ripw}),

\begin{align}
Q_{i}^{R} = \frac{1}{n-1}\sum_{j=1}^{n-1}\sum_{k=1}^{n}R_{ij}^k \label{node_ripw}
\end{align}

Let $Q^{R}(G)$ denote the RIPW of network $G$ which is computed in Eq (\ref{network_ripw}):

\begin{align}
Q^{R}(G) = \frac{1}{n}\sum_{i=1}^{n}Q_{i}^{R} \label{network_ripw}
\end{align}

A critical step to compute RIPW is to quantify reliabilities of the edges (i.e., road segments and bridges), which depends on the specific hazard of interest, and often are evaluated through fragility analyses associated with damage states of interest.

\subsection{Emergency node-weighted IPW, EIPW}
\label{subsec_eipw}

In Stage II, immediately after a disruptive event, emergency managers need to quickly identify and establish emergency routes to send rescue teams and relief resources to the affected population. The connectivity between critical facilities and the people who need to be rescued is of the paramount concern \citep{jones1995emergency, higgs2004literature}. Accordingly, it is defined that the emergency facility-weighted IPW, EIPW, as the network performance metric for Stage II.  EIPW measures the average number of independent pathways between emergency nodes and non-emergency nodes. The EIPW is highly dependent on the post-event damage condition of the network. Let $d_{ij}$ denote the damage status of edge $(i,j) \in A$, which can be measured on a 0 to 4 scale, corresponding to the damage level of none, slight, moderate, extensive and complete, respectively, as termed in \citet{scawthorn2006hazus}.  Edges with the damage conditions of extensive or complete are considered unserviceable. The serviceability of edge $(i,j) \in A$ is denoted $s_{ij}$ and defined as,

\begin{align}
s_{ij} = \begin{cases}
1, d_{ij} \leq 2\\
0, \text{otherwise}
\end{cases}. \label{edge_service}
\end{align}

Eq. \eqref{path_service} defines the serviceability of $P_{ij}^{k}$, denoted as $s_{ij}^{k}$, as the product of the serviceability of all road edges included in the path $P_{ij}^k$:

\begin{align}
s_{ij}^{k} = \prod_{l\in P_{ij}^{k}}s_{l} \label{path_service}
\end{align}

If $s_{ij}^{s}$ equals 1, the path is serviceable and if it equals 0, it is not. Let $Q_{i}^{E}$ denote the EIPW of node:

\begin{align}
Q^{E}(g) = \frac{1}{e}\sum_{j\in E}\sum_{k=1}^{K_{ij}} s_{ij}^{k} \label{node_service}
\end{align}

The EIPW of a network is defined as the average EIPW of non-emergency nodes, which is computed as,
\begin{align}
Q^{E}(G) = \frac{1}{n-e}\sum_{i\in N} Q_{i}^{E} \label{network_service}
\end{align}

If $Q^{E}(G)$ equals 0, the transportation network cannot support any emergency actions.

\subsection{Average daily traffic weighted IPW, TIPW}
\label{subsec_tipw}

In Stage III, the author uses the pre-event average daily traffic (ADT) weighted IPW, or TIPW, as the network performance metric fundamental to decisions. The ADT on roads and bridges are the field measurements routinely maintained by the Federal or State Department of Transportation.  ADT data may be more accurate than current traffic assignment models at representing real traffic patterns since the path choice of travelers are only partially rational, whereas traffic assignment models commonly assume decisions to be based on distance or travel time \citep{yagar1971dynamic, patriksson2015traffic}. It is emphasized that the work uses pre-event ADT data as the benchmark for recovery decisions, as the goal in Stage III is to resume the full pre-event functionality level of the network through a timely and effective recovery process. The service level of each edge $(i,j)$ is idealized as $1-d_{ij}/{4}$.  For example, if an edge is completely damaged, $d_{ij}$ is set to be 4, and the corresponding service level is 0.  The service level of the path $P_{ij}^{k}$ is approximated as the product of the service levels of all arcs $l \in P_{ij}^{k}$. Let $A_{l}$ denote the ADT of edge $l\in P_{ij}^{k}$. Define $A_{ij}^{k}$, the ADT of $P_{ij}^{k}$, as the minimum ADT of all edges on that pathway, i.e., 

\begin{align}
A_{ij}^{k} = \text{min} \{A_{l} | l \in P_{ij}^{k} \} \label{path_adt}
\end{align}

The normalized ADT of the path is defined as:

\begin{align}
\widetilde{A}_{ij}^{k} = \frac{K_{ij}A_{ij}^{k}}{\sum_{p=1}^{K_{ij}}A_{ij}^{p}}
\end{align}

Note that for any node pair $(i,j )\in V$, $\sum_{k=1}^{K_{ij}}\widehat{A}_{ij}^{k}=K_{ij}$. Let $Q_{i}^{A}$ denote the TIPW of node $i\in V$ and

\begin{align}
Q_i^T = \frac{1}{n-1}\sum_{j=1}^{n-1}\sum_{k=1}^{K_{ij}}\widehat{A}_{ij}^{k}\min_{l\in P_{ij}^{k}}\left(1-\frac{d_{l}}{4}\right) \label{node_tipw}
\end{align}

The TIPW of a network is further defined as:

\begin{align}
	Q^{T}(G) = \frac{1}{n}\sum_{i=1}^n Q_{i}^T \label{network_tipw}
\end{align}

For Stage III, a damaged transportation network is considered fully recovered if all the damaged roadway segments are restored, i.e., the network TIPW computed using Eq. \eqref{network_tipw} returns to its pre-disaster level (without damaged network components). 

The defined SWM - comprised of RIPW, EIPW and TIPW - are tailored and sensitive to the unique objectives of the specific decisions they are serving, at the same time, they ensure a level of consistency among the decisions in different planning stages. These IPW-based metrics can be employed individually to support decisions at each planning stage, or used together as a system in the SWD formulated in Section \ref{sec_stage_formulation}.

\section{Formulation of the stage-wise decision framework}
\label{sec_stage_formulation}

\subsection{Stage I - pre-disaster risk mitigation}
\label{subsec_ripw}

At Stage I, the decision is to make selections from a set of candidate edges to maximize the RIPW and simultaneously minimize the related investment cost. The mitigation decisions, represented as the binary decision vector $x$, is comprised of a component $x_{ij}$ for each $(i,j) \in A$ as defined as below,

\begin{align}
x_{ij} = \begin{cases}
 1, \text{edge } (i,j) \quad \text{is selected} \\
 0, \text{otherwise}
\end{cases}, \forall (i,j) \in A \label{ripw_selection}
\end{align}

The first objective of the decision process is to maximize the expected RIPW: 

\begin{align}
	\text{max} \quad Q^{R}(x) = E_{\widetilde{\xi}_{q}}\left\{\frac{1}{n}\sum_{i=1}^{n}\frac{1}{n-1}\sum_{j=1}^{n-1}\sum_{k=1}^{K_{ij}}\left[(1-x_{l})q_{l}(\widetilde{\xi_{q}}) + x_{l}q_{l}'\right]\right\} \label{max_ripw}
\end{align}

where $q_{l}(\xi_{q})$ is the realization of the random variables $q_{l}$ representing the reliabilities of edge $l\in P_{ij}^{k}$ prior to retrofit, and $q_{l}'$ is the reliability of edge after retrofit. Eq. \eqref{max_ripw} ensures that the reliability of edge $l$ is $q_{l}'$ if it is selected; otherwise, it does not change. Let $C(x)$ denote the total cost associated with decision $x$. The second objective is to minimize the expectation of total retrofit costs:

\begin{align}
\text{min} \quad C(x) = E_{\widetilde{\xi_{c}}} \left[ \sum_{(i,j)\in A} c_{ij}(\widehat{\xi_{c}})x_{ij} \right] \label{ripw_cost}
\end{align}
where $c_{ij}(\widehat{\xi}_{c})$ is the stochastic realization of random variable retrofit cost $c_{ij}$.

This decision problem is closely related to the stochastic knapsack problem \citep{ross1989stochastic}, but further complicated by the procedure of iteratively computing a series of weighted shortest paths. The non-dominated sorting genetic algorithm II (NSGA-II), successfully applied to several similar problems \citep{bertsimas2002approximate, deb2002fast}, is employed to identify the near-optimal Pareto frontier for this multi-objective optimization problem in which the two objectives conflict.   The author couples NSGA-II with Monte Carlo Simulation (MCS) and optimize the expected RIPW and total costs.

\subsection{Stage II: post-disaster emergency response}
\label{subsec_stage_ii}

In this stage, the problem is to select a number of edges from failed edges and to schedule their restoration sequence, the objective is to minimize the total time to ensure that there exists at least one path between non-emergency nodes and emergency nodes. Let $D = \{1,2,...,b\}$ denote the set of damaged edges (candidates). Let $x_{l}$ and $y_{l}$  respectively denote whether or not to select edge $l \in D$ and the time at which emergency restoration is initiated for edge $l \in D$. The network emergency recovery time associated with the selection $x$ and schedule $y$ is denoted as $t^{E}(x,y)$. Let $s_{l}'$  and $p_{l}(\widehat{\xi}_{p})$ denote serviceability for each edge $l \in D$ after emergency restoration and the stochastic variables of duration of temporal action. Let $T$ denote the set of discrete points in time: $t_{0} \cup \{t1,...,t_b\}$ which is the union of initial time $t_0$ and the intervention completion times $t_{l} = x_{l}\left[y_{l} + p_{l}(\widehat{\xi}_p)\right]$ for each edge $l\in D$. Using the above notation, the emergency accessibility recovery model is presented as follows:

\begin{align}
\text{min} \quad t^{E}(x,y) = E_{\widehat{\xi}_{p}}< \max _{l\in D} \{x_l[y_l + p_l(\widehat{\xi}_p)]\} - t_0> \label{obj_eipw}
\end{align}

subject to:

\begin{align}
\sum_{j\in E}\sum_{k=1}^{K_{ij}}\prod_{l\in P_{ij}^{k}}s_{l}' \geq 1, \forall i \in N \label{eipw_1}\\
s_{l}' = (1-x_{l})s_{l} + x_l, \forall l \in D \label{update_service}\\
\sum_{l\in D}[t \geq y_{l}][t \leq y_l + p_l (\widehat{\xi}_{p})] \leq N_{SI}^{max}, \forall t \in T \label{not_exceed_si} \\
y_{l} \leq Mx_{l}, \forall l \in D \label{big_M}\\
s_{l}' \in \{0,1\}, \forall l \in D \label{service_binary}\\
x_{l} \in \{0,1\}, \forall l \in D \label{selection_binary}\\
y_{l} \geq 0, \forall l \in D \label{schedule_time}
\end{align}

Eq. \eqref{obj_eipw} defines the total emergency accessibility recovery time as the intervention completion time of the very last scheduled edge. Eq. \eqref{eipw_1} requires each non-emergency node to have at least one IPW to an emergency node and Eq. \eqref{update_service} ensures the edge  remains at its initial serviceability unless it is selected for restoration. In Eq. \eqref{not_exceed_si}, $[P]$ is the Iverson bracket, which returns 1 if  is true, and 0 otherwise. $N_{SI}^{max}$ denotes the maximum number of simultaneous emergency restoration interventions possible based on the  the human and financial resources available in the community for emergency recovery during this stage. The parameter $M$ in Eq. \eqref{big_M} is set to a large number and is used to ensure edge $l\in D$ is scheduled for restoration only if it is selected. Note that $M$ must be a sufficiently large number (e.g., greater than any reasonable scheduling times) to not unintentionally reduce the set of feasible solutions.

The problem under investigation is closely related to the NP-hard parallel machine selection and scheduling problem \citep{cheng1996parallel, subramaniam2000machine, cao2005parallel}. It is  assumed that bridge repair scheduling is non-preemptive, that is, once a crew has begun repair on a given bridge, they must complete their work before moving to another bridge. A genetic algorithm (GA) is employed to identify the near-optimal solutions for the emergency accessibility recovery problem.  

\subsection{Stage III: Long-term recovery}
\label{subsec_stage3}

In the final stage, the objective is to restore the transportation system to the pre-disaster condition with minimized total recovery time and maximized efficiency. This chapter utilizes two criteria to evaluate the rapidity and efficiency of the long-term recovery scheduling: total recovery time (TRT) and the skewness of the recovery trajectory (SRT), which are recently introduced by \citet{zhang2017resilience}. It must be noted that the emergency restoration in Stage II is only associated with temporary interventions, that is, any bridge selected in Stage II, has not been repaired for long-term use and must be considered in Stage III. The long-term recovery problem then is to determine an optimal schedule $x=\{x_1,x_2,...,x_b\}$ for the repair of all  damaged edges. Let $r_{l}(\widehat{\xi}_{p})$ denote the realization of the random variable  denoting the duration of the restoration intervention on bridge $l$. Let $t^{R}(x)$ and $t^{s}(x)$ denote the TRT and SRT associated with schedule $x$, respectively. The author sets $T' = \{t_0', t_1',...,t_b'\}$ as the permutation of $T$ such that $t_0' \leq t_1' \leq t_2' \leq ... \leq t_b'$ The long-term recovery model is presented as below:

\begin{align}
\text{min} t^{R}(x) = E_{\widehat{\xi}_{p}}\left\{\max_{l\in B} \left[ x_l + r_l(\widehat{\xi}_p)\right] - t_0\right\} \label{tipw_obj_1}\\
\text{min} t^{S}(x) = E_{\widehat{\xi}_{p}}\left\{ \frac{\sum_{i=1}^{u}t_i' Q_{t_i'}^A (t_i' - t_{i-1}')}{ \sum_{i=1}^{u} Q_{t_i'}^A (t_i' - t_{i=1}') }\right\}
\end{align}

subject to:

\begin{align}
Q_t^A = \frac{1}{n}\sum_{1}^{n-1}\frac{1}{n-1}\sum_{j=1}^{n-1}\sum_{k=1}^{K_{ij}}\widehat{A}_{ij}^{k}\min_{l\in P_{ij}^{k}}\left(1-\frac{d_{l}}{4}\right), \forall t \in T' \label{tipw_recovery}\\
d_l^t = d_l \left[x_l + p_l(\widehat{\xi}_p > t)\right],  \forall t \in T' \\
\sum_{l\in D}[t \geq x_{l}][t \leq x_l + p_l (\widehat{\xi}_{p})] \leq N_{SI}^{max}, \forall t \in T'  \\
x_l \geq 0, \forall l \in D
\end{align}

As discussed in Section \ref{subsec_ripw}, this study uses NSGA-II to search the near-optimal non-dominated solutions for the multi-objective stochastic integer-programming problem.

\section{Stage-wise resilience planning for roadway network in Shelby County, TN}
\label{sec_stage_illustration}

The proposed SWD framework is illustrated using a skeleton highway network in the Shelby County, Tennessee, illustrated in Figure \ref{road_network_shelby}. The system includes 46 edges representing the highways and 34 nodes representing major road intersections and economic hubs; there are 10 hospitals in the study area, 8 of which are located in downtown. There are 24 bridges are considered in this skeleton network, and the detailed descriptions of those bridges are presented in Table \ref{mean_shelby_parameter}. For simplicity, this investigation only consider bridges as the vulnerable network components in the subsequent illustration. 

The chance of a moderate earthquake occurring in the New Madrid Seismic Zone (NMSZ) in the near future is high.  Scientists estimate that the probability of a magnitude 6 to 7 earthquake occurring in NMSZ within the next 50 years is higher than 90\% \citep{hildenbrand1996quantitative}. However, most civil infrastructure in the NMSZ were not seismically designed, as opposed to those in frequent earthquake regions (e.g., California, USA or Japan). The author considers a scenario earthquake with magnitude Mw equal to 7.7 and the epicenter located at 35.3N and 90.3W (on the New Madrid Fault Line) as proposed in the MAE Center study \citep{adachi2007impact}.  A selected ground motion attenuation model \citep{atkinson1995ground} is used to estimate the peak ground acceleration at the site of the bridges.  

\myfigure{
	\centering
	\includegraphics[width=\textwidth]{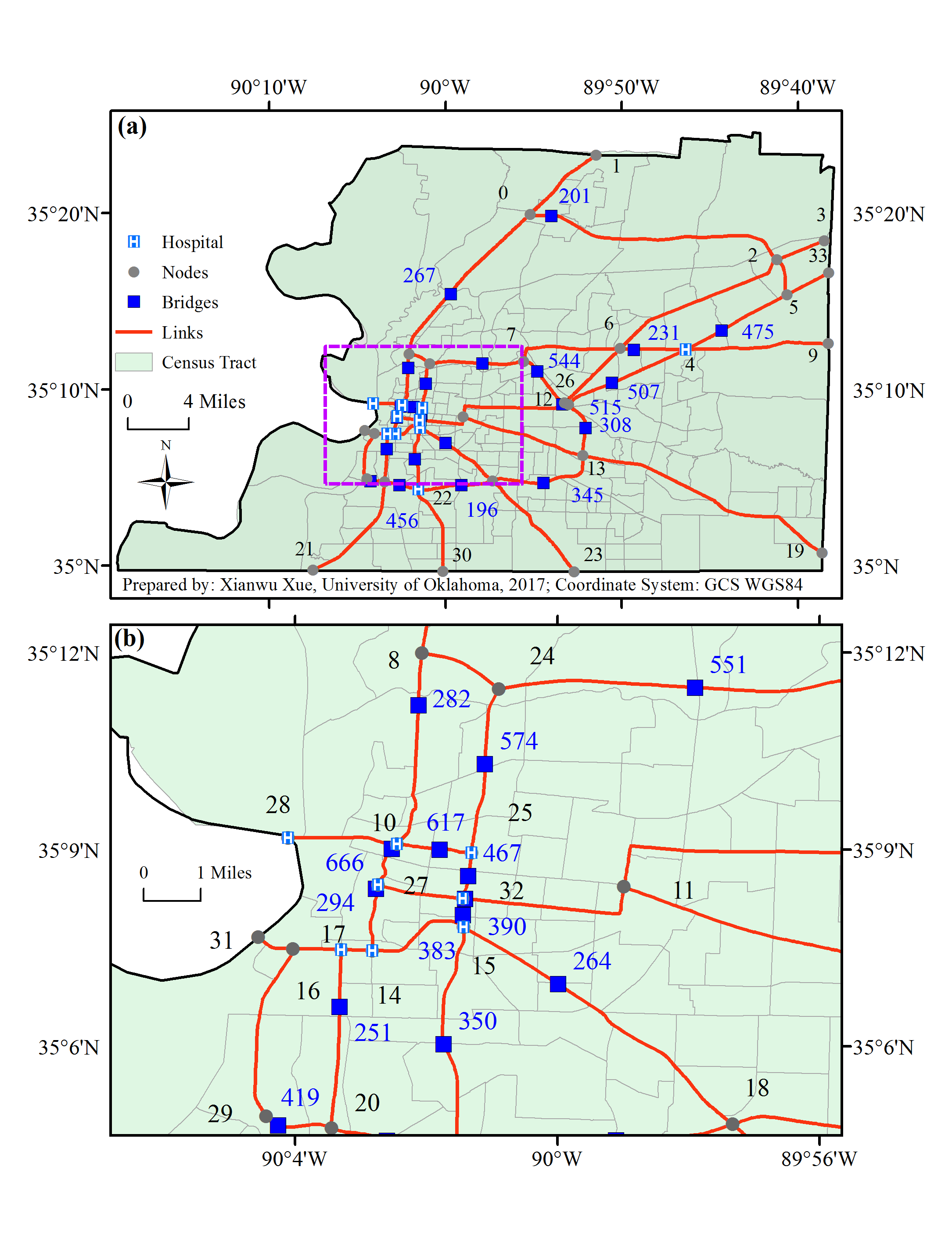}
}{(a) Road network in Shelby County, TN.  (b) Detailed network topology in downtown Memphis, TN.  [The plotting area in (b) is highlighted by the dash-line box in (a)] 
}{road_network_shelby}

The random variables, including ADT ($A_{ij}$), pre-disaster retrofit costs ($c_{ij}$) (assumed to be a function of bridge geometry, material type, year built, and deck area), reliabilities ($q_{ij}$) , expected damage ($d_{ij}$) and the corresponding restoration duration ($p_{ij}$), for each of the bridges under the considered earthquake scenario, are assume to be normal distributions, with mean values listed in Table 1 and an assumed coefficient of variation of 7\% (the implication of this assumption, often made in probabilistic risk analysis, is that the parameters of each bridge in this network can be estimated to within $\pm$ 15\% with more than 95\% confidence).  

\textbf{Stage I: Mitigation Strategy}

When all the bridges in the network are subjected to standard traffic flow (the reliabilities of bridges under service loads are assumed to be 0.99), the RIPW is 1.697, meaning on average there are 1.697 reliable independent pathways (IPW) between all O-D pairs in the network under normal operational condition. When the prescribed scenario earthquake is considered, the mean network RIPW drops to 0.635, indicating many O-D pairs become disconnected following the earthquake event.  Figure \ref{pareto_Front} shows the Pareto front of the optimal retrofit strategies mapped in the objective space with each triangular marker representing a non-dominated solution with respect to a specific combination of financial investment and corresponding RIPW; these non-dominated solutions cannot be improved with respect to one objective function without diminishing the other.

\mytable{
	\begin{tabular}{c | r r r r r}
		\toprule
		Bridge ID	&	Reliability	&	ADT 	&	Retrofit Cost	&	Damage Status	& Restoration	\\
\hline											
196	&	0.774	&	21,500	&	30,796 	&	slight	&	2	\\
201	&	0.786	&	79,910	&	49,031 	&	slight	&	4	\\
231	&	0.781	&	113,830	&	48,708 	&	slight	&	4	\\
251	&	0.74	&	94,920	&	28,799 	&	slight	&	4	\\
264	&	0.793	&	94,130	&	49,708 	&	none	&	0	\\
267	&	0.781	&	99,190	&	6,067 	&	none	&	0	\\
282	&	0.962	&	121,570	&	58,561 	&	none	&	0	\\
294	&	0.887	&	13,260	&	141,115 	&	extensive	&	82	\\
308	&	0.712	&	28,220	&	143,994 	&	moderate	&	24	\\
345	&	0.408	&	84,090	&	121,956 	&	extensive	&	108	\\
350	&	0.638	&	13,770	&	89,911 	&	slight	&	8	\\
383	&	0.716	&	13,490	&	54,971 	&	moderate	&	21	\\
390	&	0.347	&	2,040	&	78,180 	&	complete	&	287	\\
419	&	0.745	&	99,000	&	79,675 	&	complete	&	214	\\
456	&	0.841	&	39,110	&	125,180 	&	moderate	&	22	\\
467	&	0.9	&	6,450	&	44,236 	&	slight	&	6	\\
475	&	0.779	&	14,580	&	24,289 	&	extensive	&	67	\\
507	&	0.519	&	11,320	&	97,841 	&	extensive	&	95	\\
515	&	0.28	&	16,890	&	50,744 	&	complete	&	267	\\
544	&	0.835	&	7,770	&	23,443 	&	none	&	0	\\
551	&	0.336	&	14,020	&	123,765 	&	extensive	&	83	\\
574	&	0.905	&	31,660	&	19,439 	&	moderate	&	14	\\
617	&	0.855	&	16,320	&	25,996 	&	slight	&	2	\\
666	&	0.482	&	49,190	&	306,042 	&	none	&	0	\\
\bottomrule		
\end{tabular}
}{Mean values of bridge parameters } {mean_shelby_parameter}

The specifics of the four strategies on the Pareto front, highlighted in Figure 3, are summarized in Table \ref{optimal_solutions}2. For example, Strategy I in Figure 4 is located toward the bottom left of the objective space and emphasizes cost savings over RIPW.  This strategy identified three bridges (bridges 196, 390, and 515) for retrofit which are located near the center of the roadway system and likely are included in IPWs for several O-D pairs. Furthermore, bridges 390 and 515 (cf. Table \ref{mean_shelby_parameter}) are among the bridges with lowest reliability but their retrofit costs are relatively moderate.  Strategy II moves toward spending more money to improve RIPW and indicates that with a budget of \$447K, 9 bridges can be retrofitted, leading to a RIPW of 1.00.  The significance of a RIPW equal to or greater than 1 is that on average at least one reliable pathway exists between all O-D pairs. The RIPW increases for solutions along the Pareto front as moving towards the upper right of the objective space.  As costs increase, more bridges can be selected for retrofit.  However, the tradeoff between RIPW and retrofit cost is not simply one of adding more bridges.  For instance, bridge 196 is selected in Strategy I, but de-selected in Strategy II and III because neighboring bridges (bridges 345 and 350) are more cost-effective for improving the overall network RIPW; this Bridge 196 is re-selected again in Strategy IV when the retrofit budget become sufficiently high.  Additionally, some bridges are always selected in the optimal retrofit solutions, e.g., bridges 390 and 515, often because their reliability is much lower than average, or they are shared by IPWs between multiple O-D pairs.

\myfigure{
	\centering
	\includegraphics[width=\textwidth]{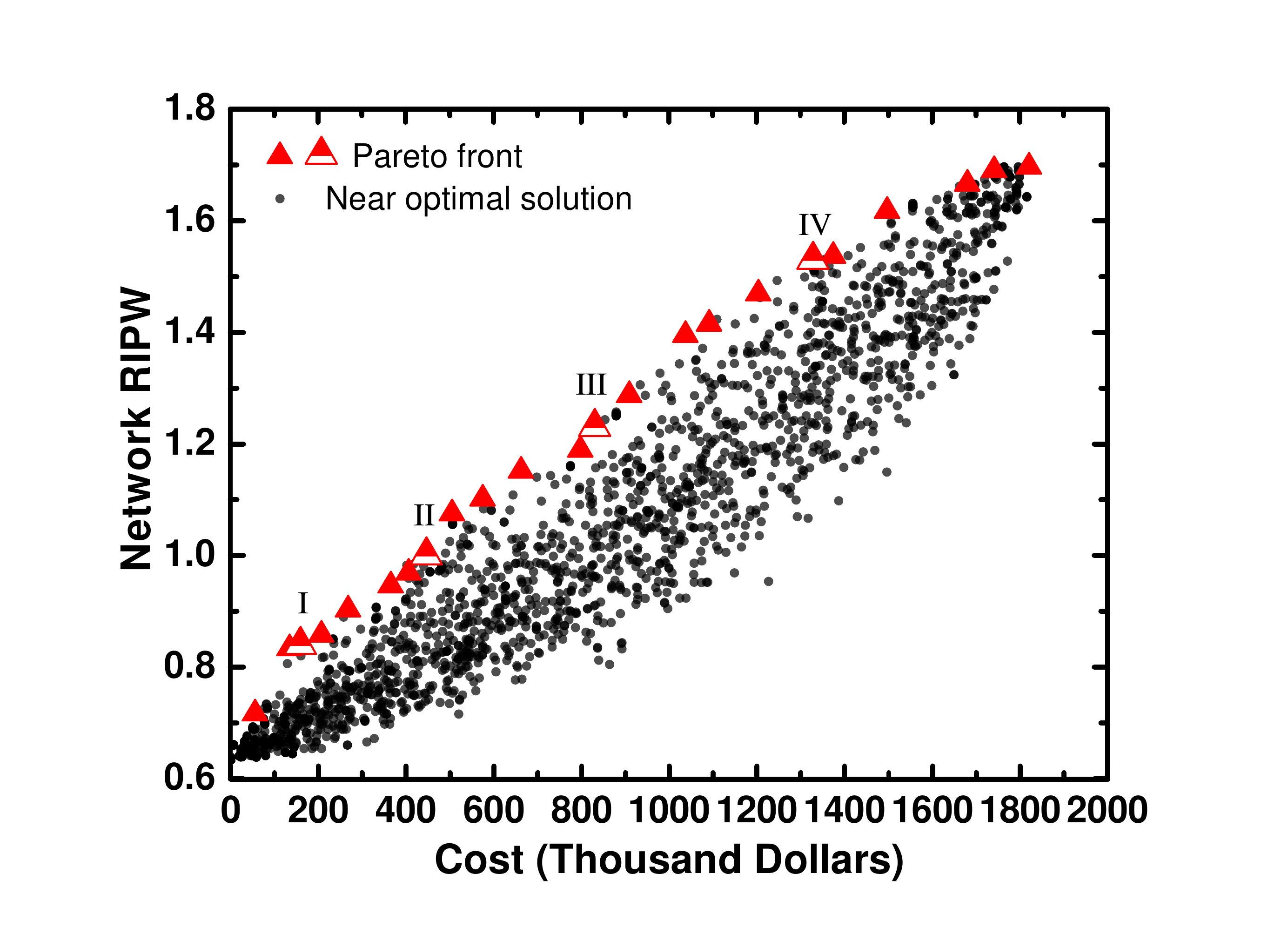}
}{Pareto front: RIPW and Cost}{pareto_Front}

\mytable{
	\begin{tabular}{c  c c c l}
		\toprule
		Strategy	&	$E(\text{cost})$	& $E(\text{RIPW})$	& $\#$ of bridges	&	Bridge selected	\\
		\hline
		I & 160K & 0.82 & 3 & 196, 390, 515\\
		\multirow{2}{*}{II} & 447K & 1.00 & 9 & 267, 350, 383,390,  \\
		&  &  & & 475, 507, 515, 574, 617\\ 
		\multirow{3}{*}{III} & 831K & 1.23 & 12 & 267, 282, 308, 345,  \\
			&  &  & & 383, 390, 467, 475, \\ 
			&  &  & & 507, 515,  551, 617\\
		\multirow{5}{*}{IV} & 1,329K & 1.46 & 20 & 196, 201, 231, 251, \\
			&  &  & & 264, 267, 282, 308,  \\ 
			&  &  & & 345, 350, 383, 390, \\
			&  &  & & 419, 456, 467,  475, \\ 
			&  &  & & 507,  515, 544, 551 \\ 
		\bottomrule		
	\end{tabular}
}{Details of four optimal solutions in Pareto front} {optimal_solutions}

\textbf{Stage II: Emergency Response}

The actual damage condition of the network immediately following a hazard event should serve as the initial state for the Stage II planning. For illustration, the mean damage states tabulated in Table 1 as the starting point are used. As shown in Figure \ref{damage_road_network}, following the scenario earthquake, among the 24 bridges, 5 bridges sustained negligible damage, 7 bridges are slightly damaged, 4 bridges have moderate damage, 5 bridges suffered extensive damage, and 3 bridges fail completely.  Prior to the scenario earthquake the EIPW is 1.89, however, immediately after it falls to 0.55. In particular, nodes 11, 13, and 19 (highlighted in Figure \ref{damage_road_network}), do not have access to any hospital in the Shelby area. it is assumed that the time to establish a temporary path is on average 20 hours and 35 hours for extensively and completely damaged bridges, respectively, with COV of 10\%.  Furthermore, it is assumed a maximum of two repair teams (i.e. $N_{SI}^{max} = 2$) are available for emergency rescue immediately following the disaster.  Our decision space is to schedule the rapid repair for the 8 severely damaged bridges in order to restore the EIPW between all non-emergency nodes and hospitals from 0.55 to a value that is greater than or equal to 1.0 as quickly as possible. 

Figure \ref{emergency_recovery} reveals the best emergency recovery time defined by Eq. \eqref{obj_eipw} at each iteration of the genetic algorithm. The GA converges at the $16^{th}$ iteration with an emergency recovery time of 20 hours. The optimal solution identifies that the bridge 345 is the most critical bridge to bring the EIPW of nodes 11, 13, and 19 up to 1.0 within the shortest time. These three nodes can all use bridge 345 (after emergency restoration) and 264 (with only negligible damage) to access hospitals. While bridge 390 is an alternative candidate (to bridge 345), it is completely damaged and will take longer for emergency restoration. The illustrated approach provides useful information for emergency managers to make informed decisions for timely response and relief efforts.

\myfigure{
	\centering
	\includegraphics[width=\textwidth]{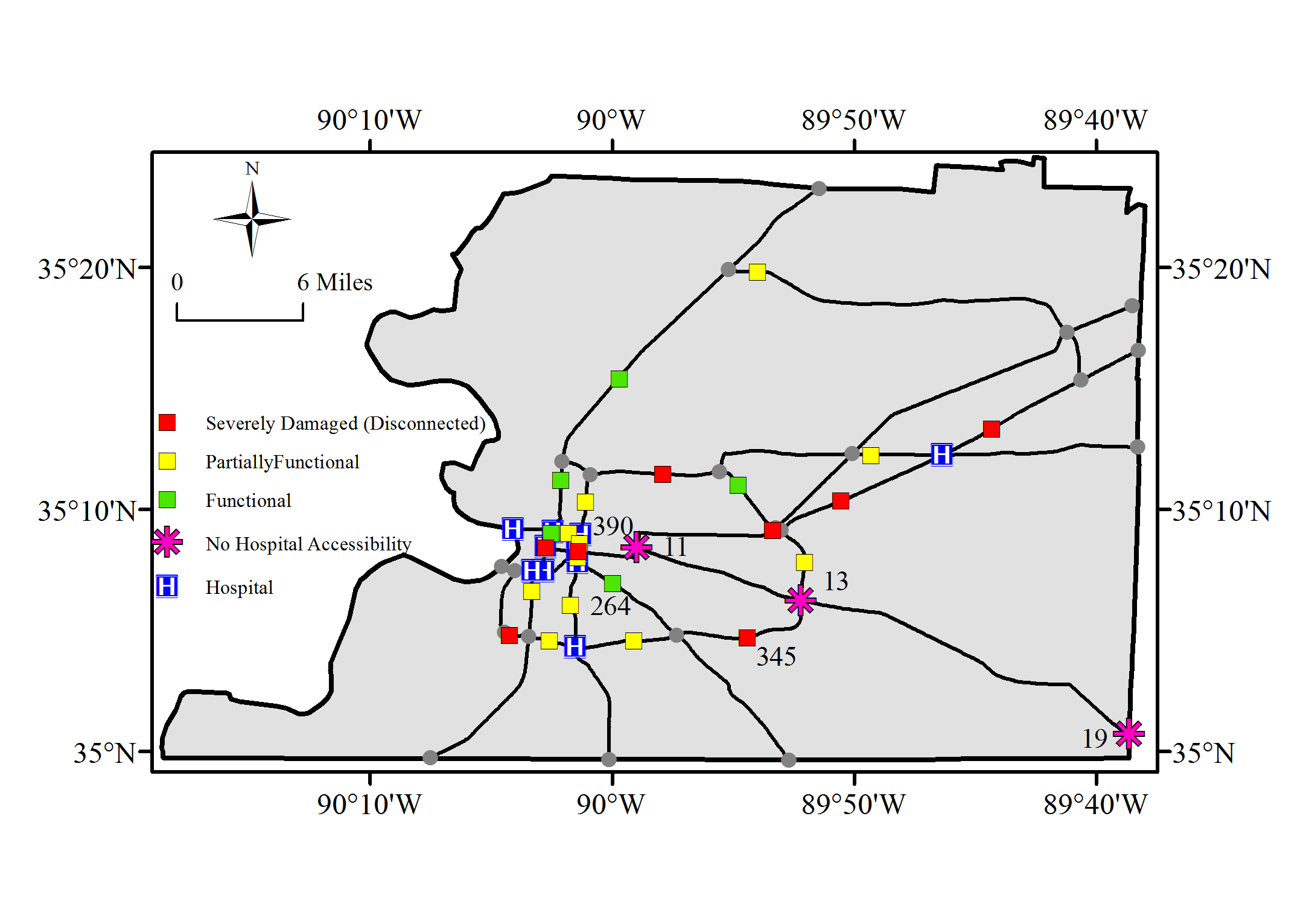}
}{Damage condition of the roadway network}{damage_road_network}

\myfigure{
	\centering
	\includegraphics[width=\textwidth]{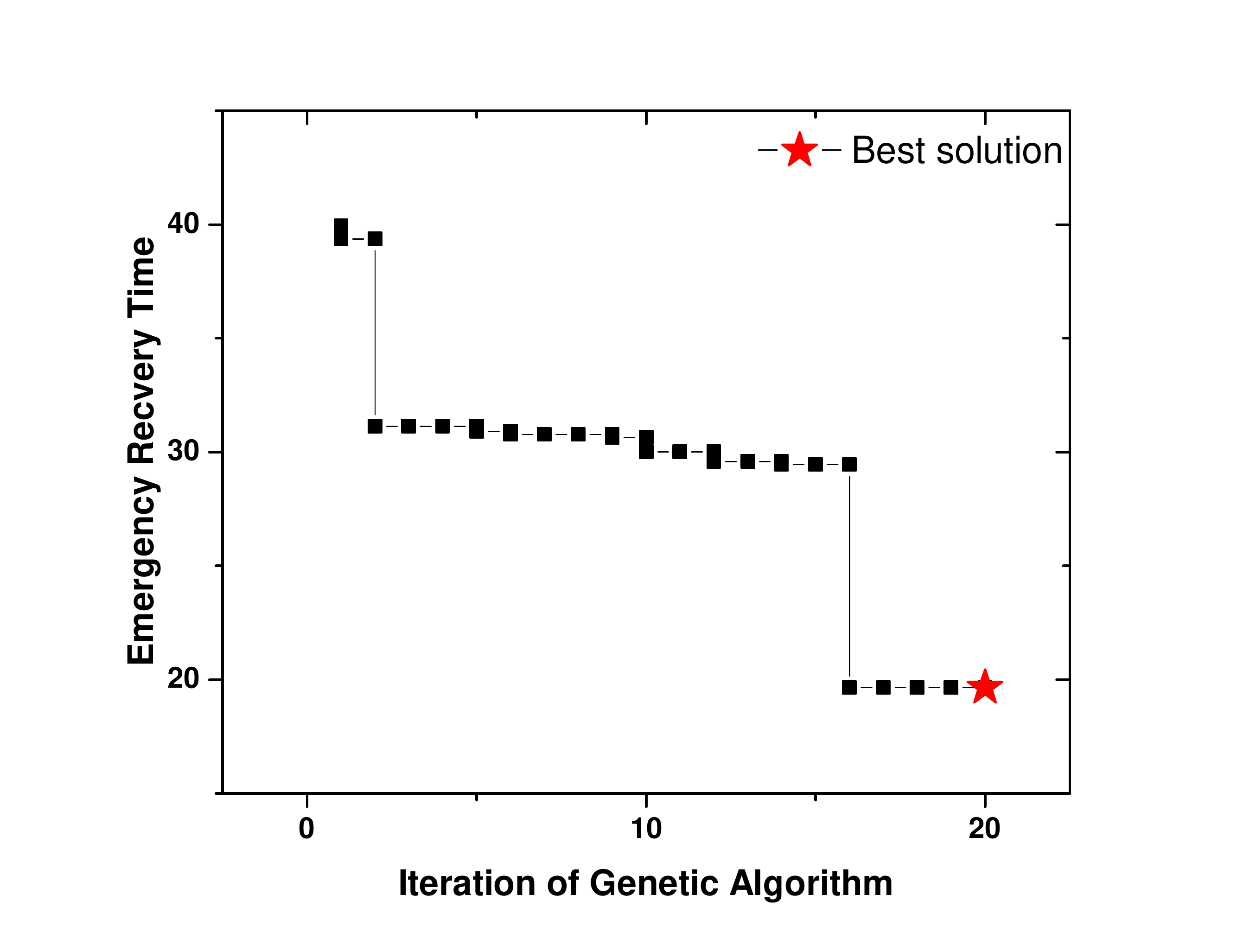}
}{Emergency recovery time over GA iterations}{emergency_recovery}

\textbf{Stage III: Long-term Recovery Scheduling}

The network performance in Stage III is evaluated using TIPW as the performance metric defined by Eq. \eqref{network_tipw}.  The TIPW equals 0.578 following the earthquake event, while the pre-hazard TIPW is 1.752 under normal traffic conditions. The decision problem is to determine the restoration schedule for all the 19 damaged bridges (including slight, moderate, extensive, and complete damage states) to minimize the TRT and SRT. It is assumed limited resources are available and only a maximum of 5 bridges can be repaired simultaneously.

The optimal recovery trajectories are depicted in Figure \ref{recovery_trajectories}. The optimal solution is clearly superior to the suboptimal solution shown, even though they both have the same total recovery time. For example, the optimal solution takes 88 days to improve the TIPW to 1.0, while it takes 178 days for the suboptimal solution to achieve the same results. This confirms that TRT, if used as the sole objective, is not sufficient to ensure the recovery schedule with the ``best'' trajectory. The restoration schedule corresponding to the optimal solution is detailed in Figure \ref{optimal_schedule_tipw}8.  The complete network recovery occurs in 300 days given that the maximum of 5 bridges can be repaired simultaneously. The optimal strategy selects in the early phase 8 slight or moderately damaged bridges associated with large traffic flow (e.g., the ADT of bridge 231 is 113K). The restoration of these 8 bridges improves the network TIPW by 34\% in 34 days as displayed in Figure \ref{recovery_trajectories}.  Bridge 390 (which takes 287 days to restore) and the other two completely damaged bridges (515 and 419) are scheduled early along the time horizon to ensure total recovery time is minimized. These results demonstrate that the optimal scheduling of bridge restoration can greatly improve the efficiency of the transportation network recovery. Although reasonable strategies can be developed by empirical analysis, the advantage of the proposed model would become more apparent when dealing with large, extensively-damaged networks where the decision variables, possible alternative strategies and constraints create a complex decision problem where intuition may fail.

\myfigure{
	\centering
	\includegraphics[width=\textwidth]{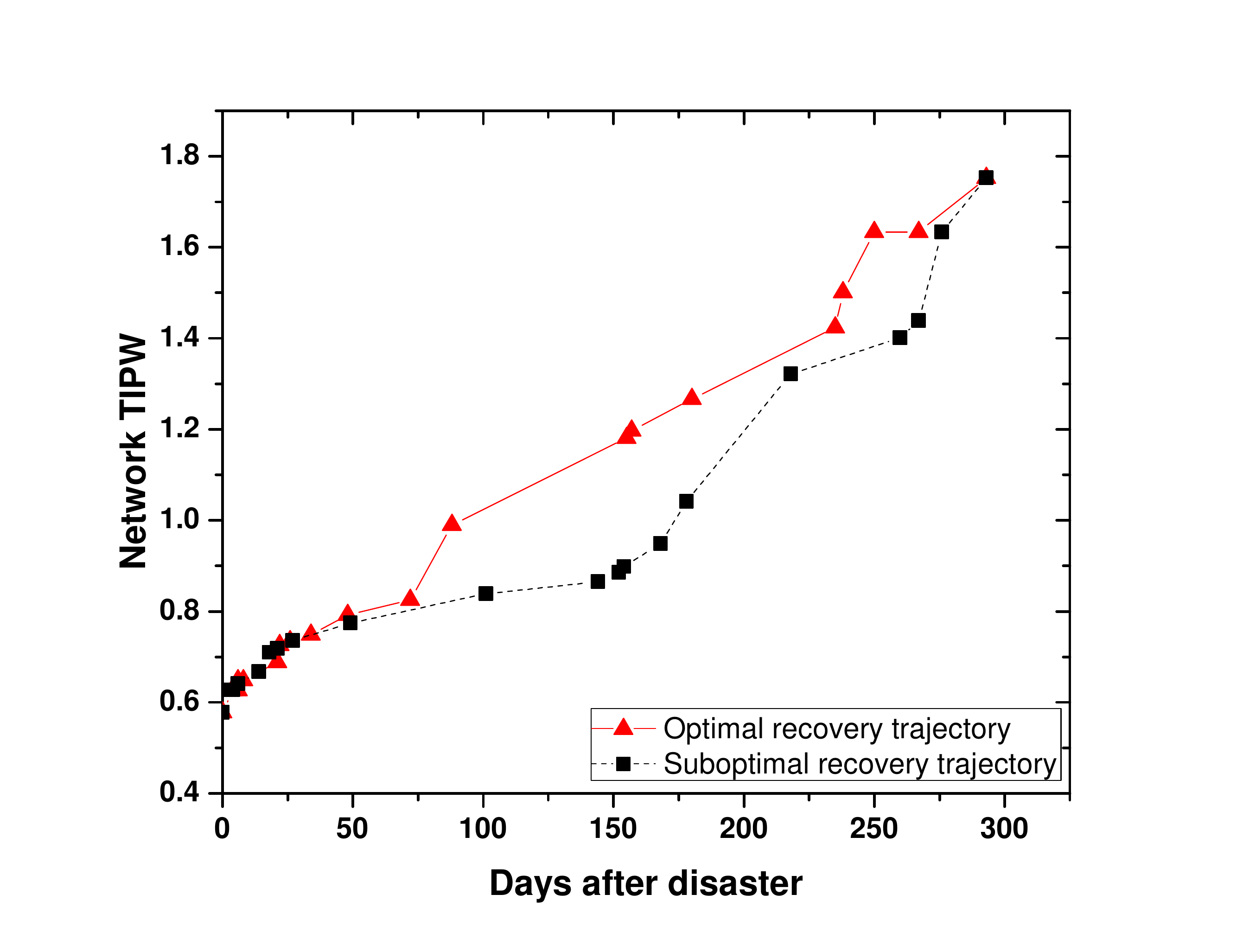}
}{Recovery trajectories with different restoration schedules}{recovery_trajectories}

\myfigure{
	\centering
	\includegraphics[width=\textwidth]{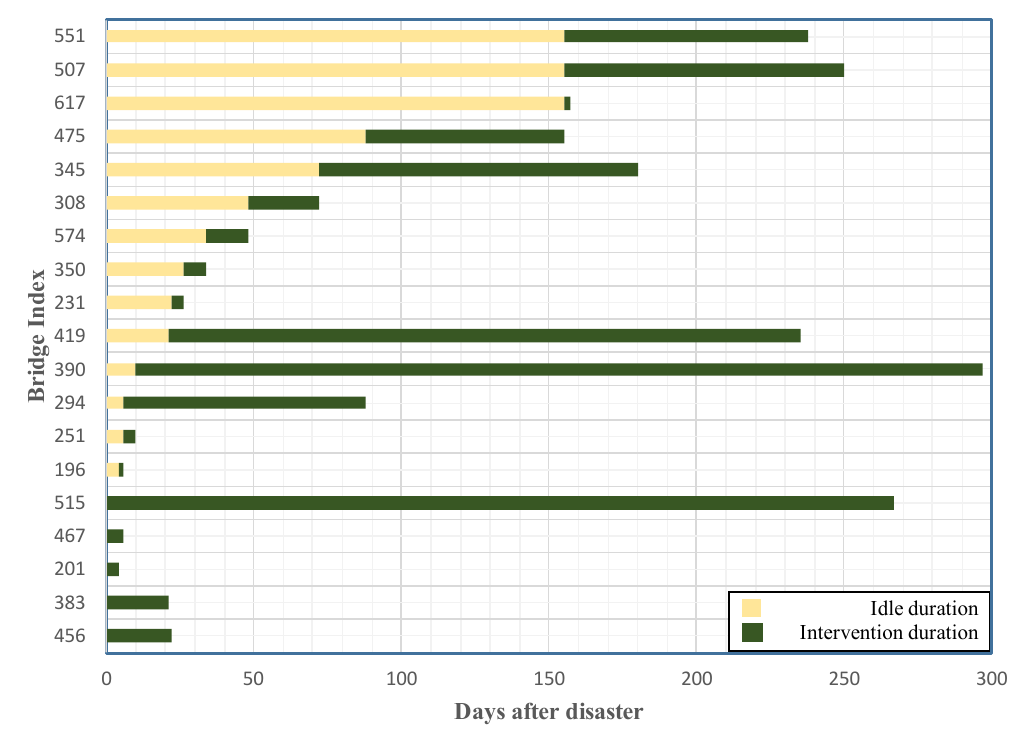}
}{The optimal scheduling for long-term recovery of the network}{optimal_schedule_tipw}

\section{Summary}
\label{sec_summary}

As more and more communities are incorporating resilience as key concept in their day-to-day practices, public decision makers and other stakeholders are likely to be required to take a system perspective concerning physical infrastructure risk management. To that end, this chapter formulates a comprehensive stage-wise decision framework to support the resilience planning activities for roadway network in the three main stages: pre-disaster mitigation (Stage I), post-disaster emergency response (Stage II), and long-term recovery (Stage III). The decision framework includes an IPW-based SWM system and a set of stochastic SWD formulations that are designed to meet the specific objectives and decision purposes of each of the planning phases. Specifically:

	Three quantitative resilience-based, decision-specific network performance metrics: RIPW, EIPW and TIPW are proposed.  RIPW is formulated to measure the network robustness to extreme natural hazards. EIPW is designed to quantify the critical facility accessibility in damaged roadway networks. TIPW represents the efficacy of the network in carrying traffic flow.  These specifically defined SWMs are used in formulating the specific objectives of the decisions in each of the planning phases.

	Three stochastic optimization models are formulated to support decisions in each planning stage. In Stage I, the author develops a multi-objective stochastic binary integer programming problem to select bridges for retrofit with minimized total cost and maximized RIPW. In Stage II,  the author employs a stochastic integer programming problem to select and schedule emergency restoration (temporary path) for damaged bridges constrained by limited resources to ensure all residents have access to hospitals within the shortest time period following extreme events. In Stage III, a multi-objective stochastic integer programming problem is built to optimize the schedule of network recovery interventions by minimizing the total recovery time (TRT) and maximizing the efficiency (SRT) during the recovery period under resource constraints. All these decision problems are NP-hard; the NSGA-II and GA are designed to identify near optimal solutions efficiently and are scalable to handle large and complex networks.

\chapter{Conclusions}
\label{ch-conlusion}

The goal of this dissertation was to develop a systematic methodology to model risk from catastrophic events, explore the performance and disaster resiliency for transportation infrastructure systems, and provide decision support tools for stakeholders. The principal goals listed in Chapter \ref{sec_goals} are accomplished by developing a comprehensive resilience-based metrics system for evaluating the performance of transportation systems, providing effective intervention strategies and recovery schedules to mitigate potential losses from earthquakes, and presented stage-wise metrics and decision frameworks. The major conclusions and contributions from the dissertation are summarized below.

First, network resilience can be improved by increasing reliabilities of critical bridges/roads through appropriate retrofitting, optimizing network redundancy through new construction, by altering traffic flows through appropriate routing policies, and by strategically allocating emergency response facilities.   Stakeholders of communities of different sizes, populations and social-economic vulnerabilities and support systems are likely to reveal different values and preferences in evaluating resilience of their transportation systems since the corresponding network topology, redundancy level, traffic patterns, location of community emergency response facilities as well as failure probability of individual bridges is unique for each community. This study has formulated a quantitative metric, named as weighted number of independent pathways (WIPW), which can not only permit multiple dimensions of resilience for any transportation networks but also allow retrofit alternatives to be evaluated and compared on a common basis. 
Using the WIPW, a decision methodology for effective risk mitigation through bridge retrofitting or new construction is formulated as a multi-objective optimization problem, which shows the Pareto frontier between competing WIPW and cost objectives. The solutions allow decision-maker to select the best strategy with the corresponding subjective preferences in decision process.

Second, this study has presented a novel and dynamic model to optimize the restoration schedules of transportation networks  over the horizon following extreme events using the WIPW as metric of evaluating resilience.  The model incorporates a network resilience-based performance metric, recovery trajectory, community resourcefulness, and uncertainties relating to damage levels and restoration duration of damaged bridges in a quantitative resilience-based decision framework for road network recovery.  In order to quantify the recovery trajectories, two-dimensional metric - total recovery time (TRT) and the skew of recovery trajectory (SRT) -  are developed to measure the restoration schedule. Both the values of TRT and SRT are correlated with the restoration schedule and the available repair crews. The model is solved approximately using the powerful metaheuristic approach of NSGA-II. The case study in Chapter \ref{ch_recovery} reveals that:
(1) Different prioritization schedules can lead to significant changes in both TRT and SRT.  The value of TRT and SRT are positive related and SRT can be useful to ensure concaveness of the recovery trajectory. This two-dimensional metric can be used collectively as the objectives in identifying the optimal post-event scheduling for the network recovery.  
(2) According to Figure \ref{recovery_trajectory} and Figure \ref{recovery_trajectoy_multi_capacity}, simple functions (linear or nonlinear) for recovery trajectory apparently do not exist in even simple networks such as the one considered herein. The dynamic nature of available resources (immediately on-hand or delayed due to outsourcing) can have a notable impact on the network recovery time and trajectory.  Even under circumstances in which the total amount of resources utilized in the recovery period is the same, the recovery trajectory is sensitive to the timing in which such resources are made available.   
(3) The distribution of recovery time under stochastic conditions, the intelligent network restoration scheduling (intervention time and completion time) and the corresponding recovery trajectory can be visualized easily, which provides a wealth of information for bridge authorities to make decisions regarding the post-disaster recovery of transportation networks.  
  
Finally, this study introduces a comprehensive stage-wise decision framework to support the resilience planning activities for roadway network in the three main stages: pre-disaster mitigation (Stage I), post-disaster emergency response (Stage II), and long-term recovery (Stage III).  The decision framework includes an IPW-based SWM system and a set of stochastic SWD formulations that are designed to meet the specific objectives and decision purposes of each of the planning phases. In specific: (a) the WIPW is decomposed to a stage-wise metric systems with three metrics, i.e., RIPW (for evaluating network robustness to extreme natural hazards before hazards), EIPW (for quantifying critical facility accessibility in damaged roadway networks), and TIPW (for measuring efficacy of the network in carrying traffic flow);
(b) the corresponding optimization models are formulated to support decisions in each planning stage. In Stage I, a multi-objective stochastic binary integer programming problem to select bridges for retrofit with minimized total cost and maximized RIPW is developed. In Stage II, this study employs a stochastic integer programming problem to select and schedule emergency restoration (temporary path) for damaged bridges constrained by limited resources to ensure all residents have access to hospitals within the shortest time period following extreme events. In Stage III, a multi-objective stochastic integer programming problem is built to optimize the schedule of network recovery interventions by minimizing the total recovery time (TRT) and maximizing the efficiency (SRT) during the recovery period under resource constraints. All these decision problems are NP-hard; the NSGA-II and GA are designed to identify near optimal solutions efficiently and are scalable to handle large and complex networks. 

The resilience assessment and modeling of transportation infrastructure systems cover a broad
spectrum of research areas and only some of the components were studied in detail in this research. The limitations of this research require additional research in the future, especially in the topics given below. First, this research assumed that bridges are the only vulnerable components in the transportation systems, but the vulnerability of other components should be involved in the future research. Second, this study focused on scenario of seismic damage. However, different types of hazards may cause different patterns of damage and requires modifying the model to accommodate it.  For example, the elevation of roadways should be incorporated in emergency response. Finally, the interdependency between transportation networks and other utility infrastructures are not considered in this study. Although transportation is normally built and operated as a stand-alone system, in reality they depend on the power and telecommunication networks. For instance, traffic lights and metro systems require the input of power and signal  to be functional. Future research could be beneficial to understand the interdependency between transportation and other infrastructures to better evaluate and improve the entire community resilience.

\newpage

\addcontentsline{toc}{chapter}{References}
\bibliographystyle{plainnat}
\bibliography{ref2}
\renewcommand{\refname}{References} 
\end{document}